\documentclass[12pt]{article}
\usepackage{pslatex}
\usepackage{amssymb}
\usepackage{graphicx}
\usepackage{color}
\makeatletter
\renewcommand{\theequation}{\thesection.\arabic{equation}}
\@addtoreset{equation}{section} \makeatother

\newcommand{\bea}{\begin{eqnarray}}
\newcommand{\eea}{\end{eqnarray}}
\newcommand{\benu}{\begin{enumerate}}
\newcommand{\eenu}{\end{enumerate}}
\newcommand{\nn}{\nonumber \\}

\def\d{\delta}
\def\dd{\rm d}

\newcommand{\bra}[1]{\langle{#1}|}
\newcommand{\ket}[1]{|{#1}\rangle}

\def\Tr{{\rm Tr}}
\def\tr{{\rm tr}}
\def\det{{\rm det}}
\def\dd{{\rm d}}

\def\bra{\langle}
\def\ket{\rangle}

\newcommand{\limit}{\rightarrow}

\def\vectt[#1,#2]{\left(%
\begin{array}{c} #1 \\ #2 \end{array} \right)}

\def\trivectt[#1,#2,#3]{\left(%
\begin{array}{c} #1 \\ #2 \\ #3 \end{array} \right)}

\def\d {{\rm d}}

\def\wh {\widehat}
\def\wt {\widetilde}
\def\ttr {{\tt R}}

\renewcommand{\theequation}{\arabic{section}.\arabic{equation}}

\newcommand{\gsim}{\mathrel{\mathop{\kern 0pt \rlap
  {\raise.2ex\hbox{$>$}}}
  \lower.9ex\hbox{\kern-.190em $\sim$}}}
\newcommand{\lsim}{\mathrel{\mathop{\kern 0pt \rlap
  {\raise.2ex\hbox{$<$}}}
  \lower.9ex\hbox{\kern-.190em $\sim$}}}

\begin{document}

\begin{titlepage}

\begin{flushright}
%ArXiV:1002.4636
\end{flushright}
%\vspace*{-0.5cm}
%
\centerline{\Large\bf A Nonperturbaive Proposal for Nonabelian Tensor Gauge Theory}
\vskip0.3cm
\centerline{\Large \bf and}
\vskip0.3cm
\centerline{\Large \bf Dynamical Quantum Yang-Baxter Maps
}
\centerline{\Large\bf} \vspace*{1cm}
\centerline{\large Soo-Jong Rey $^{a,b}$ \ \ \small and \large \ \  Fumihiko Sugino $^{c}$}
\vspace*{1cm}
\centerline{\sl $^a$ School of Physics \& Astronomy, Seoul National University, Seoul 151-747 {\rm KOREA}}
\vskip0.25cm
\centerline{\sl $^b$ School of Natural Sciences, Institute for Advanced Study, Princeton NJ 08540 {\rm USA}}
\vskip0.25cm
\centerline{\sl $^b$ Okayama Institute for Quantum Physics, Kyoyama 1-9-1, Kita-ku, Okayama 700-0015
{\rm JAPAN}}
\vskip0.25cm
\centerline{\tt sjrey@snu.ac.kr \,\,\, fumihiko-sugino@pref.okayama.lg.jp }
\vspace*{1.5cm}
\centerline{\rm ABSTRACT} \vspace*{0.5cm} \noindent
We propose a nonperturbative approach to nonabelian two-form
gauge theory. We formulate the theory on a lattice in terms of
plaquette as fundamental dynamical variable, and assign U($N$)
Chan-Paton colors at each boundary link. We show that, on 
hypercubic lattices, such colored plaquette variables constitute
Yang-Baxter maps, where holonomy is characterized by certain
dynamical deformation of quantum Yang-Baxter equations. 
Consistent dimensional reduction to Wilson's lattice gauge
theory singles out unique compactness condition. We study a class of
theories where the compactness condition is solved by Lax pair ansatz. 
We find that, in naive classical continuum limit, these theories recover 
Lorentz invariance but have degrees of freedom that scales as
$N^2$ at large $N$. This implies that nontrivial quantum continuum limit must be sought for. 
We demonstrate that, after dimensional reduction,
these theories are reduced to Wilson's lattice gauge theory.
We also show that Wilson surfaces are well-defined physical observables 
without ordering ambiguity. Utilizing lattice strong coupling expansion, 
we compute partition function and correlation functions of the Wilson surfaces. 
We discover that, at large $N$ limit, the character expansion coefficients 
exhibit large-order behavior growing faster than exponential, in
striking contrast to Wilson's lattice gauge theory. This hints a
hidden, weakly coupled theory dual to the proposed tensor gauge
theory. We finally discuss relevance of our study to topological quantum order
in strongly correlated systems.

%\vspace*{1.1cm}

\end{titlepage}

%%%%%%%%%%%%%% Section 1: Introduction  %%%%%%%%%%%%%%%%%%
\section{Pictures and Discussions}
\rightline{\sl "What is the use" {\rm thought Alice} "of a book if
it has} \rightline{\sl no picture or conversation?" --- \rm Louis
Carroll} \hfill\break
An outstanding problem in theoretical physics is a constructive
definition of $p$-form gauge theories, especially, nonabelian and
self-interacting ones. Variants of $p$-form gauge theory arise in
diverse contexts, ranging from string or M theories \cite{kalbramond, polchinski} and 
higher-dimensional integrable systems \cite{alvarez} to topological order and 
phases in strongly correlated systems \cite{stringnet} and to quantum error 
correction codes \cite{errorcorrection} in quantum information sciences. 
Of particular interest is whether a nonabelian $p$-form
gauge theory exists and, if so, what sort of self-interactions are
allowed by the gauge invariance. To the problem posed, one's first
guess is that the fundamental degrees of freedom are some sort of
nonabelian extension of the abelian $p$-form gauge theory, but
then the question is now more to "what types of nonabelian
extensions can be endowed to abelian $p$-form gauge theory?" and
to "what types of self-interaction are possible for a given
nonabelian extension?".

The ${\mathbb{Z}}_2$ Ising model provides the simplest situation
of all. Consider the model defined on a $d$-dimensional hypercubic
lattice. In dual formulation, the Ising model is mapped to a gauge
theory, where the gauge potential is a $p=(d-2)$-form and takes a
value in $G = {\mathbb{Z}}_2$. As is firmly established, the Ising
model does not exhibits any nontrivial renormalization group fixed
point for $d \ge 4$. It implies that the dual $p$-form gauge
theory in $d \ge 4$ ought to be free in the continuum limit,
yielding no obvious self-interaction among the $p$-form gauge
potentials \footnote{The Ising fixed point exhibits strong
stability under renormalization group flow. For instance, even
dense dilution over simplices of nonzero codimensions is unable to
induce a flow away from the Ising fixed-point \cite{parisi}. This
implies that the dual $p$-form gauge theory {\sl remains} free
even for randomly distributed coupling parameter.}.

In string and M theories, variety of the $p$-form gauge potentials is
rich and complex. Foremost, all known string theories, whether
supersymmetric or not, contain universally the Kalb-Ramond
two-form potential $B_2 = {1 \over 2!} B_{mn} \d x^m \wedge \d
x^n$, and the fundamental string couples minimally to it
\cite{kalbramond}. Type I and II superstring theories contain in
addition R-R(Ramond-Ramond) $p$-form potentials \cite{polchinski}. The
D$(p-1)$-branes are the charged objects coupled minimally to these
R-R potentials. Because of different chirality projection, type
IIA superstring gives rise to $p=1,3, \cdots, 9$ only, while Type
IIB superstring does so for $p=0,2, \cdots, 10$ only. These NS-NS
and R-R fields exhaust all possible $p$-form potentials permeating
through the ten-dimensional bulk spacetime. To the extent
understood so far, they are essentially abelian.

There are, in addition, $p$-form potentials residing only on the
worldvolume of string solitons. Consider first the D-branes, the
objects minimally coupled to the R-R gauge potentials. At
low-energy below the string scale, D-brane worldvolume dynamics is
dominated by the lowest excitation of open strings whose both ends
are attached to the D-brane. The excitation constitutes one-form
($p=1$) potential $A_m(x) \d x$ of gauge group U(1) and free
scalar fields. A novelty is that, when $N$ parallel D-branes stack
on top of one another, combinatorially, there are $N \times N$
possible {\sl open} strings and the lowest excitation of them
forms U($N$) matrix-valued 1-form potential \cite{wittencp}. Thus, 
for D-branes, $N^2$ variety of open strings
constitutes the microscopic {\sl field} and {\sl particle} degrees
of freedom of the D-brane worldvolume. Consider next the
NS5-brane, the magnetic dual to the fundamental string.
Worldvolume dynamics of an NS5-brane in Type IIA string theory,
equivalently, an M5-brane in M-theory is described at low-energy
by the six-dimensional ${\cal N}=(2,0)$ superconformal theory \cite{20theory}, and
the theory is known to contain a selfdual two-form ($p=2$)
potential of gauge group U(1). Again, a novelty is that, when $N$
parallel NS5-branes or M5-branes stack on top of one another, the
microscopic degrees of freedom constitute tensionless strings that
arise in M-theory from the variety of {\sl open} M2-branes
connecting all possible combinatoric pairs of the $N$ M5-branes \cite{tensionless}.
Similar to the D-branes, one might anticipate that $N^2$ variety
of open M2-branes connecting all possible pairs of M5-brane
constitute the microscopic {\sl field} and {\sl string} degrees of
freedom of the M5-brane worldvolume. However, in stark contrast, various
considerations ranging thermodynamic free energy \cite{klebanovtseytlin} 
and gravitational anomaly cancellation \cite{mooreanomaly} 
all indicate a peculiarity that the degrees
of freedom are intrinsically quantum-mechanical and scales in $N
\rightarrow \infty$ limit as $N^3$, in stark contrast to $N^2$ behavior \cite{n4thermo}
observed 
for the D3-brane worldvolume dynamics described by the ${\cal N}=4$ 
super Yang-Mills theory. 

In this paper, we study a viable extension of nonabelian gauge
invariance to tensor gauge field and put forward a 
nonperturbative approach for tensor gauge theory in $d \ge 4$
dimensions~\footnote{In this paper, we study exclusively two-form
gauge theory --- as will become evident in foregoing discussions,
the construction is extendible to higher $p$-form gauge theories
straightforwardly.} by putting the theory on a
$d$-dimensional hypercubic lattice. The lattice formulation allows
a tranparent and concerete description for origin of nonabelian gauge
symmetries and self-interactions among the $p$-form gauge
fields~\footnote{There has been in the past occasional attempt for
constructing nonabelian $p$-form gauge theories. See \cite{Baez} and 
also \cite{others}.}. By putting the theory on a
lattice, we are sidestepping from other structures that can be
endowed to the tensor gauge theory such as (extended)
supersymmetry or (anti)self-duality. Though these structures are
desirable for making contact with those arising in string theory,
in this paper, we focus primarily on the issue of nonabelian gauge 
symmetry and self-interactions thereof.

This paper is organized as follows. We begin in section 2 with
etiology of Chan-Paton factors. We assign Chan-Paton factors to
boundary links of an elementary plaquette so that it carries four
`color' indices. We take these objects as fundamental dynamical
variables and construct in section 3 a nonabelian two-form tensor
gauge theory defined on a $d$-dimensional hypercubic lattice.
Expressing the plaquette variable as $(N^2 \times N^2)$ matrices
of U($N$) gauge group, we construct action for nonabelian tensor 
gauge theory. We then study possible compactness conditions. 
We show that consistent reduction to Wilson's lattice gauge theory \cite{wilson}
by a dimensional reduction and unitarity or reflection-positivity 
singles out a unique choice of the condition.
In section 4, we study ground state of the lattice
theory. We show that ground-state configurations are provided by 
vanishing holonomy of the nonabelian tensor fields. Remarkably, 
these configurations are specified by the solution of so-called 
dynamical quantum Yang-Baxter equations, pointing to an extremely 
rich structure of the ground-state wave function. 
Nontrivial holonomies are measured by a set of possible
deformations of the dynamical quantum Yang-Baxter equations and their 
solutions. In section 6, we study Lax pair ansatz for solving the
compactness condition by parametrizing the plaquette variables in terms 
of direct product of $(N \times N)$ matrices of U($N$) group. 
We then study continuum
limit of the classical action, viz. naive continuum limit and
point out that the gauge symmetry becomes abelian in this limit.
In section 6, we show that the theory passes up an important
consistency condition: upon lattice dimensional reduction, the
theory is reduced to the Yang-Mills theory in the classical
continuum limit. In section 7, utilizing the lattice strong-coupling
expansion, we compute free energy and Wilson surface correlators. 
We find that in stark contrast to Polyakov-Wilson lattice gauge theory,
the expansion series is not Borel summable even in the large $N$ limit. 
We present an intuitive argument for the behavior and argue that the 
large-order behavior point to the existence of a weakly interacting, 
dual lattice theory. Section 8 is
devoted to a summary of this paper and discussion on remaining
issues. In Appendix A, we explain that a truncation of the
internal degrees of freedom following the dimensional reduction
leads the Wilson's lattice gauge theory at the lattice level the
plaquette variables to the Wilson's link variables. Consistency
with this procedure reduces the original four possibilities of the
theory into two. In Appendix B, we analyze the rank-$N$ theory and
see that it does not lead a physically meaningful continuum theory
under the dimensional reduction. In Appendix C, we explain
solutions of the compactness condition of the rank-$N^2$(I) which
are not expressed as a direct-product structure. Although the solutions 
have the ${\cal O}(N^3)$ degrees of
freedom, the gauge degrees of freedom reduce to $\left({\rm
U}(1)\right)^N$, which does not lead to an interesting nonabelian
tensor theory. Also, since the gauge degrees of freedom are not
enough to eliminate modes of wrong-sign kinetic terms (ghosts), the
solutions do not seem to lead physically meaningful continuum
theory at least in the classical level. In Appendix D, we present
some computation on a character expansion coefficient used in
strong coupling expansion.

\newpage
%%%%%%%% section 2: chan-paton factors %%%%%%%%%%%%%%%%%%%%%%%%%%
\section{Etiology of Chan-Paton Factors}
%%%%%%%%%%%%%%%%%%%%%%%%%%%%%%%%%%%%%%%%%%%%%%%%%%%%%%%%%%%%%%%%%
We begin with etiology of the Chan-Paton factors \cite{chanpaton}.
Originally, the Chan-Paton factors were introduced as a
prescription for introducing ``colors" to open string amplitudes.
In this section, we will adapt the notion on a lattice and
interpret it as endowing a Chan-Paton bundle over a finite-dimensional
vector space $V_{\rm CP}$ to parallel transport.

\subsection{elementary link variables}
Consider the Wegner-Wilson-Polyakov formulation of the lattice gauge
theory \cite{wilson} on a $d$-dimensional hypercubic lattice. Taking
gauge group $G$ to be compact $U(1)$, elementary degrees of freedom are
associated with link variables, $U(x, x + \hat{\mu})$. This is a dynamical
variable residing at each unit link (of length $a$) $[x;\hat{\mu}]$ of
direction $\hat{\mu}$, connecting the nearest neighbor sites $x$ and $x +
\hat{\mu}$. The link variable assigns a dynamical weight to
parallel-transporting a charged particle (charge $+1$) long the
unit link $[x;\hat{\mu}]$. The link variable $U_\mu(x)$ is parametrizable in
terms of gauge potential $A_\mu(x)$ that takes values on $\mathbb{S}^1$:
\bea U_\mu (x) \equiv U(x, x + \hat{\mu}) = \exp ( i a A_\mu(x)),
 \label{abelian1} \eea
By the exponential map defined so, the link variable satisfies the compactness condition:
\bea U^{-1}(x, x+\hat{\mu}) = U(x+\hat{\mu}, x) .
\label{abeliancompactness} \eea
As the charged particle is parallel-transported, $U_\mu(x)$
creates the electric charge $-1$ at the site $x$ (where a unit
charge is depleted) and $+1$ at the site $x + \hat{\mu}$ (where a
unit charge is deposited). For an arbitrary path on the lattice,
the corresponding parallel-transport is determined by multiplying
link variables along the path. Therefore, the set of all possible
link variables $\{ U_\mu (x) \}$ constitutes microscopic degrees
of freedom on the lattice. Of particular interest is the
parallel-transport around a unit plaquette $[x; \hat{\mu},
\hat{\nu}]$:
\bea U_{\rm P} \equiv U(x, x+\hat{\mu}) U(x + \hat{\mu}, x +
\hat{\mu} + \hat{\nu}) U(x + \hat{\mu} + \hat{\nu}, x + \hat{\nu})
U(x + \hat{\nu}, x). \nonumber \eea
This is a gauge-invariant operator, and deviation of it from unity
measures the holonomy around the unit plaquette $[x; \hat{\mu},
\hat{\nu}]$.

\begin{figure}[htbp]
    \begin{center}
    \includegraphics[width=0.5\linewidth,keepaspectratio,clip]
      {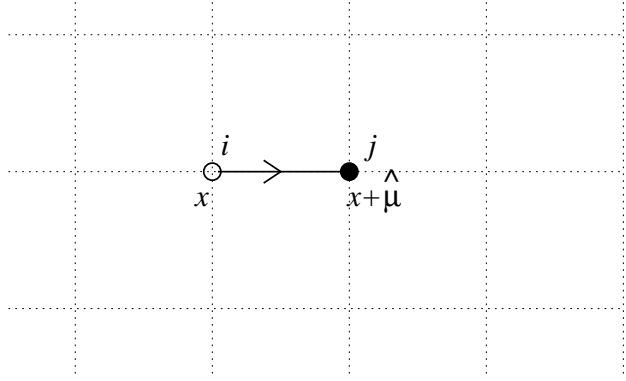}
    \end{center}
%\begin{figure}
%\epsfxsize=10cm \epsfysize=5.5cm \centerline{\epsfbox{link2.eps}}
\caption{\sl The link corresponding to the variable
$(U_{\mu}(x))_{ij}$. The Chan-Paton indices $i,j$ are assigned as
quantum numbers labelling the 0-simplices at the two ends of the
1-simplex defining the link variable.} \label{fig:link}
\end{figure}

The construction is readily extendible by adjoining Chan-Paton factors.
We promote the link variable to a matrix-valued one. It comes through
a pair of the Chan-Paton factors $i, j \in V_{\rm CP}$ attached at
the two ends of each unit link at $x$:
\bea U_\mu (x) \otimes | \, i j \rangle \equiv \Big(U_\mu(x)
\Big)_{ij} \label{nonabelian1}\eea
The $i,j = 1, \cdots, N$ indices label an orthogonal basis of the
vector space $V_{\rm CP}$, so equivalence relations and hence the
gauge group $G$ would be associated with the rational map
\bea G : V_{\rm CP} \rightarrow  V_{\rm CP}. \nonumber \eea
As such, the elementary link variables are matrix-valued in GL($N,
\mathbb{C}$). See Fig.\ref{fig:link}.

The Chan-Paton etiology can be restated as follows. The sites and
links are 0- and 1-simplices of the $d$-dimensional hypercubic
lattice. The two sites at the end of a link variable are boundary
0-simplices of a 1-simplex. To define nonabelian lattice gauge
theory, we are then assigning the Chan-Paton factors $i, j, \cdots
\in V_{\rm CP}$ at each boundary 0-simplex of 1-simplices. The
matrix-valued holonomy is measured around the boundary of each
plaquette. For the unit plaquette located at $x$ and
oriented along $[\hat{\mu}\hat{\nu}]$-directions, the holonomy
is measured by
\bea \Big( U_\mu (x) \Big)_{ij} \Big( U_\nu (x + \hat{\mu})
\Big)_{j \, k} \Big( U_{-\mu} (x + \hat{\mu} + \hat{\nu}) \Big)_{k
\ell} \Big( U_{-\nu} (x + \hat{\nu}) \Big)_{\ell m} .  \nonumber
\eea
It measures the gauge flux felt by a colored particle around
the closed loop of 1-simplices, transmuting the color through $i
\rightarrow j \rightarrow k \rightarrow \ell \rightarrow m$. In
the context of open strings ending on D-branes, the colored
particle is nothing but an endpoint of the open string.

\vskip0.3cm
%%%%%%%%%%%%%%%%%%%  Fig: link %%%%%%%%%%%%%%%%%%%%%%%%
%%%%%%%%%%%%%%%%%%%%%%%%%%%%%%%%%%%%%%%%%%%%%%%%%%%%%%%
\vskip0.3cm
The elementary link variable $(U_\mu(x))_{ij}$ evolves dynamically
and sweeps out a plaquette $P$ (of area $a^2$) inside the
$d$-dimensional hypercubic lattice. The Chan-Paton factor should
be consistent with various closure relations. First, a product of
two link variables should be isomorphic to another link variable:
\bea \sum_{j=1}^N \Big(U_\mu(x) \Big)_{ij} \cdot \Big(U_\mu(x+\wh
\mu) \Big)_{jk} := \Big( U_\mu(x) U_\mu(x+\wh\mu)\Big)_{ik}. \eea
Second, generalizing Eq.(\ref{abelianunitarity}), unitarity of the
link variable forces the compactness condition:
\bea \sum_{j=1}^N \Big(U_\mu (x) \Big)_{ij} \Big( U_\mu (x)
\Big)^*_{kj} = \delta_{ik}. \label{linkcompactness} \eea
This reduces the space of link variables from GL($N, \mathbb{C})$
to U($N$). Third, the gauge invariance requires that
\bea \Big( U_\mu (x) \Big)_{ij} \rightarrow \Big(V(x)\Big)_{ik}
\Big( U_\mu (x)\Big)_{kl} \Big(V^\dagger(x + \wh \mu)\Big)_{lj}
\eea
is an equivalence relation.

\subsection{elementary $p$-simplex variables}
The above consideration is readily generalizable to higher simplices.
For link variables, crux of the idea to the Chan-Paton factor was that `color' degrees
of freedom is attached to the boundaries
(0-simplices) of the link variables (1-simplices). As the boundary consists of two
elements, the resulting Chan-Paton gauge groups are simply matrix group GL($N$). We shall
now extend this notion to higher-dimensional simplices, viz. assign Chan-Paton indices to
the $(p-1)$-simplicial boundaries of a $p$-simplex.

Consider again $d$-dimensional hypercubic lattice, and take
$p$-simplices $\Delta_p$ ($2 \le p \le d-1$) and $p$-pairs of
$(p-1)$-simplices $\Delta_{p-1}$ instead of 1-simplices and one
pair of 0-simplices, respectively, and assign a dynamical variable
$U_{\mu_1 \cdots \mu_p}(x)$ on the $p$-simplex located at $x$ and
oriented along $[\hat{\mu}_1,\cdots,\hat{\mu}_p]$. The set of
these $p$-simplex variables constitute the microscopic degrees of
freedom of $p$-form lattice gauge theory. So, for the compact $U(1)$ gauge
theory, we may parametrize the $p$-simplex variable by a direct
extension of Eq.(\ref{abelian1}) as
\bea U_{\mu_1 \cdots \mu_p}(x) = \exp \left(i a^p A_{\mu_1 \cdots
\mu_p}(x)\right)  \nonumber \eea
viz. an exponential map for the variable $A_{\mu_1
\cdots\mu_p}(x)$, and interpret it as the dynamical weight for
parallel transporting a charged $(p-1)$-dimensional brane by
lassoing it around the $p$-simplex $\Delta_p$. Extending the
interpretation, one can now assign the Chan-Paton factors along
the boundary of the $p$-simplex variable, viz. the
$(p-1)$-simplices forming the boundary of $p$-simplices:
$\Delta_{p-1} = \partial \Delta_p$ \footnote{More generally, one
can assign $(p-1)$-dimensional ``boundary" field theory living on
$\Delta_{p-1}$, whose excitations would carry a ``continuous
color" degrees of freedom in replacement of the finite-dimensional
Chan-Paton indices. We thank E. Witten for suggesting this extension.}.
Assigned this way, gauge transformation of
the nonabelian $p$-simplex variable is defined as
\bea U_{\mu_1 \cdots \mu_p}(x) \rightarrow \Omega(x) \cdot
U_{\mu_1 \cdots \mu_p}(x) \cdot \Omega^{-1}(x), \nonumber \eea
where the multiplication is defined in terms of appropriate action
on the $p$-tuple of the Chan-Paton vector space, $V_{\rm CP}
\otimes \cdots \otimes V_{\rm CP}$. The curvature of the $p$-form
gauge potential is measured by the deviation of
\bea \Big( \prod_{\{\mu\} \in \Delta_{p+1}} U_{\mu_1 \cdots
\mu_p}(x) \Big) \Big( \prod_{\{\mu\} \in \Delta_{p+1}} U_{\mu_1
\cdots \mu_p}(x) \Big)^\dagger \nonumber \eea
from unity, where again multiplication of variables is defined
appropriately on the $p$-tubles of the Chan-Paton vector space,
$\otimes^p V_{\rm CP}$. Evidently, the holonomy and the curvature
are valued in the $(p+1)$-tubles of the Chan-Paton vector space,
$\otimes^{p+1} V_{\rm CP}$.

%%%%%%%%%%%%%%%%%%%%%%%%%%%%%%%%%%%%%%%%%%%%%%%%%%%%%%%%%%%%%%%%%%
\subsection{Universality}
%%%%%%%%%%%%%%%%%%%%%%%%%%%%%%%%%%%%%%%%%%%%%%%%%%%%%%%%%%%%%%%%%%
Having generalized as above, one immediately realizes that there
is an enormous difference between the link variables and the
higher-dimensional extensions. In discretizing the space, we have
tacitly chosen $d$-dimensional hypercubic lattice. We could have
chosen an alternative lattice leading to  a different
discretization. Then, among all $p$-simplex variables, the link
variable is special since, irrespective of the lattice type,
boundary of the link variable consists always of two points. This
is essentially a statement of the `universality' --- detailed
specification of the discretization should not matter for
long-distance and continuum limit of the theory defined so.

Thus, universally, a link variable with Chan-Paton indices would
give rise to $N^2$ degrees of freedom. For higher-dimensional
simplices, the number of boundary simplices depend explicitly on
details of the latticization. For instance, consider two types of
plaquettes: square or triangle ones. For each of them, we would be
assigning four or three Chan-Paton indices. Then, in sharp
contrast to link variables, a plaquette variable of 3 or 4
Chan-Paton indices may give to $N^3$ or $N^4$
degrees of freedom depending on the choice of latticization.
Interestingly, the degerees of freedom is at least $N^3$, as the
minimal choice is lattice with triangular plaquettes.
%This is potentially a disastrous aspect of lattice definition of  nonabelian tensor gauge theories.

Remarkably, in the next section, we will find that compactness
conditions, which are direct generalizations of the condition
Eq.(\ref{linkcompactness}), constrain it so severely that the
plaquette variable actually carries ${\cal O}(N^2)$ degrees of
freedom.

%%%%%%%% section 2: Nonabelian Tensor gauge Theory on a Lattice %%%
\section{Nonabelian Tensor Gauge Theory on a Lattice}
In this section, built upon the result of the previous section, we
construct a nonabelian tensor gauge theory defined on a
$d$-dimensional hypercubic lattice.

%%%%%%%%%%%%%%%%%%%%%%%%%%%%%%%%%%%%%%%%%%%%%%%%%%%%%%%%%%%%%%%%
\subsection{lattice action}
\begin{figure}[htbp]
    \hskip2cm \begin{center}
 \includegraphics[width=0.7\linewidth,keepaspectratio,clip]
      {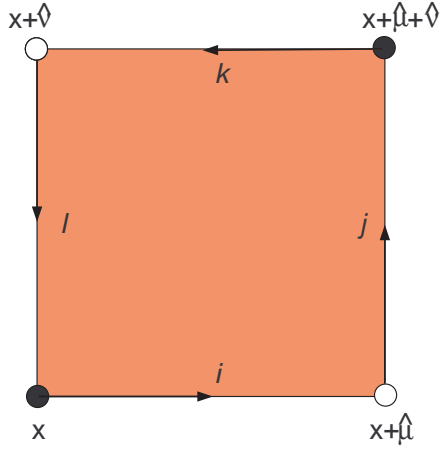}
    \end{center}
\caption{\sl The plaquette corresponding to the variable
$\Big(U_{\mu\nu}(x) \Big)_{ijkl}$ for $\mu \neq \nu$. The Chan-Paton
indices are assigned on each link (1-simplex) as $i,j,k,l = 1,
\cdots, N$.} \label{fig:plaquette}
\end{figure}

%%%%%%%%%%%%%%%%%%%%%%%%%%%%%%%%%%%%%%%%%%%%%%%%%%%%%%%%%%%%%%%%
We now utilize the observation we made concerning the Chan-Paton
prescription to a plaquette variable $U_{\mu \nu}(x)$, defined on
an elementary plaquette located at site $x$ along $\mu, \nu$
directions See Fig.\ref{fig:plaquette}. Heuristically, the
2-simplex is interpretable as the trajectory of a string on the
plaquette $[x, x+\hat{\mu}, x+\hat{\mu + \nu}, x+\hat{\nu}, x]$,
and the plaquette variable is the amplitude of
parallel-transporting a ``colored" string from two adjacent links
along $\hat{\mu}, \hat{\nu}$ directions to diagonally opposite
links. There are two possible choices of the ``move" but they are
related by the discrete subgroup ${\bf O}_d$ of O($d$) rotation
group. The boundary of the 2-simplex consists of one-simplices:
$[x, x+\hat{\mu}], [x+\hat{\mu}, x+\hat{\mu} + \hat{\nu}],
[x+\hat{\mu}+\hat{\nu}, x+\hat{\nu}], [x+\hat{\nu}, x]$. We then
attach Chan-Paton factors $i,j,k,l$ to each 1-simplex, taking
values in the newly introduced `color' space $V_{\rm color}$. This
`color' dressed 2-simplices will be the basic building blocks of
the nonabelian tensor gauge theory we are proposing. In the
followings, we take $V_{\rm color} = {\rm U}(N)$.

Consider $d$-dimensional hypercube lattice, whose sites are
labelled as $x = (x^1, \cdots, x^d)$ and unit vectors are labelled
as $\hat{\mu}, \hat{\nu}, \hat{\lambda}, \cdots$. The action we
propose is a direct generalization of the Wegner-Wilson-Polyakov
prescription:
 \bea S_{\rm tensor}  = -
\beta \sum_{\{x\}}\sum_{\mu, \nu, \lambda=1}^d \hspace{-3mm}' \,
\mbox{Re}\Big[ {\cal U}[x;\mu \nu \lambda] - N^3\Big],
\label{cubeS}\eea
where $\beta$ denotes inverse coupling parameter, and ${\cal
U}[C_{\mu\nu\lambda}(x)]$ refers to the gauge-invariant action
density for an elementary cube $[x;\hat\mu \hat\nu \hat\lambda]$,
defined as
\bea {\cal U}[x;\hat\mu \hat\nu \hat\lambda] &=& \sum_{i, \cdots,
u = 1}^N \Big(U_{\mu\nu}(x) \Big)_{ijkl}
\Big(U_{\nu\lambda}(x) \Big)_{lqun} \Big(U_{\lambda\mu}(x) \Big)_{nrmi} \nn \\
 & & \times \Big(U_{\mu\nu}(x +\hat{\lambda}) \Big)_{rstu}^*
\Big(U_{\nu \lambda}(x + \hat{\mu})\Big)_{jpsm}^*
\Big(U_{\lambda\mu}(x + \hat{\nu}) \Big)_{qtpk}^* \nn \\
&\equiv& {\rm Tr} \Big( (U \cdot U \cdot U )_{\rm corner} (x)
\cdot (U \cdot U \cdot U)_{\rm corner}^\dagger(x) \Big).
\label{UC} \eea
The last expression is to emphasize the interpretation that the
action density is a `trace' over the string holonomy, viz. phases
acquired through three adjacent plaquettes on a corner located at
$x$ and their hermitian conjugates. In Eq.(\ref{cubeS}), the $(')$
primed sum over $\mu$, $\nu$, $\lambda$ runs over the
$d$-dimensional lattice directions, where no pair of them are
allowed to coincide. The variable $(U_{\mu\nu}(x))_{ijkl}$ is
complex-valued, and lives on the plaquette encompassing the sites
$x$, $x + \hat{\mu}$, $x + \hat{\mu} + \hat{\nu}$, $x +
\hat{\nu}$, $x$. The Chan-Paton indices $i,j,k,l$ are associated
to the four links forming the boundary of the plaquette as
depicted in Fig. \ref{fig:plaquette}. The complex conjugation
`$*$' refers to simultaneous reversal of the plaquette orientation
and the Chan-Paton colors at the boundaries. As such, we adopt the
convention that the plaquette variables satisfy
\bea \Big(U_{\mu\nu}(x) \Big)_{ijkl}^* = \Big(U_{\nu\mu}(x)
\Big)_{lkji}. \label{orientation} \eea
\vskip0.3cm
%%%%%%%%%%%%%%%%%%% Fig. plaquette %%%%%%%%%%%%%%%%%%%%%%%%%%%%
%\begin{figure}
%\epsfxsize=7cm \epsfysize=5.3cm
%\centerline{\epsfbox{plaquette.eps}}
%%%%%%%%%%%%%%%%%%%%%%%%%%%%%%%%%%%%%%%%%%%%%%%%%%%%%%%%
\vskip0.3cm
Notice that ordering of both the directional indices $\mu, \nu$
and the color indices $i,j,k,l$ are reversed by the complex
conjugation.

\begin{figure}[htbp]
    \begin{center}
    \includegraphics[width=0.6\linewidth,keepaspectratio,clip]
    {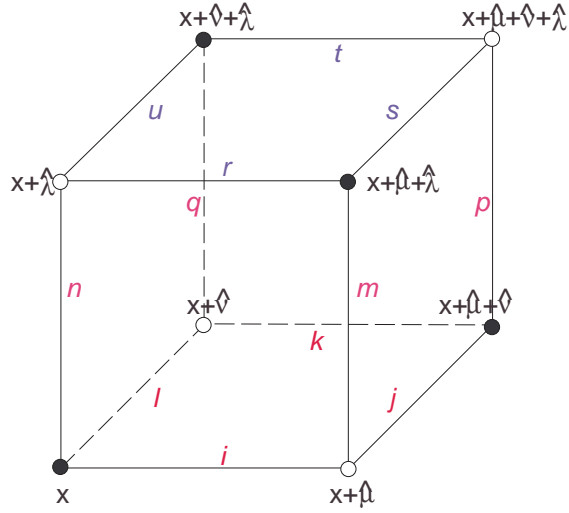}
    \end{center}
    \caption{\sl Chan-Paton factor assignment for the cube
    corresponding to ${\cal U}[C_{\mu \nu \lambda}(x)]$ in
    Eq.(\ref{UC}).
    }
\label{fig:cube}
\end{figure}

We also define a gauge transformation rule as
 \bea \Big(U_{\mu\nu}(x) \Big)_{ijkl}
\quad &\rightarrow&
  \sum_{i',j',k',l'}\Big(V_{\mu}(x)\Big)_{ii'}
\Big(V_{\nu}(x +\hat{\mu})\Big)_{jj'}
\Big(U_{\mu\nu}(x)\Big)_{i'j'k'l'} \Big(V_{\mu}(x
+\hat{\nu})^{\dagger}\Big)_{k'k}
\Big(V_{\nu}(x)^{\dagger}\Big)_{l'l} \nonumber\\
&\equiv&
 \Big((V_\mu \cdot V_\nu \cdot U_{\mu \nu} \cdot
 V_\mu^\dagger \cdot V_\nu^\dagger)(x)\Big)_{ijkl},
 \label{gaugetr} \eea
where $V_{\mu}(x)$ are gauge transformation functions valued in
U($N$) group, obeying $V_{\mu}(x)V_{\mu}^{\dagger}(x) = {\bf
1}_N$.
%associated to the link $(x, x + \hat{\mu})$.
It is easy to see that ${\cal U}[C_{\mu\nu\lambda}(x)]$ in the
action $S$ is associated to a 3-simplex (cube) as in Fig.
\ref{fig:cube}, and invariant under the gauge transformation
Eq.(\ref{gaugetr}).
\vskip0.3cm
%%%%%%%%%%%%%%%%%% Fig. cube %%%%%%%%%%%%%%%%%%%%%%%%%%%%
%\begin{figure}
%\epsfxsize=8cm \epsfysize=6cm \centerline{\epsfbox{cube.eps}}
%%%%%%%%%%%%%%%%%%%%%%%%%%%%%%%%%%%%%%%%%%%%%%%%%%%%%%%
\vskip0.3cm
\subsection{compactness conditions}
In Wegner-Wilson-Polyakov lattice gauge theory, the link variables $(U_{\mu}(x))_{ij}$ are
assumed compact, as defined via the conditions
\bea \sum_{j=1}^N \Big(U_{\mu}(x)\Big)_{ij}
\Big(U_{\mu}(x)\Big)^*_{i'j} = \delta_{ii'}  \quad {\rm and} \quad
\sum_{i=1}^N \Big(U_{\mu}(x)\Big)_{ij}
\Big(U_{\mu}(x)\Big)^*_{ij'} = \delta_{jj'}.  \label{wilsoncase}
\eea
It means that $(U_{\mu}(x))_{ij}$ are represented as $(N \times
N)$ unitary matrices of the Lie group U($N$).

Likewise, we shall be restricting the configuration space of the plaquette variables
$(U_{\mu\nu}(x))_{ijkl}$ to be compact. In the present case, however, there are four
viable definitions of the compactness for plaquette variables. The first one, which we
call as `rank-$N^3$ class' is defined by
\bea \mbox{rank-$N^3$} : \quad \sum_l\Big(U_{\mu \nu}(x) \Big)_{ijkl}
\Big(U_{\mu \nu}(x) \Big)^*_{i'j'k'l} &=& \frac1N \delta_{ii'} \delta_{jj'}
\delta_{kk'}
\nonumber \\
\sum_k\Big(U_{\mu \nu}(x) \Big)_{ijkl} \Big(U_{\mu \nu}(x)
\Big)^*_{i'j'kl'} &=& \frac1N\delta_{ii'} \delta_{jj'} \delta_{ll'}
\nonumber \\
\sum_j\Big(U_{\mu \nu}(x) \Big)_{ijkl} \Big(U_{\mu \nu}(x)
\Big)^*_{i'jk'l'} &=&\frac1N\delta_{ii'} \delta_{kk'} \delta_{ll'}
\nonumber \\
\sum_i\Big(U_{\mu \nu}(x) \Big)_{ijkl} \Big(U_{\mu \nu}(x)
\Big)^*_{ij'k'l'} &=& \frac1N\delta_{jj'} \delta_{kk'}
\delta_{ll'}. \nonumber \eea
The factor $\frac1N$ in the right-hand side was introduced by taking an appropriate
overall normalization of the plaquette variables $U_{\mu\nu}(x)$. All four conditions are
necessary, else the discrete rotational symmetry of the hypercube lattice would not be
warranted.

For `rank-$N^2$ classes', there are two possible choices. We will
label them as $N^2$(I) and $N^2$(II) classes. They are
\bea \mbox{rank-$N^2$(I)}: \quad \sum_{k,l}\Big(U_{\mu \nu}(x) \Big)_{ijkl}
\Big(U_{\mu \nu}(x)
\Big)^*_{i'j'kl} &=& \delta_{ii'} \delta_{jj'}  \label{N2-1} \\
\sum_{l,i}\Big(U_{\mu \nu}(x) \Big)_{ijkl} \Big(U_{\mu \nu}(x) \Big)^*_{ij'k'l}
&=& \delta_{jj'} \delta_{kk'} \label{N2-2} \\
\sum_{i,j}\Big(U_{\mu \nu}(x) \Big)_{ijkl} \Big(U_{\mu \nu}(x) \Big)^*_{ijk'l'}
&=& \delta_{kk'} \delta_{ll'} \label{N2-3} \\
\sum_{j,k}\Big(U_{\mu \nu}(x) \Big)_{ijkl} \Big(U_{\mu \nu}(x) \Big)^*_{i'jkl'}
&=& \delta_{ii'} \delta_{ll'}, \label{N2-4} \eea
and
\bea \mbox{rank-$N^2$(II)}: \quad \sum_{j,l}\Big(U_{\mu \nu}(x)
\Big)_{ijkl} \Big(U_{\mu \nu}(x) \Big)^*_{i'jk'l} &=&
\delta_{ii'} \delta_{kk'}
\nonumber \\
\sum_{i,k}\Big(U_{\mu \nu}(x) \Big)_{ijkl} \Big(U_{\mu \nu}(x)
\Big)^*_{ij'kl'} &=& \delta_{jj'} \delta_{ll'}, \nn \eea
respectively.

\begin{figure}[htbp]
    \begin{center}
    \includegraphics[width=0.9\linewidth,keepaspectratio,clip]
      {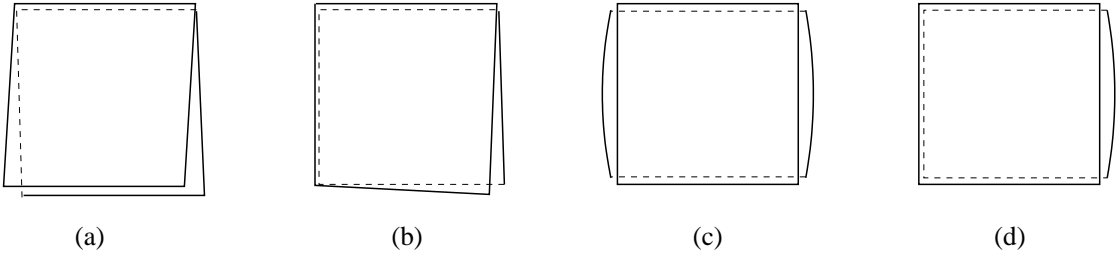}
      \end{center}
\caption{\sl Four classes of compactness conditions: (a) rank-$N^3$
conditions, (b) rank-$N^2$(I) conditions, (c) rank-$N^2$(II)
conditions, (d) rank-$N$ conditions. Horizontal and vertical
directions refer to $\hat{\mu}$ and $\hat{\nu}$, respectively.
Chan-Paton indices at each overlapping edge are summed over.}
\label{compactnesscond}
\end{figure}

Finally, there is the `rank-$N$ class', defined by
\bea \mbox{rank-$N$}: \quad  \sum_{j,k,l}\Big(U_{\mu \nu}(x) \Big)_{ijkl}
\Big(U_{\mu \nu}(x)
\Big)^*_{i'jkl} &=& N \delta_{ii'} \label{N-1} \\
\sum_{k,l,i}\Big(U_{\mu \nu}(x) \Big)_{ijkl}
\Big(U_{\mu \nu}(x) \Big)^*_{ij'kl} &=& N\delta_{jj'} \label{N-2} \\
\sum_{l,i,j}\Big(U_{\mu \nu}(x) \Big)_{ijkl}
\Big(U_{\mu \nu}(x) \Big)^*_{ijk'l} &=& N\delta_{kk'}\label{N-3} \\
\sum_{i,j,k}\Big(U_{\mu \nu}(x) \Big)_{ijkl}
\Big(U_{\mu \nu}(x) \Big)^*_{ijkl'} &=& N\delta_{ll'}. \label{N-4} \eea
Notice that each conditions are related by lattice rotational symmetry, based on relation
to the classical continuum limit. The four classes of compactness conditions are depicted
geometrically in Fig.\ref{compactnesscond}.
\vskip0.3cm
%%%%%%%%%%%%%%%%% Fig. compactness conditions %%%%%%%%%%%%%%
%\begin{figure}
%\epsfxsize=16cm \epsfysize=3.5cm \centerline{\epsfbox{cond.eps}}
%%%%%%%%%%%%%%%%%%%%%%%%%%%%%%%%%%%%%%%%%%%%%%%%%%%%%%%
\vskip0.3cm

Let us define the configuration space of the plaquette variables
obeying these conditions as ${\mathbb{X}}_{N^3},
{\mathbb{X}}_{N^2{\rm (I)}}, {\mathbb{X}}_{N^2{\rm (II)}}$, and
${\mathbb{X}}_{N}$, respectively. Obviously,
\bea {\mathbb{X}}_{N^3} \quad \subset \quad {\mathbb{X}}_{N^2{\rm
(I)}} \quad \mbox{and} \quad {\mathbb{X}}_{N^2 {\rm (II)}} \quad
\subset \quad {\mathbb{X}}_{N}. \nonumber\eea
It thus seems that four inequivalent definition of the theory
exists at the quantum level.

\begin{figure}[htbp]
    \begin{center}
    \includegraphics[width=1.0\linewidth,keepaspectratio,clip]
      {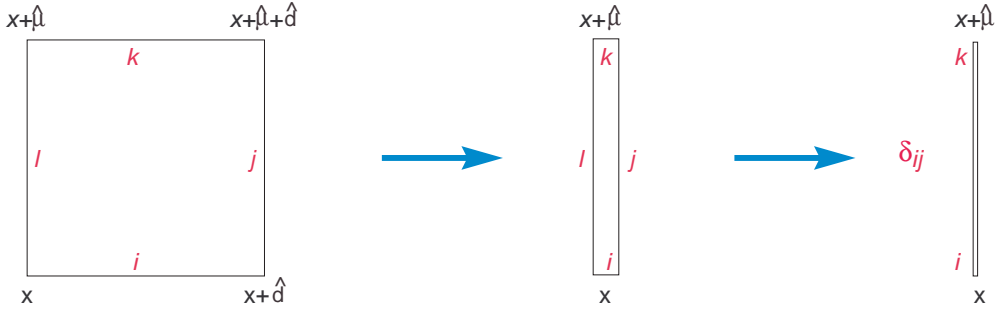}
      \end{center}
\caption{\sl The plaquette variable $(U_{d\mu}(x))_{ijkl}$ reduces a
variable on the link $(x, x+\hat{\mu})$ by truncating the degrees of
freedom of the indices $j$, $l$ after the dimensional reduction with
respect to the $d$-th direction,} \label{fig:reduction}
\end{figure}

Not all of them seem, however, physically acceptable. A condition we may impose is that
the plaquette variables with a given compactness condition are always reducible to the
link variables in a suitably prescribed limit. For example, consider `dimensional
reduction' \footnote{We will discuss more on lattice dimensional reduction in later
subsections.} of, say, the $d$-th direction with truncating the degrees of freedom of the
internal indices as
\bea & & (U_{d\mu}(x))_{ijkl} \Rightarrow
\delta_{jl}(U_{\mu}(x))_{ik}, \quad (U_{\mu
d}(x))_{ijkl}\Rightarrow\delta_{ik}(U_{\mu}(x)^\dagger)_{jl},
\quad (U_{dd}(x))_{ijkl} \Rightarrow \delta_{ik}\delta_{jl},
\qquad \label{red_to_link} \eea
where  $x = (x^1, \cdots, x^{d-1})$. This is a procedure of the
standard dimensional reduction accompanied with a truncation of
the internal degrees of freedom transforming as adjoint
representation of $V_{\mu}(x)$. Then, $(U_{\mu}(x))_{ik}$
represents a link variable in Wilson's lattice gauge theory as in
Fig. \ref{fig:reduction} obeying the standard gauge transformation
rule
\bea U_{\mu}(x) \limit V(x)U_{\mu}(x)V(x+\hat{\mu})^{\dagger},
\qquad V(x) \in {\rm U}(N). \nonumber\eea
\vskip0.3cm
%%%%%%%%%%%%%%%%%%% Fig. reduction %%%%%%%%%%%%%%%%%%%%%%%%%%%%
%\begin{figure}
%\epsfxsize=13cm \centerline{\epsfbox{reduction.eps}}
%%%%%%%%%%%%%%%%%%%%%%%%%%%%%%%%%%%%%%%%%%%%%%%%%%%%%%%
\vskip0.3cm

In the lattice dimensional reduction, the tensor gauge theory action Eq.(\ref{cubeS}) is
reduced to the Wilson's action. We require that the compactness conditions are consistent
with Eq.(\ref{wilsoncase}) under the reduction Eq.(\ref{red_to_link}). As demonstrated in
Appendix A, it is easy to see that the conditions of rank-$N^3$ and -$N^2$(II) are ruled
out, and that the remaining two, rank-$N^2$(I) and rank-$N$ conditions, are consistent.

Henceforth, in this paper, we will consider the two classes for the compactness
conditions: rank-$N^2$(I) conditions and rank-$N$ conditions. In the rest of this paper,
we consider exclusively the rank-$N^2$(I) class. As will become clearer, this class of
theory yields a physically meaningful classical continuum limit under the standard
dimensional reduction keeping all the internal indices. For the rank-$N$ case, discussions
are devoted to Appendix B. It turns out not to lead to a physically meaningful continuum
classical theory after the dimensional reduction. As such, we conclude that consistency
with the Yang-Mills theory via the dimensional reduction and the continuum limit therein
singles out the rank $N^2$(I) class as the only physically meaningful compactness
condition.

The partition function is defined as
\bea Z \equiv  \int {\cal D}U \, e^{-S}, \qquad {\cal D}U \equiv
\prod_{\{x\}}\prod_{\mu,\nu}[\dd U_{\mu\nu}(x)], \label{Z} \eea
where the measure $[\dd U_{\mu\nu}(x)]$ is induced by the norm:
\bea ||\delta U_{\mu\nu}(x)||^2 \equiv \sum_{i,j,k,l}
 \Big(\delta U_{\mu\nu}(x)\Big)_{ijkl}\Big(\delta U_{\mu\nu}(x)\Big)_{ijkl}^*,
 \nonumber
\eea
and the normalization is taken as $\int\, [\dd U_{\mu\nu}(x)] =
1$. It is readily seen that the norm is gauge invariant. So is the
measure defined from it as well. For $N=1$ case, it coincides with
the formulation of compact U(1) two-form tensor gauge theory on a
lattice \cite{Onogi}.

%%%%%%%%%%%%%%%%%%%%%%%%%%%%%%%%%%%%%%%%%%%%%%%%%%%%%%%%%%%%%%%%%%%%%%%
\section{Relation to Dynamical Quantum Yang-Baxter Maps}
%%%%%%%%%%%%%%%%%%%%%%%%%%%%%%%%%%%%%%%%%%%%%%%%%%%%%%%%%%%%%%%%%%%%%%%
In the previous section, we formulated lattice approach of nonabelian 
tensor gauge theory. The dynamical variables are elementary plaquettes, 
which carries not only spacetime orientation indices but also `color'
indices attached on four links around each elementary plaquette.  

Before proceeding further, in this section, we point out the lattice theory
is closely related to so-called dynamical quantum Yang-Baxter maps \cite{QYB}. 
To appreciate possible connection most transparently, let us first consider 
ground-states, equivalently, saddle points of the lattice action. From
Eq.(\ref{UC}), we see that the ground-states are characterized by the
condition
\bea
\Big( U_{\mu \nu} (x)  \cdot U_{\nu \lambda} (x)  \cdot U_{\lambda \mu} (x)
\Big)_{kqmjur} = \Big( U_{\lambda\mu}(x+ \hat\nu) \cdot U_{\nu\lambda} (x + \hat \mu) U_{\mu \nu} (x +
\hat\lambda) \Big)_{kqmjur} \ , 
\eea
where left- and right-hand sides are abbreviations of 
\bea  \Big(U_{\mu \nu}(x)\Big)_{ijkl} \Big(U_{\nu \lambda}(x) \Big)_{lqun}
\Big(U_{\lambda \mu}(x) \Big)_{nrmi} \nonumber \\
\Big(U_{\lambda \mu} (x +\hat\nu)\Big)_{kptq} \Big( U_{\nu \lambda} (x +
\hat\mu)\Big)_{mspj} \Big( U_{\mu \nu} (x + \hat\lambda)\Big)_{utsr} \, \eea
respectively. This is a set of {\sl cubic} algebraic relations among the elementary plaquette variables
at each lattice sites $x$ and orientations of each cube. We shall make connection with the dynamical quantum Yang-Baxter maps progressively, first taking all plaquette variables constant valued and then taking account of dependence to the lattice sites and orientations. 

Denote by
$[x;\hat{\mu}\hat{\nu}\hat{\lambda}]$ an elementary cube based at $x$ and oriented along
$\hat{\mu}, \hat{\nu}, \hat{\lambda}$ directions. Consider on the elementary cube a
colored open string whose endpoints are fixed at mutually antipodal vertices. We can
define the holonomy around the elementary cube by lassoing the string around the cube
once, viz. nonabelian holonomy under parallel transport of the string around a closed
manifold of ${\mathbb{S}}^2$ topology. The operation involves six plaquette moves on the
cube: three plaquette moves plus their conjugate ones. Each set of three moves altogether
is a map ${\cal X} \otimes {\cal X} \otimes {\cal X} \rightarrow {\cal X} \otimes {\cal X}
\otimes {\cal X}$, thus carries 6 indices. This then defines the Yang-Baxter map and the
curvature is measured by a `deformation' of the Yang-Baxter equation:
\bea U_{\mu \nu} U_{\nu \lambda} U_{\mu \lambda} = e^{ia^3H_{\mu
\nu \lambda}} U_{\mu \lambda} U_{\nu \lambda} U_{\mu \nu},
\nonumber \eea
for each elementary cube $[x;\hat{\mu}\hat{\nu}\hat{\lambda}]$. We immediately notice that
the standard quantum Yang-Baxter equation corresponds to the equation for flat two-form
connection, viz. vanishing parallel transport.

\begin{figure}[htbp]
    \begin{center}
    \includegraphics[width=0.9\linewidth,keepaspectratio,clip]
      {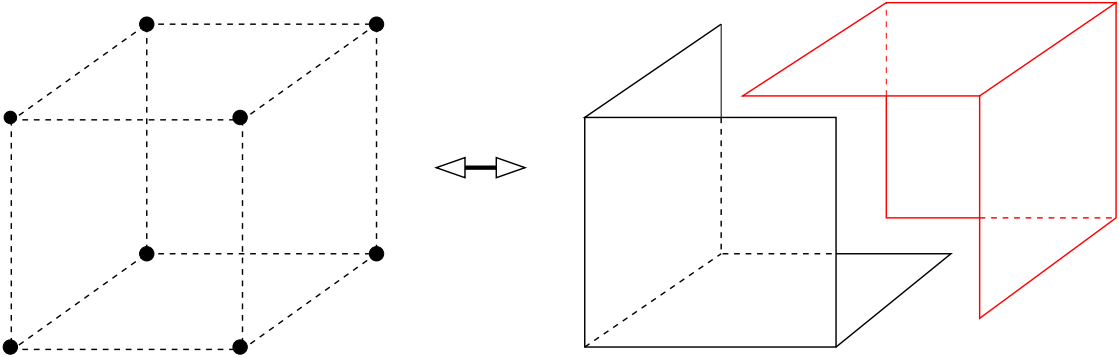}
      \end{center}
\caption{\sl Quantum Yang-Baxter interpretation of holonomy around
the cube. } \label{fig:ybe}
\end{figure}

The way the plaquette variables are ordered is not arbitrary. Rather, it is intrinsically
related to the way we define the parallel transport in terms of `lassoing' the string. To
see this, start with an `open' string whose two endpoints are held fixed at the two
oppositely diagonal vertices. It extends over three links, one per each directions of the
cube. By a forward `move', we define parallel transport of two adjacent links across the
plaquette the two links belong to. We then see that succession of the moves ought to be
such that the first three moves and the next three moves involve are oppositely ordered.
In Fig.\ref{fig:ybe}, for instance, the string initially along the {\sl lower} half cuts
(composed of three links) the cube into two pieces. We first move the string across the
lower half plaquettes of the cube: after move in the order of $[\mu \nu]$, $[\nu \lambda]$
and $[\mu \lambda]$ plaquettes, the string is now located at the along the {\sl upper} cut
(composed of three links).
\vskip0.3cm
%%%%%%%%%%%%%%%%%% Fig. cube %%%%%%%%%%%%%%%%%%%%%%%%%%%%
%\begin{figure}
%\epsfxsize=12cm \centerline{\epsfbox{yb.eps}}
%%%%%%%%%%%%%%%%%%%%%%%%%%%%%%%%%%%%%%%%%%%%%%%%%%%%%%%
\vskip0.3cm
The result of these moves is represented by
\bea U_{\mu \nu} U_{\nu \lambda} U_{\mu \lambda}. \nonumber \eea
We then continue the move through the upper half plaquettes of the
cube. In this case, the move is again in the order of $[\mu\nu]$,
$[\nu\lambda]$ and $[\mu\lambda]$, but in the transposed
orientation in each plaquette. The result of these moves is
represented by
\bea U^{\dagger}_{\mu \nu} U^{\dagger}_{\nu\lambda}
U^{\dagger}_{\mu \lambda}. \nonumber \eea
Thus, putting togeter the two set of operations, a complete
lassoing of the cube by a string is given by
\bea U_{\mu \nu} U_{\nu \lambda} U_{\mu \lambda} U^{\dagger}_{\mu
\nu} U^{\dagger}_{\nu\lambda} U^{\dagger}_{\mu \lambda}, \eea
and is precisely the holonomy of the antisymmetric tensor potential.

So far, we suppressed dependence of the elementary plaquette variables on lattice sites $x$ and
orientations. Once we reinstate their dependence, the Yang-Baxter maps become enormously complicated. 
The dependence on lattice sites $x$ and on orientations of elementary cubes is not arbitrary, however. 
In particular, orientation of elementary plaquettes is closely tied with directional dependence. Remarkably, we see that these dependences are put together into the so-called    
{\sl dynamical} quantum Yang-Baxter equations
\cite{dynamicalQYB} obeying:
\bea U_{\mu \nu} (g - h^\lambda) U_{\mu \lambda} (g) U_{\nu \lambda} (g - h^\mu) = U_{\nu
\lambda} (g) U_{\mu \lambda} (g + h^\nu) U_{\mu \nu} (g). \eea
Here, $g$ is an element of a Lie algebra ${\cal G}$ and $h$ is an element of its Cartan subalgebra. 
Then, making the expansion
\bea U_{\mu \nu} = 1 + \hbar B_{\mu \nu}(h) + {\cal O}(\hbar^2)
\eea
for an expansion parameter $\hbar$, we find that the leading-order
contribution is given by
\bea \dd_{\mu} B_{\nu \lambda} + \dd_\nu B_{\lambda \mu} +
\dd_\lambda B_{\mu \nu} - [B_{\mu \nu}, B_{\nu \lambda}] - [B_{\nu
\lambda}, B_{\lambda \mu}] - [B_{\lambda \mu}, B_{\mu \nu}] = 0,
\label{zerofs} \eea
where the derivative $\dd_\mu$ is defined as
\bea \dd_\mu = {\bf e}^{(\mu)} \cdot {\partial \over \partial {\bf
e}} \eea
for the Cartan subalgebra of the Lie algebra ${\mathcal{G}}$ over
${\mathbb{C}}$. The left-hand side of Eq.(\ref{zerofs}) takes precisely 
the structure we may interpret as  
the nonabelian tensor field strength. So, we see that the dynamical 
quantum Yang-Baxter equation is the equation for
vanishing field strength over the base manifold spanned by the
Cartan subalgebra of $\mathcal{G}$.

%%%%%%% section 3: Classical continuum limit in the Rank N^2(I) Case  %%%%%%%%
\setcounter{equation}{0}
\section{Parametrization of Plaquette Variables}
To proceed further, we will need to parametrize the plaquette variables
$\left(U_{\mu\nu}(x)\right)_{ijkl}$ that satisfies the compactness conditions. In this
section, we will construct explicitly a parametrization for the plaquette variables
defined by the rank-$N^2$(I) compactness conditions Eq.(\ref{N2-1}-\ref{N2-4}).

\subsection{Lax pair ansatz}
Among the compactness conditions Eqs.(\ref{N2-1} -- \ref{N2-4}),
Eqs.(\ref{N2-1}, \ref{N2-3}) imply that the plaquette variable
$\left( U_{\mu\nu}(x)\right)_{ijkl}$ is a U($N^2$) matrix with
respect to $(i,j)$ and $(k,l)$ pair of indices,
referred as row and column indices. Likewise, Eqs.(\ref{N2-2},
\ref{N2-4}) imply the same but now with respect to $(l,i)$ and
$(j,k)$ pairs, identified as row and column indices. From these
two requirements, we assume that $\left(
U_{\mu\nu}(x)\right)_{ijkl}$, viewed as $(N^2 \times N^2)$
matrices, are projected onto a direct product of two matrices
where the one acts on the indices, $i$ and $k$, and the other on
$j$ and $l$~\footnote{As discussed in Appendix C, there are
solutions that keep degrees of freedom of the plaquette variables
of ${\cal O}(N^3)$. However, these solutions also reduce the
U$(N)$ gauge symmetry to U$(1)^N$, which is not enough to eliminate all 
negative norm states (ghosts). Therefore, they do not appear to lead to
unitary theories.}. Therefore, the plaquette variables are
parametrizable as
\bea \Big( U_{\mu\nu}(x)\Big)_{ijkl} :=
\Big(W_{\mu\nu}(x)\Big)_{ik}
\Big(\widetilde{W}_{\mu\nu}(x)\Big)_{jl}, \eea
viz. as a direct product of $W_{\mu\nu}(x)$ and
$\widetilde{W}_{\mu\nu}(x)$ belonging to U($N)\times\wt{\rm U} (
N)$.

\begin{figure}[htbp]
    \begin{center}
    \includegraphics[width=0.5\linewidth,keepaspectratio,clip]
      {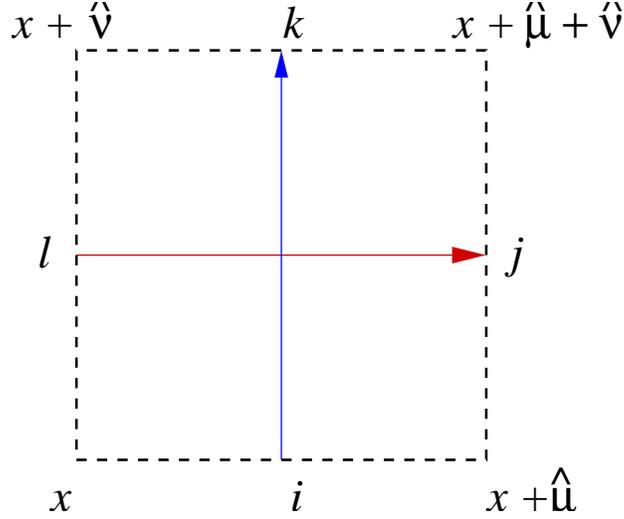}
      \end{center}
\caption{\sl Parametrization of the plaquette variable
$\left(U_{\mu\nu}(x)\right)_{ijkl}$ (the dashed lines) in terms of
`dual' links (the solid lines). The $d$-dimensional link variables
$(U^{(\mu)}_{\nu}(x))_{ik}$, $(U^{(\nu)}_{\mu}(x)^{\dagger})_{jl}$
can be associated with the `dual' links.
%$(x+ \frac12\hat{\mu}, x+\frac12\hat{\mu} + \hat{\nu})$,
%$(x+\hat{\mu} + \frac12\hat{\nu}, x + \frac12\hat{\nu})$,
%respectively.
}
\label{fig:dual_link}
\end{figure}

The two matrices are not independent. The charge-conjugation
relation Eq.(\ref{orientation}) imposes that U($N) \simeq \wt{\rm
U}(N)$, and that $\widetilde{W}_{\mu\nu}(x)=
W_{\nu\mu}(x)^\dagger$. Thus, the plaquette variable is
parametrizable in terms of U($N$) matrices $U^{(\mu)}_{\nu}(x)
\equiv W_{\mu\nu}(x)$ as
\bea \Big( U_{\mu\nu}(x)\Big)_{ijkl} =
\Big(U^{(\mu)}_{\nu}(x)\Big)_{ik} \Big(U^{(\nu)\dagger}_{\mu}(x)
\Big)_{jl}. \label{split-variable} \eea
Notice that the diagonal U(1) part of $U^{(\mu)}_{\nu}(x)$ for
either $\mu < \nu$ or $\mu > \nu$ are redundant for parametrizing
the plaquette variable. Fortuituously, these extra U(1)'s do not
interact with the rest, so they will be mod out straightforwardly.
The gauge transformation of Eq.(\ref{gaugetr}) gives rise to gauge
transformation of $U^{(\mu)}_{\nu}(x)$ as
\bea \Big(U^{(\mu)}_{\nu}(x)\Big)_{ik} \longrightarrow
\sum_{i',k'=1}^N
\Big(V_{\mu}(x)\Big)_{ii'}\Big(U^{(\mu)}_{\nu}(x)\Big)_{i'k'}
\Big(V_{\mu}^\dagger(x+\wh\nu)\Big)_{k'k}. \label{dualgt} \eea
We are thus parametrizing the plaquette variable as a square of
$d$-species of U($N$) link variables, $U^{(\mu)}_\nu (x)$ with
$\mu = 1, \cdots, d$, each of which transforms as an adjoint under
the gauge transformation $V_{\mu}(x)$. Intuitively, the newly
introduced link variables $U^{(\mu)}_\nu$ in
Eq.(\ref{split-variable}) are interpretable as the variables
residing on two `dual' links of the original plaquette, as
depicted in Fig.\ref{fig:dual_link}.

\vskip0.3cm
%%%%%%%%%%%%%%%%% Fig. dual link%%%%%%%%%%%%%%%%%%%%%%%%%%%%%
%\begin{figure}
%\epsfxsize=6cm \centerline{\epsfbox{dual_link2.eps}}
%%%%%%%%%%%%%%%%%%%%%%%%%%%%%%%%%%%%%%%%%%%%%%%%%%%%%%%
\vskip0.3cm

\begin{figure}[htbp]
    \begin{center}
    \includegraphics[width=0.7\linewidth, keepaspectratio, clip]
      {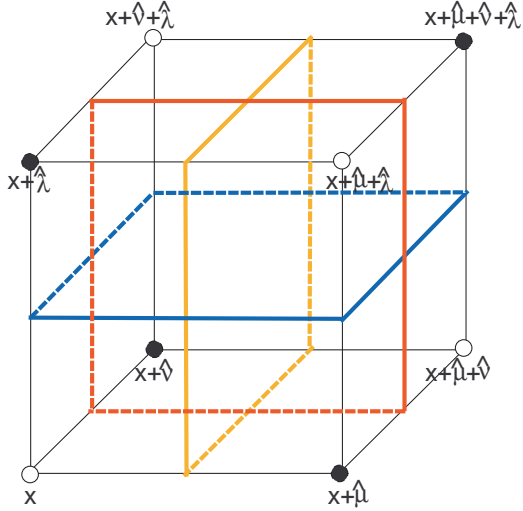}
    \end{center}
\caption{\sl The original cube corresponding to ${\cal
U}[C_{\mu\nu\lambda}(x)]$ (the dashed line) and the three `dual'
plaquettes ${\cal U}^{(\mu)}[P_{\nu\lambda}(x)]$, ${\cal
U}^{(\nu)}[P_{\lambda\mu}(x)]$, ${\cal
U}^{(\lambda)}[P_{\mu\nu}(x)]$ (the solid lines) for the case $\mu$,
$\nu$, $\lambda$ all different. } \label{fig:dual_plaquette}
\end{figure}

Actually, the dual lattice interpretation extends further. In
terms of the newly introduced link variables $U^{(\mu)}_\nu(x)$,
the gauge-invariant cube of the plaquette variables ${\cal
U}[C_{\mu\nu\lambda}(x)]$ in Eq.(\ref{UC}) are expressible as a
triple product of `dual' plaquette, as depicted in
Fig.\ref{fig:dual_plaquette}, and becomes
\bea {\cal U}[C_{\mu\nu\lambda}(x)] =  {\cal
U}^{(\mu)}[P_{\nu\lambda}(x)] \, {\cal
U}^{(\nu)}[P_{\lambda\mu}(x)] \, {\cal
U}^{(\lambda)}[P_{\mu\nu}(x)], \eea
where the $d$-species of `dual' plaquette ${\cal U}^{(\mu)}$ are
\bea {\cal U}^{(\mu)}[P_{\nu\lambda}(x)]\equiv
\tr\left[U^{(\mu)}_{\nu}(x)\,U^{(\mu)}_{\lambda}(x+\hat{\nu})\,
   U^{(\mu)^\dagger}_{\nu}(x+\hat{\lambda}) \,
   U^{(\mu)^\dagger}_{\lambda}(x) \right].
\eea
Here, `tr' refers to the trace operation with respect to $(N\times
N)$ matrix indices of the dual link variables.
\vskip0.3cm
%%%%%%%%%%%%%%%%% Fig. dual plaquette %%%%%%%%%%%%%%%%%%%%%%%%%%%%%
%\begin{figure}
%\epsfxsize=9cm \centerline{\epsfbox{dual-plaq2.eps}}
%%%%%%%%%%%%%%%%%%%%%%%%%%%%%%%%%%%%%%%%%%%%%%%%%%%%%%%
\vskip0.3cm
Putting together, the nonabelian tensor gauge theory obeying the
compactness conditions Eqs.(\ref{N2-1} -- \ref{N2-4}) is defined
by the partition function:
\bea Z_{\rm tensor} \equiv  \int [{\cal D}U] \, \exp(-S_{\rm
tensor}), \label{DPtheory} \eea
where the action is
\bea S_{\rm tensor}  = -\beta \sum_{\{x\}}{\sum}_{\mu, \,\nu,\,
\lambda=1}^{d}\hspace{-10mm}'\hspace{10mm} \mbox{Re }\Big({\cal
U}^{(\mu)}[P_{\nu\lambda}(x)]\, {\cal
U}^{(\nu)}[P_{\lambda\mu}(x)]\, {\cal
U}^{(\lambda)}[P_{\mu\nu}(x)] - N^3\Big) \label{DPactionold} \eea
and the integral measure
\bea {\cal D}U = \prod_{\{x\}} [ \d U(x)] \qquad \mbox{where}
\qquad [\d U(x)] = \prod_{\mu,\,\nu=1}^{d} \hspace{-2mm}{}^\prime
\hspace{2mm} [\dd U^{(\mu)}_{\nu}(x)] / {\rm Vol} \left({\rm
U}(1)^{d(d-1)/2} \right) \label{DPmeasureold} \eea
is given in terms of U$(N)$-invariant Haar measure~\footnote{ In
the followings, for the Haar measure, we will adopt the
normalization convention $\int [\dd U^{(\mu)}_{\nu}(x)] = 1$.}
$[\dd U^{(\mu)}_{\nu}(x)]$ of the dual link variables. The prime
$(')$ in the sum in Eq.(\ref{DPaction})  refers to summing over
all nondegenerate plaquette orientations. Likewise, the prime
$(')$ in the product in Eq.(\ref{DPmeasure}) refers to integrating
over all nondegenrate dual link variables. Notice that we
eliminated the aforementioned over-counting in
$U^{(\mu)}_{\nu}(x)$ for $\mu < \nu$ or $\mu > \nu$ by dividing
the volume of overall U$(1)^{d(d-1)/2}$ gauge groups.

%%%%%%%%%%%%%%%%%%%%%%%%%%%%%%%%%%%%%%%%%%%%%%%%%%%%%%%%%%%%%%%%%%%%%%%%%%
\subsection{classical continuum limit}
Before proceeding further, we will study classical continuum limit
of the action Eq.(\ref{DPaction}) and show that the continuum
action is manifestly $d$-dimensional Lorentz invariant.

\begin{figure}[htbp]
    \begin{center}
    \includegraphics[width=0.6\linewidth,keepaspectratio,clip]
      {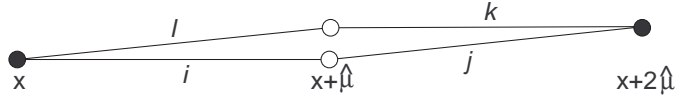}
    \end{center}
\caption{\sl Degenerate class of plaquettes corresponding to the
variable $\Big(U_{\mu\mu}(x) \Big)_{ijkl}$. }
\label{fig:col-plaquette}
\end{figure}

At this stage, we will make a slight generalization of the theory
we have constructed by incorporating into the theory the variable
$(U_{\mu \mu}(x))_{ijkl}$ which is associated a collapsed
plaquette as in Fig. \ref{fig:col-plaquette}.
\vskip0.3cm
%%%%%%%%%%%%%%% Fig. collapsed plaquette %%%%%%%%%%%%%%%%%%%%
%\begin{figure}
%\epsfxsize=9cm \epsfysize=2.5cm \centerline{\epsfbox{col-plaquette.eps}}
%%%%%%%%%%%%%%%%%%%%%%%%%%%%%%%%%%%%%%%%%%%%%%%%%%%%%%%
\vskip0.3cm
As will become clear momentarily, adding these variables are
imperative in order to obtain a Lorentz invariant theory after
taking a classical continuum limit. Still, the generalized theory
is defined by the action Eq.(\ref{DPactionold}) and the measure
Eq.(\ref{DPmeasureold}), where the sum over the Lorentz indices is
unconstrained and the volume of overall groups is mod out for
$d(d+1)$ U(1) gauge groups. Explicitly, the new action is defined
by
\bea S_{\rm tensor}  = -\beta \sum_{\{x\}}{\sum}_{\mu, \,\nu,\,
\lambda=1}^{d} \mbox{Re }\Big({\cal
U}^{(\mu)}[P_{\nu\lambda}(x)]\, {\cal
U}^{(\nu)}[P_{\lambda\mu}(x)]\, {\cal
U}^{(\lambda)}[P_{\mu\nu}(x)] - N^3\Big) \ , \label{DPaction} \eea
while the new functional integral measure is defined by
\bea {\cal D}U = \prod_{\{x\}} [ \d U(x)] \qquad \mbox{where}
\qquad [\d U(x)] = \prod_{\mu,\,\nu=1}^{d} [\dd
U^{(\mu)}_{\nu}(x)] / {\rm Vol} \left({\rm U}(1)^{d(d+1)/2}
\right) \ . \label{DPmeasure} \eea
Notice also that the gauge invariance and the parametrization of plaquette
variable into split variables are straightforwardly extendible to
degenerate variables as well.

To take continuum limit of the classical action, we expand the
dual link variable $U^{(\mu)}_{\nu}(x)$ of U($N$) gauge group
as~\footnote{As $a^2$ in the exponent of
Eq.(\ref{expandU_rankN2}) can be regarded as the area of
elementary plaquette, the expansion appears natural only for $\mu
\neq \nu$, and not for $\mu =\nu$ for which the corresponding
plaquette is degenerate and has a vanishing area. However, as we
will see, the continuum classical action is Lorentz-invariant only
if we take the same parametrization Eq.(\ref{expandU_rankN2}) for
$\mu =\nu$ as well.}
\bea U^{(\mu)}_{\nu}(x) = e^{ia^2 A^{(\mu)}_{\nu}(x)} = 1 +
ia^2A^{(\mu)}_{\nu}(x)
-\frac{a^4}{2}\left(A^{(\mu)}_{\nu}(x)\right)^2 + \cdots.
\label{expandU_rankN2} \eea
Here, $A^{(\mu)}_{\nu}(x)$ are $(N\times N)$ Hermitian matrices,
where indices $\mu$ and $\nu$ transform as Lorentz vector indices.
Accordingly, the U($N$)-invariant plaquette ${\cal
U}^{(\mu)}[P_{\nu\lambda}(x)]$ is expandable as
\bea U^{(\mu)}(P_{\nu\lambda}(x))  & = & N +ia^3
\tr\Big[f^{(\mu)}_{\nu\lambda}(x)\Big] -\frac12 a^6 \tr
\Big[\left(f^{(\mu)}_{\nu\lambda}(x)\right)^2\Big]
+ {\cal O}(a^7), \\
f^{(\mu)}_{\nu\lambda}(x) & \equiv &
\Delta_{\nu}A^{(\mu)}_{\lambda}(x) - \Delta_{\lambda}A^{(\mu)}_{\nu}(x),
\label{fmunulambda}
\eea
where $\Delta_{\nu}$ denotes the lattice difference operator:
$\Delta_{\nu}f(x) \equiv \frac{1}{a}(f(x+\hat{\nu})-f(x))$. Hence,
the classical action is expanded as
\bea S_{\rm tensor} & =& \frac{1}{2}\beta
a^6\sum_{\{x\}}\sum_{\mu, \nu, \lambda=1}^d
 \Big[N^3 H_{\mu\nu\lambda}^{\rm U(1)}(x)^2
+N^2\, \tr\left\{\tilde{f}^{(\mu)}_{\nu\lambda}(x)^2 +
\tilde{f}^{(\nu)}_{\lambda\mu}(x)^2 + \tilde{f}^{(\lambda)}_{\mu\nu}(x)^2
\right\}\Big] + {\cal O}(a^7) \nn \\
 & = &\frac{1}{2g^2} \int \dd^d x \,\sum_{\mu, \nu, \lambda=1}^d
 \Big[N^3 H_{\mu\nu\lambda}^{\rm U(1)}(x)^2
+3N^2\, \tr\left(\tilde{f}^{(\mu)}_{\nu\lambda}(x)^2 \right)\Big]
+ {\cal O}(a^7) \label{continuumS'} \eea
in the classical continuum limit:
\bea a \limit 0 \qquad {\rm and} \qquad \frac{1}{g^2} \equiv \beta
a^{6-d} = \mbox{fixed}. \label{coupling} \eea
Here, we decomposed the field strength $f^{(\mu)}_{\nu\lambda}(x)$
into diagonal U(1) and traceless parts:
\bea f^{(\mu)}_{\nu\lambda}(x) \equiv f^{(\mu)\, {\rm
U(1)}}_{\nu\lambda}(x) \mathbb{I}_N +
\tilde{f}^{(\mu)}_{\nu\lambda}(x) \qquad \tr \,
\tilde{f}^{(\mu)}_{\nu\lambda}(x)=0. \nonumber \eea
The difference operators in Eq.(\ref{fmunulambda}) should be
understood as ordinary derivatives in the continuum limit.

For the diagonal U(1) part, we have reproduced the totally
antisymmetric 3-form field strength $H^{\rm U(1)}_{\mu\nu\lambda}=
- f^{(\mu)\, {\rm U(1)}}_{\nu\lambda}(x) -f^{(\nu)\, {\rm
U(1)}}_{\lambda\mu}(x) -f^{(\lambda)\, {\rm U(1)}}_{\mu\nu}(x)$.
Thus, in the $N=1$ case, the result Eq.(\ref{continuumS'}) reduces
the previously known action of abelian tensor gauge theory. For U($N$),
the 3-form field strength is constructed as an object carrying six
color indices:
\bea \Big({\cal H}_{\mu\nu\lambda}(x)\Big)_{ijki'j'k'} \equiv
-\Big(f^{(\mu)}_{\nu\lambda}(x)\Big)_{ii'}\delta_{jj'}\delta_{kk'}
-
\Big(f^{(\nu)}_{\lambda\mu}(x)\Big)_{jj'}\delta_{kk'}\delta_{ii'}
-
\Big(f^{(\lambda)}_{\mu\nu}(x)\Big)_{kk'}\delta_{ii'}\delta_{jj'} \ .
\nonumber\eea
The field-strength is manifestly gauge invariant and has the
symmetry properties:
\bea \Big({\cal H}_{\mu\nu\lambda}(x)\Big)_{ijki'j'k'} =
-\Big({\cal H}_{\nu\mu\lambda}(x)\Big)_{jikj'i'k'} = -\Big({\cal
H}_{\mu\lambda\nu}(x)\Big)_{ikji'k'j'} = -\Big({\cal
H}_{\lambda\nu\mu}(x)\Big)_{kjik'j'i'} \, , \label{symmetry_calH}
\eea
meaning that ${\cal H}$ is antisymmetric under permutations with
respect to the three sets of indices $(\mu, i,i')$, $(\nu, j,j')$,
$(\lambda, k,k')$. Utilizing these properties, the continuum
action Eq.(\ref{continuumS'}) is rewritable compactly as:
\bea S = \frac{1}{2g^2} \int \dd^d x \,\sum_{\mu, \nu,
\lambda=1}^d \Tr_{N^3}\left( {\cal H}_{\mu\nu\lambda}(x)\right)^2,
\label{finalresultaction} \eea
where `$\Tr_{N^3}$' refers to the trace for $(N^3 \times N^3)$
matrices. It is defined for an generic element ${\cal
A}_{ijki'j'k'}$ as 
\bea
\Tr_{N^3}({\cal A}) \equiv \ \sum_{i,j,k} \ {\cal
A}_{ijkijk}.
\eea

Recall that, in defining the plaquette variables $U_{\mu \nu}(x)$,
we have included those on degenerate plaquette $\mu = \nu$. Had
one considered the lattice theory without them, the naive
continuum action would be the same as Eq.(\ref{continuumS'}) except
that the summation over $\mu$, $\nu$, $\lambda$ is now restricted
to the cases $\mu$, $\nu$, $\lambda$ all different. Then, the
first term in Eq.(\ref{continuumS'}) would still be Lorentz
invariant, but the second term would not be so because the
traceless part $\tilde{f}^{(\mu)}_{\nu\lambda}(x)$ is not totally
antisymmetric with respect to $\mu$, $\nu$, $\lambda$. We have
deliberately kept the degenerate plaquette variables so that
Lorentz invariant continuum action Eq. (\ref{continuumS'}) is
obtainable.

The action Eq.(\ref{continuumS'}) in the classical continuum limit
is purely Gaussian. It does not necessarily mean that the
corresponding quantum theory is free. For instance, as
demonstrated in \cite{rey1, rey2, Onogi}, $d=4$ tensor gauge
theory of compact U(1) gauge group leads to charge confinement as
a result of instanton effects. The result is in direct parallel to
the situation \cite{Polyakov} of $d=3$ compact U(1) gauge theory.
Such effects are, however, entirely nonperturbative, having
directly to do with the topology of the configuration space, and
the continuum theory defined by the action Eq.(\ref{continuumS'})
is perturbatively free. Given that the elementary dynamical
variables are the link variables (defined on dual lattice), it is
actually instructive to understand why the continuum theory is
noninteracting. Recall that the gauge transformation function
$V_{\mu}(x)$ in Eq.(\ref{dualgt}) is a Lorentz vector and is
expandable as
\bea V_{\mu}(x) = e^{ia\Lambda_{\mu}(x)} = 1 + ia\Lambda_{\mu}(x)
- \frac12 a^2 \left(\Lambda_{\mu}(x)\right)^2 + \cdots,
\label{expandV} \eea
with $\Lambda_{\mu}(x)$ being $(N\times N)$ Hermitian
matrix-valued. Thus, in the continuum limit, the gauge
transformation rule becomes abelian:
\bea A^{(\mu)}_{\nu}(x) \limit A^{(\mu)}_{\nu}(x)
-\partial_{\nu}\Lambda_{\mu}(x). \nonumber \eea
This conclusion, despite being formulated in terms of the `dual'
link variables, marks the significant departure from the ordinary
nonabelian lattice gauge theory. In the continuum theory, we are
exploring small neighborhood of the identity in the lattice gauge
transformation, and information of global structure of the group
is lost. The global structure can be made visible by introducing a cutoff
and considering the theory with compact variables, just like the lattice
action Eq.(\ref{cubeS}) we started with.
%%%%%%%%%%%%%% section 4: Dimensional Reduction  %%%%%%%%%%%%%%%%%%%%%%%%%%
\setcounter{equation}{0}
\section{Dimensional Reduction on the Lattice}
Before proceeding further, we check an important consistency condition. We shall take
dimensional reduction of the
$d$-dimensional lattice tensor gauge theory and show that the
theory Eq.(\ref{cubeS}) is reduced near the continuum limit to the
$(d-1)$-dimensional lattice gauge theory coupled to an adjoint
scalar field.
%%%%%%%%%%%%%%%%%%%%%%%%%%%%%%%%%%%%%%%%%%%%%%%%%%%%%%%%%%%%%%%%%%%%%
\subsection{dimensional reduction for lattice vector fields}
We first explain what we mean by 'dimensional reduction' in lattice gauge
theory. For definiteness, we shall consider the $d$-dimensional
Wilson's plaquette action:
\bea S_{\rm YM} & = & -\beta \sum_{\{x\}}\sum_{\mu,\nu =1}^d
\hspace{-2mm}' \hspace{2mm} \mbox{Re}\,\Big( {\cal
U}[P_{\mu\nu}(x)] - N\Big),
\label{Wilsonaction} \\
 {\cal U}[P_{\mu\nu}(x)] & = & \tr_N \left[ U_{\mu}(x)U_{\nu}(x + \hat{\mu})
U_{\mu}(x + \hat{\nu})^\dagger U_{\nu}(x)^\dagger \right],
\nonumber
\eea
where the summation of $\mu, \nu$ with the prime ($'$) runs over
the region $\mu \neq \nu$. The link variables $U_{\mu}(x)$ are $N
\times N$ unitary matrices belonging to the gauge group G,
satisfying Eq.(\ref{wilsoncase}). It is parametrized in terms of
Lie-algebra-valued gauge potential $A_{\mu}(x)$ as
\bea U_{\mu}(x) = e^{ia A_{\mu}(x)}, \label{DRlink} \eea
where again $a$ denotes the lattice spacing.

To consider dimensional reduction from $d$- to
$(d-1)$-dimensions, we take the $d$-th lattice anisotropic, treat
its lattice spacing $r$ much smaller than other lattice spacing $a$,
and take the limit $r \limit 0$. Then, $d$-dimensional
hypercubic lattice collapses to $(d-1)$-dimensional lattice .
Denote the $d$-th link variable $U_d (x)$ as
\bea U_d (x) = e^{ir A_d (x)}, \nonumber \eea
while all others $U_{\mu}(x)$ ($\mu =1, \cdots, d-1$) remain
unchanged in Eq.(\ref{DRlink}). Dimensional reduction then goes as
follows. One picks up the lowest Kaluza-Klein modes (zero modes)
with respect to the $d$-th lattice direction, so that $x$ now
represents a site of $(d-1)$-dimensional lattice $(x_1, \cdots,
x_{d-1})$. In the limit $r \limit 0$, $d$-th links are shrunken to
a point, and the variable $U_d (x)$ is now associated with sites.
We take the limit with the combination $\Phi(x) \equiv \frac{r}{a}
A_d (x)$ kept fixed. The scalar field $\Phi(x)$ represents an
adjoint scalar field carrying mass-dimension \footnote{Dividing by $a$ in
the definition of $\Phi(x)$ is to keep this as a canonical dimension.}.

The dimensionally reduced lattice action becomes
\bea S_{\rm YM} = -\beta \sum_{\{ x\}} \sum_{\mu,
\nu=1}^{d-1}\hspace{-2mm}' \hspace{2mm}\mbox{Re}\, \Big({\cal
U}[P_{\mu\nu}(x)] -N\Big) -2\beta \sum_{\{ x\}} \sum_{\mu=1}^{d-1}
\mbox{Re}\, \Big({\cal U}[P_{\mu d}(x)] -N\Big). \nonumber \eea
After renaming
\bea U(x) \equiv e^{ia\Phi(x)} = U_d (x), \nonumber \eea
the action in the classical continuum limit reproduces the known
result of dimensional reduction:
\bea S_{\rm YM} & = & \frac{1}{g_{\rm YM}^2} \int \dd^{d-1} x\,
\tr \left[ \sum_{\mu,\nu=1}^{d-1}\frac14
\left(F_{\mu\nu}(x)\right)^2 +\sum_{\mu=1}^{d-1}\frac12
\left(D_{\mu}\Phi(x) \right)^2\right],  \nn \\
F_{\mu\nu} & = & \partial_{\mu} A_{\nu} -\partial_{\nu}A_{\mu}
+i[A_{\mu}, A_{\nu}], \qquad  D_{\mu} \Phi =
\partial_{\mu}\Phi + i[A_{\mu}, \Phi]. \label{def_FD} \eea
The limit $a \limit 0$ is taken while holding $1/g_{\rm YM}^2
\equiv 2\beta a^{4-(d-1)}$ fixed.

%%%%%%%%%%%%%%%%%%%%%%%%%%%%%%%%%%%%%%%%%%%%%%%%%%%%%%%%%%%%%%%%%%%
\subsection{dimensional reduction for lattice tensor fields}
By taking the same procedure, we now examine the dimensional reduction 
of our
proposed lattice tensor gauge theory. Scaling the $d$-th dimension
differently, we write the elementary plaquette variables as
\bea
U_{\mu\nu}(x) & = &  I + ia^2 B_{\mu\nu}(x) + \cdots,\nn \\
U_{d\mu}(x) & = &  I + ia r B_{d\mu}(x) + \cdots, \nn \\
U_{\mu d}(x) & = &  I + ia r B_{\mu d}(x) + \cdots, \nn \\
U_{dd}(x) & = &  I + ir^2 B_{dd}(x) + \cdots, \nonumber \eea
with $\mu, \nu = 1, \cdots, d-1$ and $I_{ijkl} =
\delta_{ik}\delta_{jl}$. As $r \limit 0$, extending the reasoning
given above, we see that the variables $U_{d \mu}(x)$, $U_{\mu
d}(x)$ become associated to links, while $U_{dd}(x)$ is associated
to sites. They lead to vector and scalar fields on the
$(d-1)$-dimensional hyper-cubic lattice, respectively. Both fields
carry the mass-dimension one, and the limit $r \rightarrow 0$
ought to be taken with newly defined fields
\bea {\cal V}_{\mu}(x) \equiv r B_{d\mu}(x), \quad \tilde{{\cal
V}}_{\mu}(x) \equiv r B_{\mu d}(x), \quad \Phi(x) \equiv
\frac{r^2}{a} B_{dd}(x) \nonumber \eea 
kept finite.

After the dimensional reduction, the gauge transformation rules
become
\bea \Big( U_{\mu}(x)\Big)_{ijkl} & \equiv &
\Big( U_{d\mu}(x)\Big)_{ijkl} \nn \\
& \limit & \sum_{i'j'k'l'}
  \Big(V(x)\Big)_{ii'} \Big(V_{\mu}(x)\Big)_{jj'}
\Big(U_{\mu}(x)\Big)_{i'j'k'l'} \Big(V(x+\hat{\mu})^\dagger\Big)_{k'k}
\Big(V_{\mu}(x)^\dagger\Big)_{l'l},   \label{gaugetr_dimredUmu}
\\
\Big(\tilde{U}_{\mu}(x)\Big)_{ijkl} & \equiv &
\Big( U_{\mu d}(x)\Big)_{ijkl}  \nn \\
 & \limit & \sum_{i'j'k'l'}
\Big(V_{\mu}(x)\Big)_{ii'} \Big(V(x+\hat{\mu})\Big)_{jj'}
\Big(\tilde{U}_{\mu}(x)\Big)_{i'j'k'l'} \Big(V_{\mu}(x)^\dagger\Big)_{k'k}
\Big(V(x)^\dagger\Big)_{l'l}, \\
\Big( U(x)\Big)_{ijkl} & \equiv &
\Big( U_{dd}(x)\Big)_{ijkl} \nn \\
& \limit & \sum_{i'j'k'l'}
 \Big(V(x)\Big)_{ii'} \Big(V(x)\Big)_{jj'}
\Big(U(x)\Big)_{i'j'k'l'} \Big(V(x)^\dagger\Big)_{k'k}
\Big(V(x)^\dagger\Big)_{l'l},
\label{gaugetr_dimredU}
\eea
where $V(x)$ is a U($N$) matrix-valued field residing at the site
$x$. From the transformation rules, we see that
$(U_{\mu}(x))_{ijkl}$ is a variable defined on the link $(x,
x+\hat{\mu})$, transforming in adjoint representation of
$V_{\mu}(x)$ and carrying additional internal symmetry group
labelled by $j$, $l$ indices. Likewise, we see that 
$(\tilde{U}_{\mu}(x))_{ijkl}$ represents a variable defined on the
link $(x+ \hat{\mu}, x)$, transforming in adjoint representation
of $V_{\mu}(x)$ and carrying additional internal symmetry group
labelled by $i$, $k$ indices. The variable $(U(x))_{ijkl}$ is
defined at the site $x$, thus represents a scalar field
transforming as (adjoint)$\times$(adjoint) of the additional
internal symmetry group.

For the theory under consideration, since the plaquette variable
is given in split form Eq.(\ref{split-variable}), the fields
$U_{d\mu}$, $U_{\mu d}$, $U_{dd}$ are parametrizable as
\bea \Big(U_{d\mu}(x)\Big)_{ijkl} & = &
\Big(e^{iaA_{\mu}(x)}\Big)_{ik}\Big(e^{-iaA^{(\mu)}(x)}\Big)_{jl}
\equiv
\Big(U_{\mu}(x)\Big)_{ik}\Big(U^{(\mu)}(x)^\dagger\Big)_{jl},
\nn \\
\Big(U_{\mu d}(x)\Big)_{ijkl} & = &
\Big(e^{iaA^{(\mu)}(x)}\Big)_{ik}\Big(e^{-iaA_{\mu}(x)}\Big)_{jl}
\equiv
\Big(U^{(\mu)}(x)\Big)_{ik}\Big(U_{\mu}(x)^\dagger\Big)_{jl},
\nn \\
\Big(U_{dd}(x)\Big)_{ijkl} & = & \,
\Big(e^{+ia\varphi(x)}\Big)_{ik}\,
\Big(e^{-ia\varphi(x)}\Big)_{jl} \equiv
\Big(U(x)\Big)_{ik}\Big(U(x)^\dagger\Big)_{jl}, \nonumber \eea
and the gauge transformation rules are
\bea
U_{\mu}(x) &\limit & V(x) U_{\mu}(x) V(x + \hat{\mu})^\dagger, \nn \\
U^{(\mu)}(x)& \limit & V_{\mu}(x) U^{(\mu)}(x) V_{\mu}(x)^\dagger, \nn \\
U(x) & \limit & V(x) U(x) V(x)^\dagger. \nonumber \eea
In this case, the vector gauge fields with four indices $({\cal
V}_{\mu})_{ijkl}$, $(\tilde{{\cal V}}_{\mu})_{ijkl}$ separate into
the purely vector field degrees of freedom $(A_{\mu})_{ik}$ and
the adjoint matter one $(A^{(\mu)})_{jl}$. Also, the scalar
$(\Phi)_{ijkl}$ transforming as (adjoint)$\times$(adjoint) splits
into two $\varphi$'s. In the classical continuum limit, the gauge
transformation by $V_{\mu}(x)$ becomes invisible, so the field
$(A^{(\mu)})_{jl}$ do not transform. Notice that, to fix redundant
degrees of freedom of the overall U(1)'s in the split form, one
may take
\bea \tr \,A^{(\mu)}(x) = \tr \, \varphi(x) = 0. \label{fixU(1)}
\eea

The dimensionally reduced action is then given by
\bea S & = & -\beta \sum_{\{x\}} \sum_{\mu, \nu,
\lambda=1}^{d-1}\,
\mbox{Re}\left[{\cal U}[C_{\mu\nu\lambda}(x)]-N^3\right] \nn \\
&-&3\beta\sum_{\{x\}}\sum_{\mu, \nu=1}^{d-1} \, \mbox{Re}
\left[{\cal U}[C_{\mu\nu d}(x)]-N^3\right]  -3\beta \sum_{\{x\}}
\sum_{\mu =1}^{d-1}\, \mbox{Re}\left[{\cal U}[C_{\mu
dd}(x)]-N^3\right], \label{DRaction} \eea
where the term $({\cal U}[C_{ddd}(x)]-N^3)$ is suppressed because
it vanishes trivially as a consequence of the split form of
$U_{dd}(x)$. In the classical continuum limit, the first term
becomes Eq.(\ref{finalresultaction}), but with $d$ replaced by
$(d-1)$. For the second term, ${\cal U}[C_{\mu\nu d}(x)]$ consists
of the three factors ${\cal U}^{(\mu)}[P_{\nu d}(x)]$, ${\cal
U}^{(\nu)}[P_{d\mu}(x)]$, ${\cal U}^{(d)}[P_{\mu\nu}(x)]$. The
last factor is nothing but the Wilson's plaquette action giving
\bea {\cal U}^{(d)}[P_{\mu\nu}(x)] = N + ia^2 \tr
\left[\Delta_{\mu}A_{\nu}(x) -\Delta_{\nu}A_{\mu}(x)\right] +a^4
\tr\left(-\frac12F_{\mu\nu}(x)^2\right) + O(a^5). \nonumber \eea
The field strength $F_{\mu\nu}$ is defined as in
Eq.(\ref{def_FD}). The first factor leads
\bea {\cal U}^{(\mu)}[P_{\nu d}(x)] = N + a^4 \tr
\left[-\frac12(\Delta_{\nu}A^{(\mu)}(x))^2\right] + {\cal O}(a^5),
\nonumber \eea
where the ${\cal O}(a^2)$-contribution
$ia^2\tr(\Delta_{\nu}A^{(\mu)}(x))$ vanishes due to
Eq.(\ref{fixU(1)}). Thus, the second term in Eq.(\ref{DRaction})
becomes
\bea -3\beta\sum_{\{x\}}\sum_{\mu,\, \nu=1}^{d-1} \, \mbox{Re}
\left[{\cal U}[C_{\mu\nu d}(x)]-N^3\right] = 3\beta a^4
N^2\sum_{\{x\}} \sum_{\mu, \, \nu=1}^{d-1}\, \tr
\left[\frac{1}{2}F_{\mu\nu}(x)^2 + (\Delta_{\mu}
A^{(\nu)}(x))^2\right] + {\cal O}(a^5). \nonumber \eea
Notice that the contribution is of order ${\cal O}(a^4)$,
overwhelming the ${\cal O}(a^6)$ contribution of the first term
consisting of the tensor fields alone. The third term is similarly
computed to give the kinetic term of the adjoint matter
$\varphi(x)$ at ${\cal O}(a^4)$. It does not involve nonlinear
coupling to the gauge field $A_\mu(x)$ or the scalar field
$\varphi(x)$, since such couplings are either absent or are
higher-orders in the continuum limit. Putting them together, we
arrive at the classical continuum action
\bea S = \frac{3N^2}{g_{\rm YM}^2}\int \dd^{d-1}x \, \tr\left[
\sum_{\mu,\,\nu=1}^{d-1}\left\{\frac12 \Big(F_{\mu\nu}(x)\Big)^2
+\Big(\partial_{\mu}A^{(\nu)}(x) \Big)^2\right\}
+\sum_{\mu=1}^{d-1} \Big(D_{\mu}\varphi(x) \Big)^2\right],
\label{continuumS_rankN2} \eea
with $g^{-2}_{\rm YM} \equiv \beta a^{5-d}$ and the covariant
derivative $D_{\mu}\varphi$ is defined as in Eq.(\ref{def_FD}).
The action Eq.(\ref{continuumS_rankN2}) describes the U($N$) gauge
theory with adjoint matter, accompanied with $(d-1)$ copies of
free decoupled fields.

The classical continuum action Eq.(\ref{continuumS_rankN2}) is Lorentz invariant. 
We emphasize that this result is far
from being obvious. Since the action Eq.(\ref{finalresultaction})
emerges from the ${\cal O}(a^6)$ part of the lattice action, we
need to keep the expansion Eq.(\ref{expandU_rankN2}) up to the
$(A^{(\mu)}_{\nu})^3$ terms. On the other hand, to get the action
Eq.(\ref{continuumS_rankN2}) of the order ${\cal O}(a^4)$, we
need to expand $U_{\mu}$, $U^{(\mu)}$, $U$ up to $(A_{\mu})^4$,
$(A^{(\mu)})^4$, $\varphi^4$, respectively. Because the mass
dimension of fields is modified from taking the dimensional
reduction (which involves taking factors of lattice spacing $a$ to the fields), 
the latter case needs information of one higher-order compared to the former
case of $A^{(\mu)}_{\nu}$. Thus, it is highly nontrivial 
that the continuum limit Eq.(\ref{continuumS_rankN2}) yields Lorentz
invariant action Eq.(\ref{finalresultaction}).

%%%%%%%%%%%%%%%%%%% section 5: Strong Coupling Expansions  %%%%%%%%%%
\setcounter{equation}{0}
\section{Strong-Coupling Expansion}
An attractive and promising feature of lattice formulation is the
feasibility of exploring nonperturbative physics such as dynamical
mass generation and confinement etc. One such method is the
strong-coupling expansion, which was applied successfully for the
Wilson's lattice gauge theory, and amounts to expansions in powers
of the inverse coupling. Strong-coupling expansions are defined
intrinsically on a lattice and cannot be derived directly for the
continuum counterpart. As such, one typically supplements the
strong-coupling expansions with a suitable methods for
extrapolating the results to the continuum limit. Nevertheless,
even for finite lattice spacing, the strong-coupling expansions
can lead to new insights by revealing dynamical mechanisms which
are typical for strongly interacting continuum theories. With such
motivation, in this section, we develop the method of
strong-coupling expansions for the lattice tensor gauge theory.

Much as in Wilson's lattice gauge theory, the lattice tensor gauge
theory admits gauge-invariant, nonabelian Wilson surface operators
-- a direct counterpart of the Wilson loop operators. Correlation
functions involving these Wilson surface operators are the main
interest to us. Hence, we shall apply the strong-coupling
expansion analysis to correlators involving $n$ Wilson surface
operators for $n=0,1,2$, and extract information regarding free
energy, internal energy, and surface tension.

Remarkably, we will find very distinctive behavior of the strong
coupling expansion. For ordinary lattice gauge theory, it is well known 
that the
strong-coupling expansion has a finite radius of convergence
\cite{strongcoupling}. Here, for lattice tensor gauge theory, we
will find that the strong-coupling expansion is not absolutely 
convergent but an asymptotic series in the large $N$ limit. 
As weak-coupling perturbation theories give rise to asymptotic series, 
we conjecture that strong-coupling expansion of the
lattice tensor gauge theory at large $N$ is dual to an another, weakly
coupled lattice theory. 
%%%%%%%%%%%%%%%%%%%%%%%%%%%%%%%%%%%%%%%%%%%%%%%%%%%%%%%%%%%%%
\subsection{gauge-fixing
%$U^{(\mu)}_{\mu}=1$
}
On a lattice of finite volume and finite spacing, the total number
and the domain of integration for plaquette variables are finite.
Therefore, the functional integral is well defined even without
gauge fixing. Still, for a given computation of physical
quantities, it is often advantageous to fix a suitable gauge. We
will make use of this gauge-fixing freedom such that some of the
plaquette variables are set equal to a prescribed value. Recall
that we have included in the set of elementary plaquette variables
those associated with degenerate plaquette, $U^{(\mu)}_\mu (x)$.
We found it convenient to use the gauge freedom and set them to
unity, $U^{(\mu)}_\mu(x) = \mathbb{I}_N$ for all $x$. To show that
this procedure is always possible, it suffices to find an
appropriate gauge transformation function $V_{\mu}(x)$ such that
\bea V_{\mu}(x)U^{(\mu)}_{\mu}(x)V_{\mu}(x+\mu)^\dagger = \mathbb{I}_N. \nonumber \eea
We find that
\bea V_{\mu}(x) = \left\{\begin{array}{ll}
\mathbb{I}_N & \mbox{for $ x_{\mu} = 0$} \\
\prod_{y=0}^{x_{\mu}}U^{(\mu)}_{\mu} (x_1, \cdots,
x_{\mu-1},y,x_{\mu+1}, \cdots, x_d) & \mbox{for $x_{\mu} > 0$}
\\
\prod_{y=-1}^{x_{\mu}}U^{(\mu)}_{\mu} (x_1, \cdots,
x_{\mu-1},y,x_{\mu+1}, \cdots, x_d)^{\dagger} & \mbox{for $x_{\mu}
< 0$}. \end{array} \right. \nonumber \eea
In the gauge choice $U^{(\mu)}_{\mu}(x) = 1$, the lattice action
Eq.(\ref{DPaction}) reduces to
\bea S_{\rm tensor} & = &  -\beta
\sum_{\{x\}}\sum_{\mu,\,\nu,\,\lambda=1}^d\hspace{-3mm}'
\,\,\mbox{Re }\Big({\cal U}^{(\mu)}[P_{\nu\lambda}(x)]\, {\cal
U}^{(\nu)}[P_{\lambda\mu}(x)]\, {\cal
U}^{(\lambda)}[P_{\mu\nu}(x)] - N^3 \Big), \label{DPaction'} \eea
where the primed sum over the Lorentz indices runs over $\mu$,
$\nu$, $\lambda$ all different.

Notice that the gauge-fixing we have made is partial: the action
is still invariant under a class of gauge transformations
$V_\mu(x)$ that leaves the gauge choice $U_\mu^{(\mu)} = 1$
intact. Such gauge transformations are the ones independent of
$x^\mu$ coordinates, since
\bea V_\mu(x) V_\mu (x + \mu)^\dagger = 1 \quad \mbox{viz.} \quad
V_\mu(x) = V_\mu (x + \mu).  \eea
Notice also that the gauge-fixing does not introduce any
nontrivial Jacobian or Faddeev-Popov ghosts either.
%%%%%%%%%%%%%%%%%%%%%%%%%%%%%%%%%%%%%%%%%%%%%%%%%%%%%%%%%%%%%%%%%%%%%
\subsection{character expansion}
%%%%%%%%%%%%%%%%%%%%%%%%%%%%%%%%%%%%%%%%%%%%%%%%%%%%%%%%%%%%%%%%%%%%%
To proceed further for the strong-coupling expansion, we perform
the character expansion \cite{Drouffe}. The gauge-fixed action
Eq.(\ref{DPaction'}) is a triple product of U($N$) characters in
fundamental representation. So, for $U, V, W \in$ U($N$), we expand
as
\bea \exp \Big(3 \beta\,[\chi_\Box (U)\chi_\Box (V)\chi_\Box (W)]
+ {\rm c.c.}\Big) = \sum_{\tt R_1, R_2, R_3}C_{\tt R_1, R_2,
R_3}(\beta) \chi_{\tt R_1} (U)\chi_{\tt R_2}(V)\chi_{\tt R_3}(W),
\label{character} \eea
where $\chi_{\tt R}(U)$ refers to the character of the irreducible
representation $\tt R$ of gauge group U($N$). The character of the
complex conjugate representation $\tt \bar{R}$ is related to it as
$\chi_{\tt \bar{R}}(U) = \chi^*_{\tt R}(U) = \chi_{\tt R}
(U^\dagger)$. For the trivial representation ${\tt R} = 0$,
$\chi_0 (U) = 1$, and, for the fundamental representation ${\tt R}
= \Box$, $\chi_\Box (U) = \tr\, U$. The sum over $\tt R_1, R_2,
R_3$ in Eq.(\ref{character}) is for all unitary irreducible
representations of the gauge group U($N$). The expansion
Eq.(\ref{character}) shows that the expansion coefficient $C_{\tt
R_1, R_2, R_3}(\beta)$ is totally symmetric under permutations of
$\tt R_1, R_2, R_3$, and that $C_{\tt R_1, R_2, R_3}(\beta) =
C_{\tt \bar{R}_1,\bar{R}_2,\bar{R}_3}(\beta)$.

Recall that the characters can split or join under the U($N$)
group integrals as
\bea \int [\dd U]\, \chi_{\tt R}(UAU^\dagger B) & = &
\frac{1}{d_{\tt R}}\,\chi_\ttr (A)\,\chi_\ttr (B), \label{split}\\
\int [\dd U]\, \chi_{\ttr_1} (UA)\, \chi_{\ttr_2} (U^\dagger B) &
= & \delta_{\ttr_1, \ttr_2}\,\frac{1}{d_{\ttr_1} }\,\chi_{\ttr_1}
(AB). \label{join} \eea
Here, $d_\ttr=\chi_\ttr(1)$ denotes the dimension of the
representation $\ttr$. For example, $d_0=1$, $d_\Box =d_{\frac{}{
\Box}}=N$. The orthogonality relations of characters
(the $A=B=1$ case in Eq.(\ref{join})), the inverse relation of
Eq.(\ref{character}) is readily obtainable:
\bea C_{\ttr_1 \ttr_2 \ttr_3} = \int [\dd U][\dd V][\dd W] \,
\chi_{\ttr_1} (U)\, \chi_{\ttr_2}(V)\, \chi_{\ttr_3}(W) \, \exp
\Big( 3\beta\,[\chi_\Box (U)\chi_\Box (V)\chi_\Box (W)] + \, {\rm
c.c.}\Big). \label{CRRR} \eea
Factoring out the contribution $C_{000}(\beta)$, we express
Eq.(\ref{character}) in a more convenient form
\bea \exp \Big(\! 3\beta\,[\chi_\Box (U)\chi_\Box (V)\chi_\Box (W)] +
 {\rm c.c.} \!\Big) = C_{000}(\beta) \left[1+ \hspace{-3mm} \sum_{\ttr_1, \ttr_2,
\ttr_3}\hspace{-3mm}' \widetilde{C}_{\ttr_1 \ttr_2
\ttr_3}(\beta)\chi_{\ttr_1}
(U)\chi_{\ttr_2}(V)\chi_{\ttr_3}(W)\!\right] \hspace{-1mm} ,
\label{character2} \eea
where $\widetilde{C}_{\ttr_1 \ttr_2 \ttr_3}(\beta)\equiv C_{\ttr_1
\ttr_2 \ttr_3}(\beta)/C_{000}(\beta)$, and the primed sum runs
over all unitary irreducible representations except $(\ttr_1,
\ttr_2, \ttr_3) = (0,0,0)$.

We have computed the character expansion coefficient
$C_{000}(\beta)$ from Eq.(\ref{character2}) in Appendix D. The
result is in power-series of $\beta$:
\bea C_{000}(\beta) =  \sum_{n=0}^N \,n!\, (3\beta)^{2n} + \sum_{n
= N+1}^{\infty}\, (v_n)^3 \,n!\, (3\beta)^{2n}. \label{resultC000}
\eea
It shows that the expansion coefficient grows as $n!$ up to the
${\cal O}(\beta^{2N})$, and then decreases by the factor
$0<v_n<1$. In Appendix D, we computed the first 20 and 50 terms
for $N=2$ and 3 cases, respectively, and concluded from the result
that the suppression by $(v_n)^3$ is sufficient to render the
power-series Eq.(\ref{resultC000}) convergent for {\sl finite}
$N$. It implies that the power-series expansion of
$\widetilde{C}_{\Box\Box\Box}(\beta)$ is convergent as well:
\bea \widetilde{C}_{\Box\Box\Box}(\beta) & := &
\frac12\frac{\partial}{\partial (3\beta)}
\ln C_{000}(\beta) \label{Cfff}  \\
 & = & (3\beta)\left[1 + 3(3\beta)^2 + 13(3\beta)^4 + 71(3\beta)^6
+461(3\beta)^8 +\cdots \right].  \label{Cfffexp}  \eea
For the second equality, we assumed $N\geq 5$.

Interestingly, the large-order behavior of the character expansion
coefficients is quite different from that of Wilson's lattice
gauge theory. For the latter, the character expansion yields
\bea \exp \Big( \gamma\,\chi_\Box (U) + {\rm c.c.} \Big) =
\sum_{\ttr}C_{\ttr}(\gamma) \chi_\ttr (U), \qquad C_\ttr (\gamma)
= \int [\dd U] \, \chi_\ttr (U) \exp\Big( \gamma\chi_\Box (U) +
{\rm c.c.}\Big). \nonumber \eea
The character expansion coefficient $C_0(\gamma)$, which is the same as
$z(\gamma)$ defined in Eq.(\ref{zgammadef}), can be computed explicitly. We relegate 
details to appendix D and quote here the result from Eq.(\ref{zgamma}):
\bea C_0(\gamma) = z(\gamma) =
\sum_{n=0}^N\,\frac{\gamma^{2n}}{n!} + \sum_{n=N+1}^{\infty}\,
\frac{v_n}{n!}\, \gamma^{2n}. \label{gaugecase} \eea
Here, the first $N$ terms coincide with those of $e^{\gamma^2}$,
and the power-series converges for any value of $\gamma$.

Of notable situation is the large-$N$ limit. For Wilson's lattice
gauge theory, as is evident from Eq.(\ref{gaugecase}), the first
term with $N=\infty$ yields a convergent large-order behavior. For
the lattice tensor gauge theory, however, Eq.(\ref{resultC000})
with $N=\infty$ is obviously divergent and is not even Borel
summable. This imparts a significant departure of our lattice
tensor gauge theory from Wilson's lattice gauge theory. We will
dwell on this issue further later in subsection 7.5.

%%%%%%%%%%%%%%%%%%%%%%%%%%%%%%%%%%%%%%%%%%%%%%%%%%%%%%%%%%%%%%%%%%
\subsection{partition function and free energy}
For the gauge $U^{(\mu)}_{\mu}=1$, the gauge-fixed
partition function is given by
\bea Z & = &  e^{-\beta N^3 d(d-1)(d-2){\cal N}_s} \, {\cal Z}
\, , \nonumber \\
{\cal Z} & = & \int {\cal D}'U \, \prod_{\{x\}}\prod_{\mu < \nu <
\lambda} \exp\Big( 3\beta\, {\cal U}^{(\mu)}[P_{\nu\lambda}(x)]\,
{\cal U}^{(\nu)}[P_{\lambda\mu}(x)]\, {\cal
U}^{(\lambda)}[P_{\mu\nu}(x)] +\, {\rm c.c.}\Big), \label{calZ}
\eea
where ${\cal N}_s$ refers to the total number of lattice sites,
and the measure ${\cal D}'U$ is the U($N$)-Haar measure for the
regular plaquette variables $U^{(\mu)}_{\nu}(x)$ $(\mu\neq\nu)$.
Here, we do not consider the volume factor of U(1) groups as in
Eq.(\ref{DPmeasure}) since it merely produces an irrelevant
constant factor independent of $\beta$ and $N$.

Once the expansion Eq.(\ref{character2}) is made for each cube,
Eq.(\ref{calZ}) can be written as
\bea & & \hspace{-1cm} {\cal Z}
=
\Big(C_{000}(\beta)\Big)^{\frac{1}{3!}d(d-1)(d-2){\cal N}_s} \nn \\
& & \hspace{-1.3cm} \times \int {\cal D}'U \, \prod_{\{x\}}
\prod_{\mu<\nu<\lambda} \left[1+ \!\!\! \sum_{\ttr_1, \ttr_2 ,
\ttr_3}\hspace{-3mm}' \,\, \tilde{C}_{\ttr_1 \ttr_2
\ttr_3}(\beta)\,
\chi_{\ttr_1}\left(P^{(\mu)}_{\nu\lambda}(x)\right)\,
\chi_{\ttr_2}\left(P^{(\nu)}_{\lambda\mu}(x)\right)\,
\chi_{\ttr_3}\left(P^{(\lambda)}_{\mu\nu}(x)\right)\right].
\label{calZexp} \eea
Here, $P^{(\mu)}_{\nu\lambda}(x)$ denotes the plaquette
$P_{\nu\lambda}(x)$ formed by the dual link variables carrying
superscript $(\mu)$, viz. $P^{(\mu)}_{\nu\lambda}(x)=
U^{(\mu)}_{\nu}(x)\,U^{(\mu)}_{\lambda}(x+\hat{\nu})\,
   U^{(\mu)}_{\nu}(x+\hat{\lambda})^{\dagger}\,
   U^{(\mu)}_{\lambda}(x)^{\dagger}$.
The triple product of the characters in Eq.(\ref{calZexp})
corresponds to the elementary cube or triple product among
its`dual' plaquettes (depicted in Fig. \ref{fig:dual_plaquette})
carrying the representations $\ttr_1, \ttr_2, \ttr_3$. The
integrals in Eq.(\ref{calZexp}) is carried out by making use of
the integration formulas Eqs.(\ref{split}, \ref{join}). Nontrivial
contributions come from situations that elementary cubes, each of
which are labelled by the representations $\ttr_1, \ttr_2, \ttr_3$
are glued together into three-dimensional closed manifolds on the
lattice. Consider the simplest one of such cases. It is that eight
cubes are glued together to form a manifold of ${\mathbb{S}}^3$
topology. An example of such a configuration consists of the
following elementary cubes:
%
%\beas
\bea (1): & & \chi_{\ttr_1} \left(P^{(1)}_{23}(x)\right)\,
         \chi_{\ttr_2}\left(P^{(2)}_{31}(x)\right)\,
         \chi_{\ttr_3}\left(P^{(3)}_{12}(x)\right) \\
(2): & & \chi_{\ttr_1} \left(P^{(1)}_{32}(x+\hat{4})\right)\,
         \chi_{\ttr_2}\left(P^{(2)}_{13}(x+\hat{4})\right)\,
         \chi_{\ttr_3}\left(P^{(3)}_{21}(x+\hat{4})\right) \\
(3): & & \chi_{\ttr_2}\left(P^{(2)}_{43}(x)\right)\,
         \chi_{\ttr_3}\left(P^{(3)}_{24}(x)\right)\,
         \chi_{\ttr_4}\left(P^{(4)}_{23}(x)\right) \\
(4): & & \chi_{\ttr_2}\left(P^{(2)}_{34}(x+\hat{1})\right)\,
         \chi_{\ttr_3}\left(P^{(3)}_{41}(x+\hat{1})\right)\,
         \chi_{\ttr_4}\left(P^{(4)}_{32}(x+\hat{1})\right) \\
(5): & & \chi_{\ttr_1}\left(P^{(1)}_{34}(x)\right)\,
         \chi_{\ttr_3}\left(P^{(3)}_{41}(x)\right)\,
         \chi_{\ttr_4}\left(P^{(4)}_{31}(x)\right) \\
(6): & & \chi_{\ttr_1}\left(P^{(1)}_{43}(x+\hat{2})\right)\,
         \chi_{\ttr_3}\left(P^{(3)}_{14}(x+\hat{2})\right)\,
         \chi_{\ttr_4}\left(P^{(4)}_{13}(x+\hat{2})\right) \\
(7): & & \chi_{\ttr_1}\left(P^{(1)}_{42}(x)\right)\,
         \chi_{\ttr_2}\left(P^{(2)}_{14}(x)\right)\,
         \chi_{\ttr_4}\left(P^{(4)}_{12}(x)\right) \\
(8): & & \chi_{\ttr_1}\left(P^{(1)}_{24}(x+\hat{3})\right)\,
         \chi_{\ttr_2}\left(P^{(2)}_{41}(x+\hat{3})\right)\,
         \chi_{\ttr_4}\left(P^{(4)}_{21}(x+\hat{3})\right),
\eea
%\eeas
%
where each cube is represented in terms of characters of `dual'
plaquettes, and hermitian conjugation relations such as
$P^{(\mu)}_{\nu\lambda}(x)^\dagger = P^{(\mu)}_{\lambda\nu}(x)$
are used repeatedly. The group integration then yields %
\bea \int [\dd U]
\,\mbox{(1)}\times\mbox{(2)}\times\mbox{(3)}\times
\mbox{(4)}\times\mbox{(5)}\times\mbox{(6)}\times\mbox{(7)}\times
\mbox{(8)} = \frac{1}{(d_{\ttr_1} d_{\ttr_2} d_{\ttr_3}
d_{\ttr4})^4}. \nonumber \eea
Intuitively, the result can be understood in terms of the
resulting three manifolds. Rewriting contribution of each
representation as $d_\ttr^{-4} = d_\ttr^2\times d_\ttr^{-6}$, the
power `2' in the first factor is interpretable as the Euler
characteristic of ${\mathbb{S}}^2$ made of six `dual' plaquettes
corresponding to $\chi_{\tt R}$, and the power `$-6$' in the
second factor is the correct normalization of $\chi_\ttr$ that
permits 't Hooft's large-$N$ power counting. Summing over all
possible representations of the cubes, the total contribution is
given by
\bea \sum_{\ttr_1, \ttr_2, \ttr_3, \ttr_4}\hspace{-3mm}'
\hspace{6mm} \left(\frac{1}{d_{\ttr_1} d_{\ttr_2} d_{\ttr_3}
d_{\ttr_4}}\right)^4 \, \left(\widetilde{C}_{\ttr_1 \ttr_2
\ttr_3}(\beta)\widetilde{C}_{\ttr_2 \ttr_3 \ttr_4}(\beta)
\widetilde{C}_{\ttr_3 \ttr_4 \ttr_1}(\beta)\widetilde{C}_{\ttr_4
\ttr_1 \ttr_2}(\beta)\right)^2, \eea
where the prime ($'$) of the summation stands for excluding the
term $(\ttr_1, \ttr_2, \ttr_3, \ttr_4) = (0,0,0,0)$. The leading
nonzero term comes from the case $\ttr_1 = \ttr_2 = \ttr_3 =
\ttr_4 = \Box$ and its complex conjugate, yielding
$2(3\beta)^8/N^{16}$. The next order contribution comes from the case 
the representations involve identity or adjoint. Because $C_{0,0,{\rm
ad}}(\beta)$, $C_{0, {\rm ad}, {\rm ad}}(\beta)$, $C_{{\rm ad},
{\rm ad}, {\rm ad}}(\beta)$ all start with $(3\beta)^2$, it gives
the contribution of ${\cal O}(\beta^{12})$. Therefore, we find the
partition function in strong-coupling expansion as:
\bea {\cal Z} & = &
%e^{-\beta N^3 d(d-1)(d-2){\cal N}_s}\,
\Big(C_{000}(\beta)\Big)^{\frac{1}{3!}d(d-1)(d-2){\cal N}_s} \nn \\
 &  & \times   \left[1 + \frac{1}{4!}d(d-1)(d-2)(d-3){\cal N}_s
\cdot \frac{2}{N^{16}}
\,\Big(\widetilde{C}_{\Box\Box\Box}(\beta)\Big)^8 + {\cal
O}(\beta^{12})\right]. \label{resultcalZ} \eea
Here, as explained already, the series-expansion in $C_{\ttr_1
\ttr_2 \ttr_3}(\beta)$ represents sum over closed three-manifolds.
Since terms higher-order in $C_{\ttr_1 \ttr_2 \ttr_3}(\beta)$ come
from contractions among the same variables repeated many times,
the power series of $3\beta$ in $C_{\ttr_1 \ttr_2 \ttr_3}(\beta)$
is interpretable as contributions of singular, degenerate closed
three-manifolds.

%%%%%%%%%%%%%%%%%%%%%%%%%%%%%%%%%%%%%%%%%%%%%%%%%%%%%%%%%%%%%%%%%%%
\subsection{Wilson surface observables}
%%%%%%%%%%%%%%%%%%%%%%%%%%%%%%%%%%%%%%%%%%%%%%%%%%%%%%%%%%%%%%%%%%%
\subsubsection{Nonabelian Wilson surfaces}
We begin with a digression regarding nonabelian Wilson surfaces.
It is normally considered that, for nonabelian 2-form gauge
theory, Wilson surfaces are ill-defined. We now show that the
no-go theorem is evaded for the class of nonabelian tensor gauge
theory studied in this paper.

Consider the trace of parallel transport around a closed surface
$\Sigma$ ($\partial \Sigma = 0$):
\bea {\rm Tr} U(\Sigma) = \prod_{P \in \Sigma} U_P. \nonumber \eea
In general, the surface $\Sigma$ would be self-intersecting. We
will call expectation value of such variables:
\bea {\cal W}(\Sigma) \equiv \langle {\rm Tr} U(\Sigma) \rangle,
\eea
as the Wilson surface observables, and expectation value of their
products:
\bea {\cal W}(\Sigma_1, \cdots, \Sigma_n) \equiv \langle {\rm Tr}
U(\Sigma_1) \, {\rm Tr} U(\Sigma_2) \, \cdots {\rm Tr} U(\Sigma_n)
\rangle \eea
as Wilson surface correlators. The Wilson surface observables
measure {\sl internal energy} of the system, and the (connected
components of) the correlators measure correlation length, mass
gap, spectrum, etc.

Wilson surface observables constitute the fundamental basis of the
system. Indeed, extending the argument of \cite{durhuus}, one can
assert that every gauge-invariant operator ${\cal O}$, which
depends continuously on the plaquette variables can be
approximated arbitrarily well by a power-series of the Wilson
surface variables:
\bea {\cal O} \simeq  \sum_{n \ge 0} \sum_{\{\Sigma_n\}}
c(\Sigma_1, \cdots, \Sigma_n) \, {\rm Tr} U(\Sigma_1) \cdots {\rm
Tr} U(\Sigma_n). \eea

\begin{figure}[htbp]
    \begin{center}
    \includegraphics[width=0.8\linewidth,keepaspectratio,clip]
      {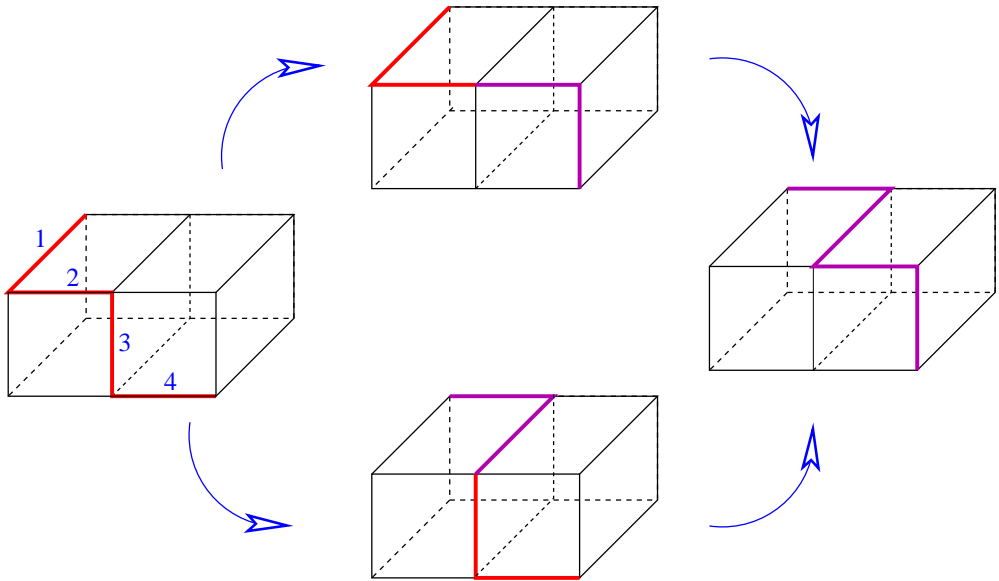}
    \end{center}
\caption{\sl Two alternative routes around two disjoint plaquettes
between two string configurations.} \label{fig:move}
\end{figure}

An important feature of proposed nonabelian Wilson surface
observables is that it bypasses folklore that such observables are
afflicted by ordering ambiguity. For a Wilson surface observable
of minimal size, viz. the one defined on a cube, we have already
shown that there is no ambiguity. Could there be any ambiguity
when the observables encompass cubes more than one? We will now
argue that there is no ambiguity by illustrating a few nontrivial
cases. The first one involves two cubes, as shown in the figure
\ref{fig:move}. Although there are two possible routes of color
move across the plaquettes involved,
\vskip0.3cm
%%%%%%%%%%%%%%%%%% simple move %%%%%%%%%%%%%%%%%%%%%%%%%%%%
%\begin{figure}
%\epsfxsize=13cm \epsfysize=9cm \centerline{\epsfbox{move.eps}}
%%%%%%%%%%%%%%%%%%%%%%%%%%%%%%%%%%%%%%%%%%%%%%%%%%%%%%%
\vskip0.3cm

\begin{figure}[htbp]
    \begin{center}
    \includegraphics[width=0.7\linewidth,keepaspectratio,clip]
      {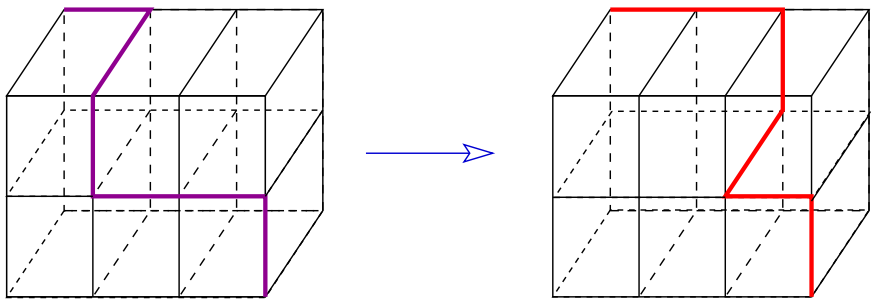}
    \end{center}
\caption{\sl Two alternative routes around an elementary cube
between two string configurations.} \label{fig:move2}
\end{figure}

We can also illustrate our claim from more sophiscated string move
in a given Wilson surface observable, as depicted in
Fig.\ref{fig:move2}.
\vskip0.3cm
%%%%%%%%%%%%%%%%%% cube move %%%%%%%%%%%%%%%%%%%%%%%%%%%%
%\begin{figure}
%\epsfxsize=15cm \epsfysize=4cm \centerline{\epsfbox{move2.eps}}
%%%%%%%%%%%%%%%%%%%%%%%%%%%%%%%%%%%%%%%%%%%%%%%%%%%%%%%
\vskip0.3cm
%

%%%%%%%%%%%%%%%%%%%%%%%%%%%%%%%%%%%%%%%%%%%%%%%%%%%%%%%%%%%%%%%
\subsubsection{One-point correlator and internal energy}
Having shown that nonabelian Wilson surface operators are well-defined
physical observables, we now consider the simplest of these
operators, taking a rectangular shape: $W_x(I,J,K)$ which
represents a box with the positive orientation, composed by the
three edges $(x, x+ I \hat{1})$, $(x, x+ J \hat{2})$, $(x, x+
K\hat{3})$. It is constructed by tiling faces of the box with the
plaquette variables $U_{\mu\nu}$. In computation of the vacuum
expectation value $\bra W_x(I,J,K)\ket$ in the strong coupling
expansion, contributions correspond to various three-dimensional
manifolds bounded by the box $W_x(I,J,K)$.

Let us begin with $\bra W_x(1,1,1)\ket$. It is nothing but the
internal energy and is computable as
\bea \bra { W}_x(1,1,1) \ket & = &
\frac{3}{d(d-1)(d-2)}\frac{1}{{\cal N}_s}
\,\frac{\partial}{\partial (3\beta)}\,\ln {\cal Z} \\
 & = & \widetilde{C}_{\Box\Box\Box}(\beta) + 2(d-3)\frac{1}{N^{16}}
\left(\widetilde{C}_{\Box\Box\Box}(\beta)\right)^7 + {\cal
O}(\beta^9), \eea
where the first term comes from the minimal volume configuration,
and the second from elementary fluctuations consisting of seven
cubes.

Extending the computation to general Wilson surface $\bra
W_x(I,J,K)\ket$, we find
\bea \bra { W}_x(I,J,K)\ket & = &
\left(\frac{\widetilde{C}_{\Box\Box\Box}(\beta)}{N^3}\right)^{IJK}\hskip-0.6cm
\times N^{I+J+K} \left[1+2(d-3)IJK
\frac{(\widetilde{C}_{\Box\Box\Box}(\beta))^6}{N^{16}} + {\cal
O}(\beta^8)\right]. \hskip5mm \label{WIJK} \eea
The leading contribution gives the volume-law, indicating that
colored strings are confined at strong coupling. The hypersurface
tension ${\cal M}$ represents strength of the volume-law and is
defined by the strong-coupling behavior of the one-point
correlator for large $I,J,K$:
\bea \bra {W}_x(I,J,K)\ket \simeq \exp \Big(-IJK{\cal M}
-(IJ+JK+KI){\cal A} -(I+J+K){\cal P}\Big) \nn \eea
with ${\cal A}$ and ${\cal P}$ being some nonuniversal constants.
It is evident from the first term that the one-point correlator
decays with the volume $IJK$ times the hypersurface (membrane)
tension ${\cal M}$. From the result Eq.(\ref{WIJK}), the
hypersurface tension ${\cal M}$ is extracted as
\bea {\cal M} = \ln \frac{N^3}{3\beta} -3(3\beta)^2
-\frac{17}{2}(3\beta)^4 -\left\{41 +
\frac{2(d-3)}{N^{16}}\right\}(3\beta)^6 + {\cal O}(\beta^8) \nn
\eea
for $N\geq 4$.

%%%%%%%%%%%%%%%%%%%%%%%%%%%%%%%%%%%%%%%%%%%%%%%%%%%%%%%%%%%%%%%%%%%
\subsubsection{Two-point correlators and excitation spectrum}
\begin{figure}[htbp]
    \begin{center}
    \includegraphics[width=0.7\linewidth,keepaspectratio,clip]
      {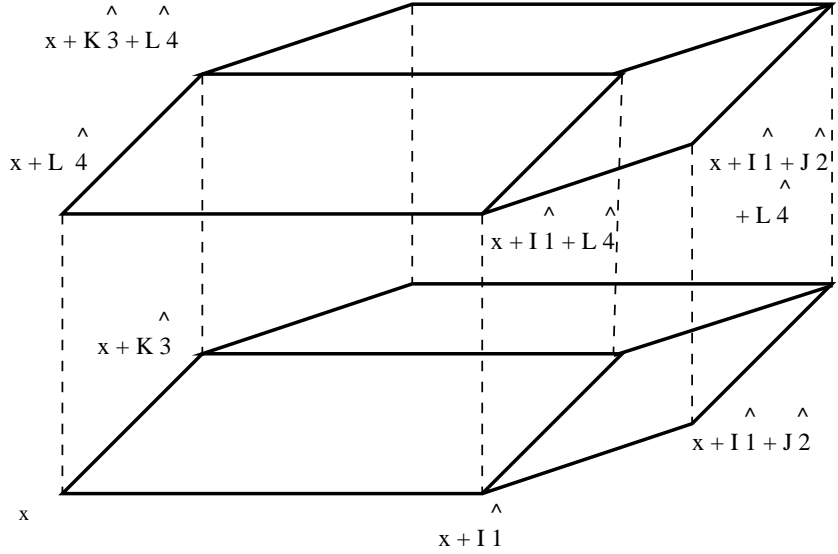}
    \end{center}
\caption{\sl Schematic view of a configuration of the case (I) for
the two-point correlator $\bra {\cal W}_x(I,J,K) {\cal W}_{x+L
\hat{4}}(I,J,K)^*\ket_C$. Elementary cubes from the action are
contracted to fill the space between ${\cal W}_x(I,J,K)$ and ${\cal
W}_{x + L\hat{4}}(I,J,K)$ indicated by the dashed lines. }
\label{fig:caseIL}
\end{figure}

For the connected two-point function $\bra { W}_x(I,J,K) {
W}_{x+L\hat{4}}(I,J,K)^*\ket_{\rm conn}$, there are two candidates
giving the leading contribution. The first one is the case that
cubes from the action are all used to fill the space between
${ W}_x(I,J,K)$ and ${W}_{x+L \hat{4}}(I,J,K)$. There are
no cubes filling inside each Wilson surface. We call it the
case (I) (See Fig. \ref{fig:caseIL}). The contribution amounts to
\bea (\mbox{Case (I)}) =
\left(\frac{\widetilde{C}_{\Box\Box\Box}(\beta)}{N^3}
\right)^{2(IJ+JK+KI)L} \times \Big[1+ {\cal O}(\beta^6)\Big], \nn
\eea
where the ${\cal O}(\beta^6)$ contribution is from elementary
fluctuations.

\vskip0.3cm
%%%%%%%%%%%%%%%%%% Fig. case (I) %%%%%%%%%%%%%%%%%%%%%%%%%%%%
%\begin{figure}
%\epsfxsize=9cm \epsfysize=6cm \centerline{\epsfbox{caseIL.eps}}
%%%%%%%%%%%%%%%%%%%%%%%%%%%%%%%%%%%%%%%%%%%%%%%%%%%%%%%
\vskip0.3cm

\begin{figure}[htbp]
    \begin{center}
    \includegraphics[width=0.7\linewidth,keepaspectratio,clip]
      {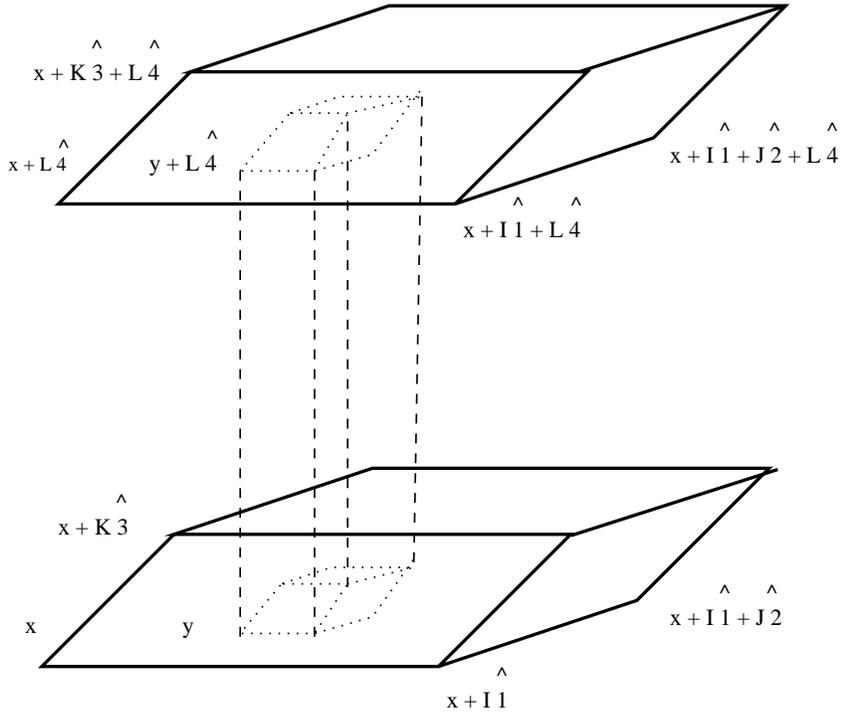}
    \end{center}
\caption{\sl Schematic view of a configuration of the case (II) for
the two-point correlator $\bra {\cal W}_x(I,J,K) {\cal W}_{x+L
\hat{4}}(I,J,K)^*\ket_{\rm conn}$. The two surfaces ${\cal
W}_x(I,J,K)$ and ${\cal W}_{x + L \hat{4}}(I,J,K)$ are connected by
the shortest tubular stack of elementary cubes (the dashed lines)
attaching at the holes $y$ and $y + L\hat{4}$. } \label{fig:caseIIL}
\end{figure}

The second possibility is that the elementary cubes first fill up
each interior of the two Wilson surfaces and leave holes of the
minimal size located at $y$ and $y + L\hat{4}$, respectively, and
then they are connected at the holes via a stack of minimal number
of elementary cubes ($6L$ cubes). See Fig. \ref{fig:caseIIL}. The
contribution is given by
\bea (\mbox{Case (II)}) = IJK \,
\left(\frac{\widetilde{C}_{\Box\Box\Box}(\beta)}{N^3}\right)^{2(IJK
+ 3L -1)} \times N^{2(I+J+K+L-3)} \times \Big[1+ {\cal
O}(\beta^6)\Big], \eea
where the overall factor $IJK$ comes from the sum with respect to
the location of $y$.

\vskip0.3cm
%%%%%%%%%%%%%%%%%% Fig. case (II) %%%%%%%%%%%%%%%%%%%%%%%%%%%%
%\begin{figure}
%\epsfxsize=9cm \epsfysize=6cm \centerline{\epsfbox{caseIILnew.eps}}
%%%%%%%%%%%%%%%%%%%%%%%%%%%%%%%%%%%%%%%%%%%%%%%%%%%%%%%
\vskip0.3cm

\begin{figure}[htbp]
    \begin{center}
    \includegraphics[width=0.9\linewidth,keepaspectratio,clip]
      {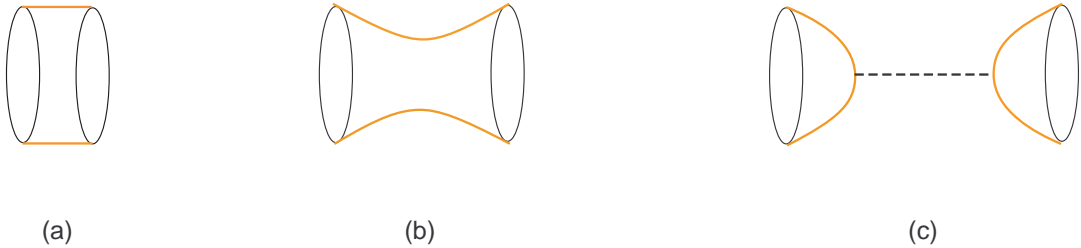}
    \end{center}
\caption{\sl Behavior of the Wilson loop two-point correlator in
gauge theory. At small separation (a), the first Wilson loop is
joined by a minimal surface to the second Wilson loop. At moderate
separation (b), the minimal surface deflects in the middle. At large
separation (c), each Wilson loop is filled up by respective minimal
surface, joined by a thin tubular column.} \label{fig:screening}
\end{figure}

Comparing the power of $\widetilde{C}_{\Box\Box\Box}(\beta)$ for
the two cases, we can see which configuration dominates in the
strong coupling. For the case $I\sim J\sim K$, the case (II)
dominates if $L \gsim \frac16 I$ i.e.  the separation is larger
than the scale of the surfaces. The case (I) dominates if $L \lsim
\frac16 I$. For generic $I, J, K$, we can draw a similar
conclusion. From these considerations, we find that the theory
develops a mass gap and screening in that the two-point correlator
undergoes a `phase transition' as the separation is varied. The
situation is rather analogous to the transition taking place for
Wilson loop correlators in gauge theory, as depicted in
Fig.\ref{fig:screening}.

\vskip0.3cm
%%%%%%%%%%%%%%%%%% Fig. case (II) %%%%%%%%%%%%%%%%%%%%%%%%%%%%
%\begin{figure}
%\epsfxsize=13cm \epsfysize=4cm \centerline{\epsfbox{screening.eps}}
%%%%%%%%%%%%%%%%%%%%%%%%%%%%%%%%%%%%%%%%%%%%%%%%%%%%%%%
\vskip0.3cm
%

%%%%%%%%%%%%%%%%%%%%%%%%%%%%%%%%%%%%%%%%%%%%%%%%%%%%%%%%%%%%%%%%%%%%%
\subsection{Large-$N$ Reduction and Asymptotic Behavior}
In lattice gauge theory, it is well known that large $N$ limit 
exhibit reduction of degrees of freedom, so-called Eguchi-Kawai reduction.
In this section, we will find an indication that a similar reduction
takes place in the large $N$ limit of the lattice tensor gauge theory. 

To observe an indication, consider again the strong coupling expansion of 
the partition function Eq.(\ref{resultcalZ}). There, all the correction terms 
in the square bracket are suppressed by some power of $1/N$ as the 
$(\widetilde{C}_{\Box\Box\Box}(\beta))^8$ term. Under $N\limit \infty$ limit with $\beta$ fixed, we
%naively 
can discard the correction terms, at least at leading order in $1/N$ expansion. We then see that
the partition function ${\cal Z}$ is reduced to the following zero-dimensional, 
unitary three-matrix model:
\bea {\cal Z} & \Rightarrow & \left(Z_{\rm MM}
\right)^{\frac{1}{3!}d(d-1)(d-2){\cal N}_s}
\eea
where
\bea
Z_{\rm MM} & = & C_{000}(\beta) = \int [\dd U][\dd V][\dd W] \,
e^{3\beta\,(\tr\,U)\,(\tr \,V)\, (\tr \,W) + \, {\rm c.c.}}. \eea
In this limit, the elementary Wilson surface operator ${\cal
W}_x(1,1,1)$ alone yields a nonvanishing vacuum expectation value
$\widetilde{C}_{\Box\Box\Box}(\beta)$. The connected two-point
correlators vanish, and thus the large $N$ factorization
\bea \bra {\cal W}_x(1,1,1) {\cal W}_{x'}(1,1,1)\ket = \bra {\cal
W}_x(1,1,1) \ket \, \bra {\cal W}_{x'}(1,1,1) \ket \eea
holds. 

In the strong coupling expansion, the matrix model partition function 
$Z_{\rm MM}$ still captures the divergent behavior of perturbation series. 
As $N \limit \infty$, the $n!$ growth in the series $C_{000}(\beta)$ continues to 
infinite orders. Therefore, the perturbation series is divergent and not Borel summable.
Likewise, for the free energy $\ln Z_{\rm MM}$, the perturbation series behaves
asymptotically as $n!\, (3\beta)^{2n}$. 

The Borel non-summability of the strong-coupling expansion implies that the entropy in 
defining the lattice partition function of the tensor gauge theory grows much faster than that 
of the Yang-Mills theory. Intuitively, we can understand this by the following geometric considerations.
At large $N$ limit, the strong coupling expansion of Yang-Mills partition function is interpretable as sum over random surfaces. Likewise, at large $N$ limit, we found above that the strong coupling expansion of nonabelian tensor partition function is interpretable as sum over random volumes. Thus, we would like to compare the entropy of random volumes in comparison with the entropy of random surfaces.
So, consider the partition function of random surfaces $\Sigma$ of area $A$. For a fixed area $A$, 
the partition function takes the form
\bea Z(A, \mu, N) = \sum_{\chi} e^{-\mu A} A^{-\chi} N^{-\chi} \eea
for a fixed area $A$. The sum is over all possible topologies of the random surface $\Sigma$. 
We are interested in the density of states, equivalently, density of states. At fixed $N$, it is 
known that the partition function scales as \cite{randomsurfaceentropy}
\bea Z \sim \exp (A \log A). \label{surface} \eea
This implies that the entropy $S$ of the random surface for a spherical topology must behave as
\bea S(\mathbb{S}^2) \sim e^{c A} A^\nu \eea
for some numerical factors $c, \nu$.  that the entropy of random
volume of $S^3$ topology behaves as
\bea S (\mathbb{S}^3) \sim \exp (\alpha A \log A). \eea
The proof goes as follows. Consider the map from $\mathbb{R}^3$ to itself. The map may be
represented by a vector ${\bf X} = {\bf X}(\xi^1, \xi^2, \xi^3)$. The map has folds at
critical points where
\bea \det \left( {\partial {\bf X} \over \partial \xi} \right) = 0. \eea
But, the boundaries of the folds are nothing but 2-dimensional random surfaces. For the
latter, we already know that the partition function behaves as Eq.(\ref{surface}). Hence,
we know that the entropy of random volume ought to behave the same way, viz.
\bea Z (\mathbb{S}^3) \sim \exp ( V_3 \log V_3). \eea

The reduction implies ultra-local nature of the theory, with no degrees of freedom propagating.
In Wilson's lattice gauge theory case, such a reduction is
different from the large $N$ reduction of Eguchi-Kawai \cite{EguchiKawai},
and does not lead an interesting theory.
In fact, the free energy simply gives $\ln z(\gamma) = \gamma^2$ in the limit
$N\limit \infty$ with $\gamma$ fixed.
In the tensor theory case, however some nontrivial pieces corresponding
to singular configurations of three-manifolds seem to
remain after the reduction.

Of course, this argument on the large $N$ limit is rather formal.
Discarding the correction terms needs to be justified by a careful treatment.
About this, we will report elsewhere.

%%%%%%%%%%%%%%%%%%% section 6: Discussion  %%%%%%%%%%%%%%%%%%%%%%%%%%
\setcounter{equation}{0}
\section{Discussions}

We constructed a lattice model for a nonabelian generalization of
two-form tensor gauge theory starting from an observation of the
Wilson's lattice theory. Requirement of consistent dimensional
reduction to lower dimensional Yang-Mills theory uniquely
determines the theory, which satisfies compactness conditions of
rank-$N^2$(I). Also, computation of strong coupling expansion was
done for the partition function and some Wilson surface
observables. There, the character expansion coefficients much more
rapidly grow as a power series of the coupling $\beta$ compared to
the ordinary lattice gauge theory case. It possibly suggests that
singular contributions to random three-dimensional geometry are
much more dominant than those to two-dimensional one.

There are many interesting and important  related to this work.
We will mention some directly related problems of them.

\begin{itemize}

\item The continuum limit at quantum level needs to be understood better.
In the classical continuum limit, we found that
the tensor gauge theory becomes purely Gaussian. However, there is a
possibility that the lattice theory we formulated may have a nontrivial
ultraviolet fixed point, where an interacting quantum theory of nonabelian
tensor gauge fields can be defined. It would be very interesting to explore 
the possibility via numerical simulation. \hfill\break

\item Universality, which is also related to a suitable continuum limit, 
need to be understood better. Namely, is the continuum theory independent of the
latticization? Here, we constructed the model on a hypercubic
lattice, where the plaquette variables $(U_{\mu\nu}(x))_{ijkl}$
encode the square lattice structure (and thus the hypercubic one)
to the four color indices $i,j,k,l$. Since the dynamical variables
explicitly depend on information of the latticization, at a glance
the universality issue seems to be problematic. (In the ordinary
lattice gauge theory case, variables are assigned on links. Note
that each link variable does not explicitly depend on a lattice
structure --- triangular lattice, square lattice, and so on.) By
solving the rank $N^2$(I) conditions, however we can rewrite them
as `dual' link variables. Now, the hypercubic structure is not
explicitly visible any longer in each `dual' link variable! Thus,
the theory with rank $N^2$(I) conditions seems hopeful also from
the viewpoint of universality. Anyway, check of the universality
is an important issue related to the Lorentz invariance of the
resulting continuum theory. \item The behavior $n!\,
(3\beta)^{2n}$ in the character expansion coefficients reminds us
of higher order behavior of weak coupling perturbation series in
quantum field theory, rather than the strong coupling. It might
suggest an interesting possibility that strong coupling region of
the large-$N$ theory is dual to some perturbative field theory.
\item We discussed a large $N$ reduction in somewhat speculative
way. Related to the duality, it would be interesting to
investigate weak coupling phase in the unitary three-matrix model
similar to Ref. \cite{GrossWitten}. \hfill\break

$\bullet$ Supersymmetric extension of the lattice tensor gauge theory
is very important, especially, in the context of $(5+1)$-dimensional
(2,0) theory. In recent years, there has been enormous progress in 
formulating supersymmetric Yang-Mills theory on a lattice. In the present
context, however, there are further stumbling blocks that need to be
overcome. In (2,0) theory, degrees of freedom involves not the whole
of tensor gauge field but only self-dual part of it. Self-dual tensor field 
in (5+1) dimensions
is chiral, described by field equations which is first-order in time.
Therefore, once put on a lattice naively, the self-dual tensor field would faces 
the problem of species doubling --- the tensor field on the lattice would 
involve not only self-dual part but also anti-self-dual part. The situation is
exactly the same as chiral bosons and chiral fermions in (1+1) dimensions. 
\hfill\break

\end{itemize}

\section*{Acknowledgement}
We are grateful to D.J. Gross and A.M. Polyakov
for numerous insightful discussions and suggestions. We also acknowledge N. Arkani-Hamed, P. Etingof, A. Gustavssson, H. Kawai, A. Kirillov, J.M. Maldacena, T. Suyama and E. Witten for discussions and correspondences. SJR also thanks the Institute for Advanced Study for generous financial support to his membership through the U.S. Department of Energy Grant DE-FG02-90ER40542. SJR was supported by the National Research Foundation of Korea Grants KRF 2005-084-C00003, KOSEF 2009-008-0372 and EU-FP Marie Curie Research \& Training Networks HPRN-CT-2006-035863 (K209090-00001-09B1300-00110). FS was supported in part by the Grant-in-Aid of Japan for Scientific Research (C) 21540290.

\medskip

%%%%%%%%%%%%%%%%%%%%% Appendix A %%%%%%%%%%%%%%%%%%%%%%%%%%%%%%%%%%%%%
\section*{Appendix}
\setcounter{equation}{0}
\renewcommand{\theequation}{\Alph{section}.\arabic{equation}}
\appendix
\section{Reduction of plaquette variables to link variables}
In this appendix, we show that the nonabelian tensor gauge theory
proposed in this paper is reduced consistently to Wilson's lattice
gauge theory by truncating internal degrees of freedom, which
transform as adjoint representation of $V_{\mu}(x)$, according to
the standard dimensional reduction Eq.(\ref{red_to_link}).

Under the dimensional reduction, the action density of an
elementary cube along $d$-th direction is reduced to that of an
elementary plaquette:
\bea
{\cal U}[C_{\mu\nu d}(x)]& \Rightarrow &  N^2 \,
{\cal U}[P_{\mu\nu}(x)], \nonumber \\
{\cal U}[P_{\mu\nu}(x)] & \equiv &
\tr\,\left(U_{\mu}(x)U_{\nu}(x+\mu)
U_{\mu}(x+\hat{\nu})^{\dagger}U_{\nu}(x)^{\dagger}\right),
\nonumber \eea
where we have used the unitarity relation
\bea \sum_{i,j,k,l}
\Big(U_{\mu\nu}(x)\Big)_{ijkl}\Big(U_{\mu\nu}(x)\Big)_{ijkl}^* =
N^2.
%\label{compact} \eea
\nonumber \eea
Both of ${\cal U}[C_{\mu dd}(x)]$ and ${\cal U}[C_{ddd}(x)]$
reduce to $N^3$, but do not contribute to the action.

Thus, after the lattice dimensional reduction, the action becomes
\bea S_{\rm reduced} = -3\beta N^2
\sum_{\{x\}}\sum_{\mu,\,\nu=1}^{d-1}\hspace{-2mm}'
\hspace{2mm}\mbox{Re}\,\Big({\cal U}[P_{\mu\nu}(x)] - N\Big)
-\beta \sum_{\{x\}}\sum_{\mu,\,\nu,\,\lambda=1}^{d-1}
\,\mbox{Re}\,\Big({\cal U}[C_{\mu\nu\lambda}(x)] - N^3\Big). \nn
\eea
Here the prime ($'$) refers to sum over $\mu, \nu$ omitting
$\mu=\nu$ terms, viz. ${\cal U}[P_{\mu\mu}(x)]=N$ contribution.
Evidently, the resulting action describes Wilson's lattice gauge
theory together with the nonabelian lattice tensor gauge theory,
both in $(d-1)$ dimensions.

As shown in section 3, in the classical continuum limit, the
second term in the reduced action scales as ${\cal O}(a^6)$. On
the other hand, expanding the link variable $U_{\mu}(x) =
e^{iaA_{\mu}(x)}$, the first term produces the Yang-Mills action
at ${\cal O}(a^4)$. Thus, in the classical continuum limit, the
first term dominates over the second, and the system is reduced to
the ordinary U($N$) gauge theory. This conclusion is valid for
theories defined by both rank-$N^2$(I) and rank-$N$ compactness
conditions.

%%%%%%%%%%%%%%%%%%%%%%%% Appendix B %%%%%%%%%%%%%%%%%%%%%%%%%%%%%%%%%
\section{Alternative nonabelian tensor gauge theory}
\setcounter{equation}{0} In this Appendix, we shall consider
alternative proposal for the nonabelian tensor gauge theory and
study its properties. The alternative one is defined in terms of
the rank-$N$ Eqs.(\ref{N-1} -- \ref{N-4}). We shall construct the
theory explicitly, work out classical continuum limit and
dimensional reduction thereof, and demonstrate that this
alternative theory does not lead to a physically meaningful
theory.
%%%%%%%%%%%%%%%%%%%%%%%%%%%%%%%%%%%%%%%%%%%%%%%%%%%%%%%%%%%%%%%%%
\subsection{polar parametrization of plaquette variable}
We will first develop an explicit parametrization of the plaquette
variables that solves the compactness conditions Eqs.(\ref{N-1} --
\ref{N-4}).

We find it convenient to introduce a set $M$ of $(N^2 \times N^2)$
matrices, and treat the plaquette variable
$\left(U_{\mu\nu}(x)\right)_{ijkl}$ as an element of $M$ where
$(i,j)$ and $(k,l)$ are interpreted as column and row indices. For
general elements $A_{ijkl}$ and $B_{ijkl}$ in $M$, we define a
matrix multiplication by $(AB)_{ijkl} =
\sum_{m,n}A_{ijmn}B_{mnkl}$. Evidently,
$I_{ijkl}=\delta_{ik}\delta_{jl}$ is an identity element of the
product. We define trace of matrices as $\Tr\, A \equiv
\sum_{ij}A_{ijij}.$ We also introduce a `twist' matrix $C_{ijkl}
\equiv \delta_{il}\delta_{jk}$ acting on a matrix $A_{ijkl}$ as
\bea (CA)_{ijkl} = A_{jikl}, \quad (AC)_{ijkl}= A_{ijlk}.
\nonumber \eea
Using these notations, the condition Eq.(\ref{orientation}) is
written as
\bea CU_{\mu\nu}(x)^\dagger C = U_{\nu\mu}(x),
\label{orientation2} \eea
where the `$\dagger$' refers to hermitian conjugation for the
$N^2\times N^2$ matrices.

In the $(N^2 \times N^2)$ matrix representation, the compactness
conditions Eq.(\ref{N-1} -- \ref{N-4}) are expressible as
\bea
\sum_k \Big( U_{\mu\nu}(x) U_{\mu\nu}(x)^\dagger \Big)_{iki'k}
& = & N \delta_{ii'}, \label{N-5}\\
\sum_k \Big( U_{\mu\nu}(x) U_{\mu\nu}(x)^\dagger \Big)_{kjkj'}
& = & N \delta_{jj'}, \label{N-6}\\
\sum_k \Big( U_{\mu\nu}(x)^\dagger U_{\mu\nu}(x) \Big)_{iki'k}
& = & N \delta_{ii'}, \label{N-7}\\
\sum_k \Big( U_{\mu\nu}(x)^\dagger U_{\mu\nu}(x) \Big)_{kjkj'}
& = & N \delta_{jj'}. \label{N-8}
\eea
Since $U_{\mu\nu}(x)$ is a matrix over $\mathbb{C}$, we
parametrize it via (a symmetric version of) the polar
decomposition:
\bea U_{\mu\nu}(x) = e^{\frac{i}{2}a^2 H_{\mu\nu}(x)}\,
R_{\mu\nu}(x)\, e^{\frac{i}{2}a^2 H_{\mu\nu}(x)} \nonumber \eea
for every $\mu, \nu$, where $a$ is a lattice spacing, and
$H_{\mu\nu}(x)$ is a $(N^2 \times N^2)$ Hermitian matrix
parametrizing the U($N^2$) subset of the configuration space
${\mathbb{X}}_N$. $R_{\mu\nu}(x)$ is a positive semi-definite
$(N^2 \times N^2)$ Hermitian matrices. The condition
Eq.(\ref{orientation}) implies that
\bea CH_{\mu\nu}(x)C = - H_{\nu\mu}(x), \qquad CR_{\mu\nu}(x)C =
R_{\nu\mu}(x). \nonumber \eea
Also, the compactness conditions Eq.(\ref{N-5} -- \ref{N-8}) read
\bea \sum_k \Big(e^{\frac{i}{2}a^2 H_{\mu\nu}(x)} R_{\mu\nu}(x)^2
e^{-\frac{i}{2}a^2 H_{\mu\nu}(x)}\Big)_{ikjk} = \sum_k
\Big(e^{\frac{i}{2}a^2 H_{\mu\nu}(x)} R_{\mu\nu}(x)^2
e^{-\frac{i}{2}a^2 H_{\mu\nu}(x)}\Big)_{kikj} = N\delta_{ij}.
\label{N-9} \eea

In the polar parametrization adopted above, the field
$R_{\mu\nu}(x)$ is expandable around the identity $I$ as
\bea R_{\mu\nu}(x)^2 = I -2a^2 A_{\mu\nu}(x) + a^4 c_2
A_{\mu\nu}(x)^2 + a^6 c_3 A_{\mu\nu}(x)^3 + \cdots .
\label{expansionR2} \eea
Here, $A_{\mu\nu}(x)$ are $(N^2 \times N^2)$ Hermitian matrices,
and $c_2, c_3, \cdots$ are real-valued constants. Taking the trace
of Eq.(\ref{N-9}), we have
\bea \Tr \left[R_{\mu\nu}(x)^2\right] = N^2. \nonumber \eea
At order ${\cal O}(a^{2m})$ ($m$: even), it leads
\bea c_m \Tr \left[A_{\mu\nu}(x)^m \right] = 0, \nonumber \eea
and implies that $A_{\mu\nu}(x) = 0$ when $c_m \neq 0$. Thus, for
$A_{\mu\nu}(x)$ to be nontrivial, we will need to set $c_m=0$ for
$m$ even.

The coefficients $c_3, c_5, \cdots$ are not fixable by the
conditions Eq.(\ref{N-9}) alone. Here, we will consider the
simplest case $c_3 = c_5 = \cdots = 0$ and discuss the classical
continuum limit in the next subsection. As will be shown there,
the final form of the continuum action does not change even when
$c_3, c_5, \cdots$ are kept nonzero. Thus, the parametrization of
$U_{\mu\nu}(x)$ is given by
\bea U_{\mu\nu}(x) = e^{\frac{i}{2}a^2 H_{\mu\nu}(x)} \, \sqrt{I
-2a^2 A_{\mu\nu}(x)} \, e^{\frac{i}{2}a^2 H_{\mu\nu}(x)},
\label{U_rankN} \eea
with $(N^2 \times N^2)$ Hermitian matrices $H_{\mu\nu}(x)$,
$A_{\mu\nu}(x)$ satisfying the constraints
\bea & & \sum_k \left(A_{\mu\nu}(x)\right)_{ikjk} =
\sum_k \left(A_{\mu\nu}(x)\right)_{kikj} = 0, \label{A} \\
& &  \sum_k \Big(\underbrace{[H_{\mu\nu}(x), \cdots , [H_{\mu\nu}(x)}_n,
A_{\mu\nu}(x)]\cdots]\Big)_{ikjk} \nn \\
& & = \sum_k \Big(\underbrace{[H_{\mu\nu}(x), \cdots ,
[H_{\mu\nu}(x)}_n, A_{\mu\nu}(x)]\cdots]\Big)_{kikj} = 0 \quad
\mbox{for } n > 0, \label{HA} \eea
as well as
\bea & & CH_{\mu\nu}(x)C = -H_{\nu\mu}(x) \qquad \mbox{and} \qquad
    CA_{\mu\nu}(x)C = A_{\nu\mu}(x).
\label{orientation3} \eea
Due to the constraints, in general, $A_{\mu\nu}(x)$ cannot move
independently of $H_{\mu\nu}(x)$.

%%%%%%%%%%%%%%%%%%%%%%%%%%%%%%%%%%%%%%%%%%%%%%%%%%%%%%%%%%%%%%%%%%%%
\subsection{classical continuum limit}
The plaquette variable $U_{\mu\nu}(x)$ in Eq.(\ref{U_rankN}) is
expanded around the identity $I$ as
\bea U_{\mu\nu}(x) = I & &+ a^2 \Big[iH_{\mu\nu}(x) - A_{\mu\nu}(x)\Big] \nn \\
& & + a^4\left[-\frac12 H_{\mu\nu}(x)^2 -\frac12 A_{\mu\nu}(x)^2
  -\frac{i}{2}\left(H_{\mu\nu}(x)A_{\mu\nu}(x)+ A_{\mu\nu}(x)H_{\mu\nu}(x)
              \right)\right] \nn \\
& & +a^6 \left[-\frac{i}{6}H_{\mu\nu}(x)^3 -\frac12 A_{\mu\nu}(x)^3
            -\frac{i}{4}\left(H_{\mu\nu}(x)A_{\mu\nu}(x)^2
                        + A_{\mu\nu}(x)^2H_{\mu\nu}(x)\right) \right. \nn \\
& & \hspace{1cm}\left.
 +\frac18 \left(H_{\mu\nu}(x)^2A_{\mu\nu}(x) +A_{\mu\nu}(x)H_{\mu\nu}(x)^2
       + 2 H_{\mu\nu}(x)A_{\mu\nu}(x)H_{\mu\nu}(x)\right)\right]  \nn \\
& & + {\cal O}(a^8). \label{expand_UN} \eea
From the gauge transformation rule Eq.(\ref{gaugetr}) and the
gauge function expanded as in Eq.(\ref{expandV}), we obtain
infinitesimal gauge transformation rules for $H_{\mu\nu}(x)$ and
$A_{\mu\nu}(x)$
\bea
\Big(H_{\mu\nu}(x)\Big)_{ijkl} & \limit &
\Big(H_{\mu\nu}(x)\Big)_{ijkl}
 + \delta_{ik}\Big(\partial_{\mu}\Lambda_{\nu}(x)\Big)_{jl}
 - \delta_{jl}\Big(\partial_{\nu}\Lambda_{\mu}(x)\Big)_{ik},
 \nn \\
\Big(A_{\mu\nu}(x)\Big)_{ijkl} & \limit &
\Big(A_{\mu\nu}(x)\Big)_{ijkl}, \nonumber \eea
and observe that $A_{\mu\nu}(x)$ is gauge invariant. Again, as for
the theory considered in the text, nonabelian interactions at
finite lattice spacing disappears in the classical continuum limit
and the gauge transformation rules are reduced to abelian ones.
So, in the continuum limit, gauge invariant field strength is
\bea \Big({\cal H}_{\mu\nu\lambda}(x)\Big)_{ijki'j'k'} \equiv
\Big(\partial_{\lambda}H_{\mu\nu}(x)\Big)_{iji'j'}\delta_{kk'} +
\Big(\partial_{\mu}H_{\nu\lambda}(x)\Big)_{jkj'k'}\delta_{ii'}
 + \Big(\partial_{\nu}H_{\lambda\mu}(x)\Big)_{kik'i'}\delta_{jj'},
\nonumber \eea
with the same symmetry property of indices as
Eq.(\ref{symmetry_calH}).

Substituting the expansion Eq.(\ref{expand_UN}) into
Eq.(\ref{cubeS}), after some algebra, we arrive at the continuum
action
\bea S_{\rm tensor} & = &\frac12\beta
a^6\sum_{\{x\}}\sum_{\mu,\nu,\lambda=1}^d \Big\{
\Tr_{N^3}\left[{\cal H}_{\mu\nu\lambda}(x)
                      {\cal H}_{\mu\nu\lambda}(x)\right]
+3N\,\Tr\left[\left(\partial_{\lambda}A_{\mu\nu}(x)\right)^2\right]
 \nn \\
 & &  \hspace{2.3cm} \frac{}{}
+16 \,T\left[A_{\mu\nu}(x), A_{\nu\lambda}(x),
A_{\lambda\mu}(x)\right]\Big\} \, +{\cal O}(a^7).
\label{continuumSN} \eea
Here, `$\Tr_{N^3}$' refers to the trace for $(N^3 \times N^3)$
matrices, which is defined for a generic element ${\cal
A}_{ijki'j'k'}$ as $\Tr_{N^3}({\cal A}) \equiv \sum_{i,j,k}{\cal
A}_{ijkijk}$. Also, $T\left[A_{\mu\nu}(x), A_{\nu\lambda}(x),
A_{\lambda\mu}(x)\right]$ expresses a trilinear interaction term,
which is defined for $(N^2 \times N^2)$ matrices $X$, $Y$, $Z$ as
\bea T[X,Y,Z] \equiv \sum_{i,j,k,l,m,n}\, X_{ijkl}Y_{lmjn}Z_{nkmi}
\eea
with the cyclic property $T[X,Y,Z] = T[Y,Z,X] = T[Z,X,Y]$. The
classical continuum limit is taken as in Eq.(\ref{coupling}), and
the result is Lorentz invariant and gauge invariant.

A remark is in order concerning the remainders in the small
lattice spacing expansion. One might wonder if higher-order terms
may yield nontrivial contributions. Even if we keep the next order
terms and consider
\bea a^6 c_3 A_{\mu\nu}(x)^3 + a^{10} c_5 A_{\mu\nu}(x)^5 + \cdots
\label{c3c5} \eea
in Eq.(\ref{expansionR2}), we arrive at the same continuum action.
In this case, in solving the conditions Eq.(\ref{N-9}),
Eq.(\ref{A}) remains the same but Eq.(\ref{HA}) is replaced by
\bea & &
\sum_k\Big([H_{\mu\nu}(x),A_{\mu\nu}(x)]\Big)_{ikjk}=
\sum_k\Big([H_{\mu\nu}(x), A_{\mu\nu}(x)]\Big)_{kikj}=0, \\
& & \sum_kc_3\Big(A^3_{\mu\nu}(x)\Big)_{ikjk}=
-\frac14\sum_k\Big([H_{\mu\nu}(x),[H_{\mu\nu}(x),A_{\mu\nu}(x)]]
\Big)_{ikjk},
\label{c3-1}\\
& &  \sum_kc_3\Big(A^3_{\mu\nu}(x)\Big)_{kikj}=
-\frac14\sum_k\Big([H_{\mu\nu}(x),[H_{\mu\nu}(x),A_{\mu\nu}(x)]]
\Big)_{kikj},
\label{c3-2}\\
& & \cdots. \nonumber \eea
In the small lattice spacing expansion, only the first term of
Eq.(\ref{c3c5}) is relevant, and it leads to $a^6
\Tr(c_3A_{\mu\nu}(x)^3)$. Using Eqs.(\ref{c3-1}, \ref{c3-2}),
however, the contribution vanishes. Therefore, again, we arrive at
the same action as Eq.(\ref{continuumSN}).

%%%%%%%%%%%%%%%%%%%%%%%%%%%%%%%%%%%%%%%%%%%%%%%%%%%%%%%%%%%%%%%%%%%%%%
\subsection{lattice dimensional reduction}
Following the procedure of section 4.2, we express the
dimensionally reduced plaquette variables $U_{d\mu}(x)$, $U_{\mu
d}(x)$, $U_{dd}(x)$ in $(N^2\times N^2)$ matrix notation as
\bea U_{\mu}(x) & \equiv & U_{d\mu}(x) =
e^{i\frac{a}{2}H_{\mu}(x)}\,
\sqrt{I-2aA_{\mu}(x)} \,e^{i\frac{a}{2}H_{\mu}(x)}, \nn \\
\tilde{U}_{\mu}(x) & \equiv & U_{\mu d}(x) =
e^{i\frac{a}{2}\tilde{H}_{\mu}(x)}\,
\sqrt{I-2a\tilde{A}_{\mu}(x)}\, e^{i\frac{a}{2}\tilde{H}_{\mu}(x)}, \nn \\
U(x) & \equiv & U_{dd}(x) = e^{i\frac{a}{2}\Phi(x)}\,
\sqrt{I-2a\eta(x)}\, e^{i\frac{a}{2}\Phi(x)}, \nonumber \eea
where the continuum fields of the mass dimension one are defined
as
\bea & & H_{\mu}(x) \equiv r H_{d\mu}(x), \qquad A_{\mu}(x) \equiv
r A_{d\mu}(x), \nn \\
& & \tilde{H}_{\mu}(x) \equiv r H_{\mu d}(x), \qquad
\tilde{A}_{\mu}(x) \equiv r A_{\mu d}(x), \nn \\
& & \Phi(x) \equiv \frac{r^2}{a}H_{dd}(x), \qquad \eta(x) \equiv
\frac{r^2}{a}A_{dd}(x), \nonumber \eea
satisfying the conditions corresponding to Eqs.(\ref{A} --
\ref{orientation3}):
\bea
 & & \sum_k \left(A_{\mu}(x)\right)_{ikjk} =
     \sum_k \left(A_{\mu}(x)\right)_{kikj} = 0, \nn \\
& &  \sum_k \Big(\underbrace{[H_{\mu}(x), \cdots , [H_{\mu}(x)}_n,
A_{\mu}(x)]\cdots]\Big)_{ikjk} \nn \\
& & =
 \sum_k \Big(\underbrace{[H_{\mu}(x), \cdots , [H_{\mu}(x)}_n,
A_{\mu}(x)]\cdots]\Big)_{kikj} = 0 \quad \mbox{for } n > 0, \nn \\
& & CH_{\mu}(x)C = -\tilde{H}_{\mu}(x),
\qquad CA_{\mu}(x)C = \tilde{A}_{\mu}(x),  \\
& & \sum_k \left(\eta(x)\right)_{ikjk} =
     \sum_k \left(\eta(x)\right)_{kikj} = 0, \nn \\
& &  \sum_k \Big(\underbrace{[\Phi(x), \cdots , [\Phi(x)}_n,
\eta(x)]\cdots]\Big)_{ikjk} \nn \\
& & =
 \sum_k \Big(\underbrace{[\eta(x), \cdots , [\eta(x)}_n,
\Phi(x)]\cdots]\Big)_{kikj} = 0 \quad \mbox{for } n > 0, \nn \\
& & C\Phi(x) C = -\Phi(x), \qquad C\eta(x) C = \eta(x). \nonumber
\eea
The gauge transformation rules are unchanged from
Eqs.(\ref{gaugetr_dimredUmu} -- \ref{gaugetr_dimredU}).
Introducing the $(N\times N)$ matrix notation as
\bea \Big( H_{\mu}(x)\Big)_{ijkl} &  = &
\Big(h_{\mu}(x)\Big)_{ik}\delta_{jl} +
\Big(H_{\mu}^{(jl)}(x)\Big)_{ik},
\qquad \sum_j H_{\mu}^{(jj)}(x) = 0, \nn \\
\Big( A_{\mu}(x)\Big)_{ijkl} & = &
\Big(A_{\mu}^{(jl)}(x)\Big)_{ik}, \qquad \sum_j A_{\mu}^{(jj)}(x)
= \tr\, A_{\mu}^{(jl)}(x) = 0, \nonumber \eea
it turns out that $h_{\mu}$ transforms as a vector gauge field and
$H_{\mu}^{(jl)}$, $A_{\mu}^{(jl)}$ as an adjoint matter field:
\bea h_{\mu}(x) & \limit & V(x)\,[h_{\mu}(x) -i\partial_{\mu}]\,
V(x)^\dagger,
\nn \\
H_{\mu}^{(jl)}(x) & \limit & V(x) H_{\mu}^{(jl)}(x) V(x)^\dagger, \nn \\
A_{\mu}^{(jl)}(x) & \limit & V(x) A_{\mu}^{(jl)}(x) V(x)^\dagger.
\label{gauge_hHA} \eea
Indices in the superscript $(jl)$ do not transform in the
continuum limit. Also, both  $\Phi$ and $\eta$ transform as
(adjoint)$\times$(adjoint):
\bea \Big(\Phi(x)\Big)_{ijkl} & \limit & \sum_{i'j'k'l'}
 \Big(V(x)\Big)_{ii'} \Big(V(x)\Big)_{jj'}
\Big(\Phi(x)\Big)_{i'j'k'l'} \Big(V(x)^\dagger\Big)_{k'k}
\Big(V(x)^\dagger\Big)_{l'l}, \nn \\
\Big(\eta(x)\Big)_{ijkl} & \limit & \sum_{i'j'k'l'}
 \Big(V(x)\Big)_{ii'} \Big(V(x)\Big)_{jj'}
\Big(\eta(x)\Big)_{i'j'k'l'} \Big(V(x)^\dagger\Big)_{k'k}
\Big(V(x)^\dagger\Big)_{l'l}.
\label{gauge_phieta}
\eea

The dimensionally reduced action now reads
\bea S & = & -\beta \sum_{\{x\}} \sum_{\mu,\nu,\lambda=1}^{d-1}\,
\mbox{Re}\left[{\cal U}[C_{\mu\nu\lambda}(x)]-N^3\right]
-3\beta\sum_{\{x\}}\sum_{\mu,\nu=1}^{d-1} \, \mbox{Re}
\left[{\cal U}[C_{\mu\nu d}(x)]-N^3\right] \nn \\
 & &  -3\beta \sum_{\{x\}} \sum_{\mu =1}^{d-1}\,
\mbox{Re}\left[{\cal U}[C_{\mu dd}(x)]-N^3\right] -\beta
\sum_{\{x\}} \, \mbox{Re}\left[{\cal U}[C_{ddd}(x)]-N^3\right].
\label{DRaction2} \eea
The first term gives rise to the same contribution as
Eq.(\ref{continuumSN}) but with the Lorentz indices running over
$1, \cdots, d-1$. The second term yields the ${\cal O}(a^4)$
contribution:
\bea & & -3\beta \sum_{\{x\}}\sum_{\mu,\nu=1}^{d-1} \, \mbox{Re}
\left[{\cal U}[C_{\mu\nu d}(x)]-N^3\right] \nn \\
&=& \beta a^4 \sum_{\{ x\}} \sum_{\mu, \nu=1}^{d-1}\left\{ \tr
\left[\frac32N^2 f_{\mu\nu}(x)^2 \right. \right.
  +3N\sum_{k,l} \left( D_{\mu}H^{(kl)}_{\nu}(x)D_{\mu}H^{(lk)}_{\nu}(x)
+  D_{\mu}A^{(kl)}_{\nu}(x)D_{\mu}A^{(lk)}_{\nu}(x)\right) \nn \\
& &\hspace{1cm} -\frac32\sum_{i,j,k,l}\left([H_{\mu}^{(ij)}(x), H_{\nu}^{(kl)}(x)]
 [H_{\mu}^{(ji)}(x), H_{\nu}^{(lk)}(x)]+
[A_{\mu}^{(ij)}(x), A_{\nu}^{(kl)}(x)]
 [A_{\mu}^{(ji)}(x), A_{\nu}^{(lk)}(x)] \right. \nn \\
& & \hspace{1cm}\left.+2[H_{\mu}^{(ij)}(x), A_{\nu}^{(kl)}(x)]
 [H_{\mu}^{(ji)}(x), A_{\nu}^{(lk)}(x)] \right)\left. \frac{}{}\right] \nn \\
& &\hskip2cm +24\sum_{i,j,k,l}\Big(A_{\mu\nu}(x)\Big)_{ijkl}\tr
\left( A_{\mu}^{(ki)}(x)A_{\nu}^{(lj)}(x)\right)\left.
\frac{}{}\right\} + {\cal O}(a^5), \label{munuD} \eea
where the field strength and the covariant derivatives are defined by
\bea
f_{\mu\nu} & \equiv & \partial_{\mu} h_{\nu} - \partial_{\nu} h_{\mu}
+i[h_{\mu}, h_{\nu}], \nn \\
D_{\mu}H_{\nu}^{(ij)} & \equiv & \partial_{\mu} H_{\nu}^{(ij)}
+i[h_{\mu}, H_{\nu}^{(ij)}], \nonumber\eea
and similarly for $D_{\mu}A_{\nu}^{(ij)}$.

However, contributions from the third and fourth terms start with
the ${\cal O}(a^3)$ terms as
\bea & &  -3\beta \sum_{\{x\}} \sum_{\mu =1}^{d-1}\,
\mbox{Re}\left[{\cal U}[C_{\mu dd}(x)]-N^3\right] \nn \\
&=& \sum_{\{ x\}} \sum_{\mu=1}^{d-1}\Big( \left[24\beta a^3
\eta_{lmjn}(A_{\mu}^{(ik)})_{jl}(A_{\mu}^{(ki)})_{nm} +12\beta a^4
(\partial_{\mu}\eta_{lmjn})(A_{\mu}^{(ik)})_{jl}
(A_{\mu}^{(ki)})_{nm}\right] \nn \\
& & \hskip1.5cm + \beta a^4 \Big\{ \frac32N\,\Tr\left[
(D_{\mu}\Phi)^2 + (D_{\mu}\eta)^2\right] \nn \\
& &\hspace{2.5cm}
-6\left[(H_{\mu}^{(ik)})_{nm}(H_{\mu}^{(ki)})_{jl} +
(A_{\mu}^{(ik)})_{nm}(A_{\mu}^{(ki)})_{jl}\right] \nn \\
& &\hspace{2.7cm} \times\left[\Phi_{lqun}\Phi_{umjq} +
\eta_{lqun}\eta_{umjq} + \Phi_{lqnu}\Phi_{mujq} +
\eta_{lqnu}\eta_{mujq} - (\Phi^2 + \eta^2)_{lmjn}
\right] \nn \\
& &\hspace{2.5cm} +6\left(H_{\mu}^{(ik)}H_{\mu}^{(ki)}+
A_{\mu}^{(ik)}A_{\mu}^{(ki)}\right)_{mn}
(\Phi^2 + \eta^2)_{nqmq} \nn \\
& &\hspace{2.5cm} +3\, \tr\left(H_{\mu}^{(ik)}H_{\mu}^{(kj)} +
A_{\mu}^{(ik)}A_{\mu}^{(kj)}
\right)\tr\left(H_{\mu}^{(jl)}H_{\mu}^{(li)} +
A_{\mu}^{(jl)}A_{\mu}^{(li)}
\right) \nn \\
& &\hspace{2.5cm} -3\, \tr\left(H_{\mu}^{(ik)}H_{\mu}^{(lj)} +
A_{\mu}^{(ik)}A_{\mu}^{(lj)}
\right)\tr\left(H_{\mu}^{(ji)}H_{\mu}^{(kl)} +
A_{\mu}^{(ji)}A_{\mu}^{(kl)}
\right) \nn \\
& &\hspace{2.5cm} -3\, \tr\left(H_{\mu}^{(ik)}A_{\mu}^{(lj)} -
A_{\mu}^{(ik)}H_{\mu}^{(lj)}
\right)\tr\left(H_{\mu}^{(ji)}A_{\mu}^{(kl)} -
A_{\mu}^{(ji)}H_{\mu}^{(kl)}
\right) \nn \\
& &\hspace{2.5cm} +12\left[\Phi_{mlnj}(H_{\mu}^{(qi)})_{nm} +
\eta_{mlnj}(A_{\mu}^{(qi)})_{nm}\right] \left([H_{\mu}^{(ik)},
H_{\mu}^{(kq)}] + [A_{\mu}^{(ik)}, A_{\mu}^{(kq)}]
\right)_{jl}\Big\}\Big) \nn \\
& + & {\cal O}(a^5), \label{muDD} \eea
and as
\bea
 & & -\beta \sum_{\{x\}} \,
\mbox{Re}\left[{\cal U}[C_{ddd}(x)]-N^3\right] \nn \\
&=& 8\beta a^3 \sum_{\{ x\}} T[\eta(x), \eta(x), \eta(x)]  \nn \\
& + & \beta a^4 \sum_{\{ x\}} \left\{\frac{}{}
-6\left(\Phi_{nkmi}\Phi_{jilk}
+\eta_{nkmi}\eta_{jilk}\right)\right.
%& & \hspace{3cm}\times
\left[\Phi_{lqun}\Phi_{umjq}+\eta_{lqun}\eta_{umjq}
-(\Phi^2 + \eta^2)_{lmjn}\right] \nn \\
& &\hspace{2cm}
-3\left(\Phi_{nkmi}\Phi_{jilk}+\eta_{nkmi}\eta_{jilk}\right)
\left(\Phi_{lqnu}\Phi_{mujq}+\eta_{lqnu}\eta_{mujq}\right) \nn \\
& &\hspace{2cm} +3\left(\Phi^2 + \eta^2 \right)_{mknk}
\left(\Phi^2 + \eta^2 \right)_{nqmq}  \nn \\
& &\hspace{2cm}
-\left(\Phi_{imkn}\eta_{lnjm}-\eta_{imkn}\Phi_{lnjm}\right)
\left(\Phi_{jpiq}\eta_{kqlp}-\eta_{jpiq}\Phi_{kqlp}\right)
\left.\frac{}{}\right\} \nn \\
&+& {\cal O}(a^5), \label{DDD} \eea
respectively. Here, the covariant derivative for $\Phi$ is defined
according to its transformation property as
(adjoint)$\times$(adjoint) under U($N$):
\bea \left(D_{\mu}\Phi\right)_{ijkl} & \equiv
&\partial_{\mu}\Phi_{ijkl}
+i\left(h_{\mu}\right)_{ii'}\Phi_{i'jkl}
+i\left(h_{\mu}\right)_{jj'}\Phi_{ij'kl}
-i\Phi_{ijk'l}\left(h_{\mu}\right)_{k'k}
 -i\Phi_{ijkl'}\left(h_{\mu}\right)_{l'l}, \qquad  \label{def_Dphi}
\eea
and similarly for $D_{\mu}\eta$. In Eqs.(\ref{muDD} --
\ref{def_Dphi}), Einstein convention for Latin indices is assumed
for simplicity of the notation.

All the expressions in Eqs.(\ref{munuD}, \ref{muDD}, \ref{DDD})
are manifestly gauge invariant except the first line in the
right-hand-side of Eq.(\ref{muDD}), where $\eta$-field is acted by
the ordinary derivative instead of the covariant derivative. To
see that it is nevertheless gauge invariant, we will need to
consider the ${\cal O}(a)$-correction contributions to the
transformation rules Eq.(\ref{gauge_hHA}). One can solve
Eq.(\ref{gaugetr_dimredUmu}) iteratively with respect to $a$ and
get
\bea h_{\mu}(x) & \limit & V(x)\left(h_{\mu}(x)
-i\partial_{\mu}\right)V(x)^\dagger
+a\, G_1(h_{\mu}(x)) + \cdots, \nn \\
H_{\mu}^{(jl)} & \limit & \sum_{j',l'}\Big(V_{\mu}(x)\Big)_{jj'}
\left[V(x)H_{\mu}^{(j'l')}(x)V(x)^\dagger + a\, G_1(H_{\mu}^{(j'l')}(x))
\right]\Big(V_{\mu}(x)^\dagger\Big)_{l'l} + \cdots, \nn \\
A_{\mu}^{(jl)} & \limit & \sum_{j',l'}\Big(V_{\mu}(x)\Big)_{jj'}
\left[V(x)A_{\mu}^{(j'l')}(x)V(x)^\dagger + a\,
G_1(A_{\mu}^{(j'l')}(x)) \right]\Big(V_{\mu}(x)^\dagger\Big)_{l'l}
+ \cdots, \label{oa_correction} \eea
where the ellipses denote ${\cal O}(a^2)$-terms and $G_1$'s are
${\cal O}(a)$ corrections originating from the
$V(x)$-transformations:
\bea
G_1(h_{\mu}(x)) & \equiv & \frac12\Big[
\left(\partial_{\mu}V(x)\right)h_{\mu}(x)V(x)^\dagger +
V(x)h_{\mu}(x)\partial_{\mu}V(x)^\dagger
-i\partial_{\mu}\left(V(x)\partial_{\mu}V(x)^\dagger\right)\Big], \nn \\
G_1(H_{\mu}^{(jl)}(x))& \equiv & \frac12\Big[
\left(\partial_{\mu}V(x)\right)H_{\mu}^{(jl)}(x)V(x)^\dagger +
V(x)H_{\mu}^{(jl)}(x)\partial_{\mu}V(x)^\dagger \Big], \nn \\
G_1(A_{\mu}^{(jl)}(x))& \equiv & \frac12\Big[
\left(\partial_{\mu}V(x)\right)A_{\mu}^{(jl)}(x)V(x)^\dagger +
V(x)A_{\mu}^{(jl)}(x)\partial_{\mu}V(x)^\dagger \Big]. \nn \eea
For the transformations generated by $V_{\mu}(x)=
e^{ia\Lambda_{\mu}(x)}$, Eq.(\ref{oa_correction}) is exact and
produces no higher order corrections of $a$. The transformation
rules of $\Phi$ and $\eta$, Eq.(\ref{gauge_phieta}), are also
exact. Under the gauge transformation Eq.(\ref{oa_correction})
with the ${\cal O}(a)$-corrections retained, one can check that
the first line in the right-hand side of Eq.(\ref{muDD}) is gauge
invariant (up to irrelevant ${\cal O}(a^5)$-terms).

In the continuum limit, the ${\cal O}(a^3)$ terms in
Eq.(\ref{muDD}, \ref{DDD}) dominate over the Yang-Mills
contribution Eq.(\ref{munuD}) of ${\cal O}(a^4)$. The ${\cal
O}(a^3)$ terms originate from the trilinear coupling of
Eq.(\ref{continuumSN}). To try to resolve it, if we assumed the
mass dimension of $\eta$ being two and rescaled as $\eta \limit
a\eta$ in Eqs.(\ref{muDD}, \ref{DDD}), the contributions would
start from ${\cal O}(a^4)$ and $\eta_{ijkl}$ would become an
auxiliary field imposing the constraint $(A_{\mu}^{(jl)})_{ik} =
0$. In this case, however, the theory is not Lorentz invariant
because of the quartic interactions of $H_{\mu}^{(jl)}$ that
remain in Eq.(\ref{muDD}). Therefore, the classical continuum
limit of the dimensionally reduced action does not yield a
physically meaningful theory.

%%%%%%%%%%%%%%%%%%%%%%%% Appendix C %%%%%%%%%%%%%%%%%%%%%%%%%%%%%%%%
\section{$ {\cal O}(N^3)$ solution for rank-$N^2$(I) compactness condition }
\setcounter{equation}{0}
Is it possible to find a solution for the rank-$N^2$(I)
compactness conditions Eqs.(\ref{N2-1}-\ref{N2-4}), whose degrees
of freedom are of order ${\cal O}(N^3)$ at large $N$? In this
appendix, we demonstrate that such a solution can be constructed,
and hence demonstrating that Eqs.(\ref{N2-1}-\ref{N2-4}) by
themselves are not so restrictive by themselves. What actually
renders the dynamical degrees of freedom reduced further to ${\cal
O}(N^2)$ is the orientation condition (\ref{orientation}). We also
demonstrate that, in the naive continuum limit, the gauge degrees
of freedom is again reduced to $\left({\rm U}(1)\right)^N$.

Two of the compactness conditions, Eqs.(\ref{N2-1}, \ref{N2-3}),
suggests that the plaquete variables are parametrizable as
$U_{\mu\nu}(x) = \exp[ia^2 T_{\mu\nu}(x)]$, where $T_{\mu\nu}(x)$
are $N^2\times N^2$ hermitian matrix-valued tensor fields. For the
moment, for notational simplicity, we shall suppress the indices
$\mu, \nu$ and $x$. Then, the hermiticity of the tensor field $T$
is expressed as
\bea
T_{ijkl}^*=T_{klij}.
\label{T_zeroth}
\eea
The other two of the compactness conditions, Eqs.(\ref{N2-2},
\ref{N2-4}), put further constraints, which we shall consider
order-by-order in the power-series expansion of the plaquette
variables:
\bea U_{ijkl}= \delta_{ik}\delta_{jl} +(ia^2)T_{ijkl}
+\frac12\left(ia^2\right)^2\left(T^2\right)_{ijkl} +\cdots.
\nonumber \eea
Begin with the condition Eq.(\ref{N2-2}). At the order ${\cal
O}(a^2)$, it is satisfied by Eq.(\ref{T_zeroth}). The order ${\cal
O}(a^4)$ puts the conditions
\bea
\sum_k \Big[\, T^{(ik)}, T^{(ki')}\,\Big] =0.
\label{T_4th}
\eea
Here, each of $T^{(ik)}$ is a $(N\times N)$ matrix-valued tensor
field with the notation
\bea \left(T^{(ik)}\right)_{jl}\equiv T_{ijkl}. \nonumber \eea
Then the hermiticity condition, Eq.(\ref{T_zeroth}), is expressed
as
\bea
T^{(ik) \dagger}=T^{(ki)}.
\label{T_zeroth2}
\eea
At next order, ${\cal O}(a^6)$, no new constraints arise since
Eqs.(\ref{T_zeroth2}, \ref{T_4th}) solve automatically up to this
order. At the order ${\cal O}(a^8)$, we get the constraints:
\bea
 & & \sum_{m,n,k}\left( T^{(im)}T^{(mn)}T^{(nk)}T^{(ki')} -2T^{(ki')}T^{(im)}T^{(mn)}T^{(nk)}
\right. \nn \\
 & & \, \left. -2T^{(mn)}T^{(nk)}T^{(ki')}T^{(im)} + 3T^{(nk)}T^{(ki')}T^{(im)}T^{(mn)}\right)
=0.
\label{T_8th}
\eea
We did not find general solutions for Eq.(\ref{T_8th}).
Nevertheless, it is easy to see that the following two cases solve
Eq.(\ref{T_8th}) as well as the conditions for the all orders of
$a^2$:
\bea
({\rm A}): & &
%\Big[\, T^{(ik)}, \, T^{(mn)}\, \Big] =0 \quad
%\mbox{for} \quad ^\forall i,k,m,n \nn \\
% & & \mbox{i.e. }
T^{(ik)}=
V\left(\begin{array}{ccc}t^{(ik)}_1 &  & \\ & \ddots &  \\ &  &  t^{(ik)}_N \end{array}
\right)V^{-1}
\label{case_A} \\
({\rm B}): & & T^{(ik)}= \sum_m V_{im}t^{(m)} (V^{-1})_{mk},
\label{case_B}
\eea
where in the case (A) $(t^{(ik)}_n)^*= t^{(ki)}_n$, and in the
case (B) each of $t^{(m)}$ is a $(N \times N)$ hermitian matrix.
The case (A) means that $(T^{(ik)})_{jl}$ are simultaneously
diagonalizable with respect to the indices $j, l$, while the case
(B) requires the diagonalizable structure for the indices $i, k$.

Conditions that arise at higher order ${\cal O}(a^{2K})$ take the
form
\bea
 & & \sum_{p_1, \cdots, p_{K-1}}T^{(ip_1)}T^{(p_1p_2)}\cdots T^{(p_{K-1}i')}
+(-1)^K\sum_{q_1, \cdots,q_{K-1}}T^{(iq_1)}T^{(q_1q_2)}\cdots T^{(q_{K-1}i')} \nn \\
 & & +\sum_{\scriptsize{\begin{array}{c} n,m\geq 1 \\ n+m=K \end{array}}}
(-1)^m\frac{K!}{n!m!}\sum_{\scriptsize{\begin{array}{r} k,q_1,\cdots, q_{m-1},\\ p_1, \cdots, p_{n-1}
\end{array}}}
T^{(kq_1)}T^{(q_1q_2)}\cdots T^{(q_{m-1}i')}T^{(ip_1)}T^{(p_1p_2)}\cdots T^{(p_{n-1}k)}  \nn \\
 & & =0.
\label{T_Kth}
\eea
It is a straightforward calculation to see that both of the cases
(A) and (B) solve Eq.(\ref{T_Kth}). Similarly, it is easy to check
that the cases (A) and (B) are solutions for Eq.(\ref{N2-4}) at
the same time. Note that in the cases (A) and (B), $t^{(ik)}_n$
and $t^{(m)}_{ij}$ have the ${\cal O}(N^3)$ degrees of freedom
respectively. ($V$ has the degrees of freedom of sub-leading order
${\cal O}(N^2)$.)

The plaquette variables are expressed as
\bea U_{\mu\nu}=\exp\left[ia^2B_{\mu\nu}\right], \qquad
B_{\mu\nu}=\sum_A B_{\mu\nu}^A T^A, \qquad B_{\mu\nu}^A\in {\bf
R}, \nonumber \eea
where $T^A$ are linearly independent basis taking the form
Eq.(\ref{case_A}) or Eq.(\ref{case_B}). The orientation condition
Eq.(\ref{orientation}) means
\bea -\sum_A B_{\mu\nu}^A\left(T^{(ki) A}\right)_{lj} = \sum_A
B_{\nu\mu}^A\left(T^{(lj) A}\right)_{ki}. \nn\eea
Note that it relates the upper indices $k, i$ to the lower indices $l, j$ for
$\left(T^{(ki) A}\right)_{lj}$.
For the case (A), it reads
\bea -\sum_A B_{\mu\nu}^A\, \sum_mV_{lm}(V^{-1})_{mj}\,t^{(ki)
A}_m = \sum_A B_{\nu\mu}^A\, \sum_mV_{km}(V^{-1})_{mi}\,t^{(lj)
A}_m \nn\eea
which means the diagonal structure also for the indices $k, i$:
\bea t^{(ki) A}_n \propto \sum_m t_{nm}^A V_{km}
\left(V^{-1}\right)_{mi} \nn\eea
with $t^A_{nm}=t^A_{mn}$. The degrees of freedom reduce to ${\cal O}(N^2)$.
Similarly, for the case (B), it requires the diagonal structure for the lower indices $l, j$
reducing the degrees of freedom to ${\cal O}(N^2)$.

Finally, because the gauge transformation rule Eq.(\ref{gaugetr})
ought to keep the diagonal structure in the cases (A) and (B), we
should consider the gauge group $({\rm U}(1))^N$, a subgroup of
${\rm U}(N)$. Thus, the solutions do not lead to an interesting
nonabelian tensor theory. Also, since the gauge degrees of freedom
do not have enough degrees of freedom to kill the modes of
wrong-sign kinetic term (ghosts), the solutions do not seem to
lead physically meaningful continuum theory at least in the
classical level.

%%%%%%%%%%%%%%%%%%%%%%%% Appendix D %%%%%%%%%%%%%%%%%%%%%%%%%%%%%%%%%
\section{Character expansion coefficient $C_{000}(\beta)$}
\setcounter{equation}{0}

\subsection{evaluation of generating function}
Here, we compute $C_{000}(\beta)$ we needed in the text for
obtaining results for the partition function and the correlators
among Wilson surface operator. Using Eq.(\ref{CRRR}), we can
express $C_{000}(\beta)$ as
\bea C_{000}(\beta) =
\sum_{n=0}^{\infty}\frac{(3\beta)^{2n}}{(n!)^2} \left[\int [\dd U]
\left(\tr\,U\tr\,U^\dagger\right)^n \right]^3. \label{C000} \eea
To compute the U($N$) group integral in the right-hand-side,
consider the following generating function:
\bea
z(\gamma) & \equiv & \int [\dd U]\, e^{\gamma\, \tr (U+U^\dagger)} \nn \\
 & = & \sum_{n=0}^{\infty}\frac{\gamma^{2n}}{(n!)^2}\,
\int [\dd U]\,\left(\tr\,U\tr\,U^\dagger\right)^n.
\label{zgammadef} \eea
The large-$N$ behavior of this integral, with the 't Hooft
coupling $\lambda \equiv N/\gamma$ fixed, was investigated in
\cite{GrossWitten}. Here, we will extend it to finite $N$.
%(For discussions about finite $N$ corrections, see Refs. \cite{Bars}.).
%
Making use of the character expansion:
\bea e^{\gamma\, \tr\,U} = \sum_\ttr \,\zeta_\ttr
(\gamma)\,\chi_\ttr (U) \qquad \mbox{and} \qquad e^{\gamma\,
\tr\,U^\dagger} = \sum_\ttr \,\zeta_\ttr (\gamma)\,\chi_\ttr
(U^\dagger) \label{zeta} \eea
and the character orthogonality relations, we find that $
z(\gamma) = \sum_\ttr \,\zeta_\ttr (\gamma)^2. $ We shall first
evaluate $\zeta_\ttr (\gamma)$ by carrying out the integral
\bea \zeta_\ttr (\gamma) = \int [\dd U]\, \chi_\ttr (U^\dagger)\,
e^{\gamma \,\tr \, U}. \label{zeta_inv} \eea
Any unitary irreducible representation is labelled by the ordered
integers $m_1 \geq m_2 \geq \cdots \geq m_N$, running over all
integer values while keeping the order. Note, however, that the
representations that appear in Eq.(\ref{zeta}) with nonzero
$\zeta_R(\gamma)$ are restricted to nonnegative $m_i$'s, i.e. $m_1
\geq m_2 \geq \cdots \geq m_N \geq 0$. When $U$ is diagonalized in
the form
\bea U=\Omega \,\left(\begin{array}{ccc}
e^{i\phi_1} &        &          \\
            & \ddots &          \\
            &        & e^{i\phi_N}
\end{array}\right)\, \Omega^\dagger, \qquad \Omega \in {\rm SU}(N)
\nn \eea
in terms of the shifted weights $h_i \equiv N-i +m_i$, the
character $\chi_\ttr (U)$ is expressed as
\bea \chi_\ttr (U) =
\frac{\det_{j,k}\left(e^{ih_j\phi_k}\right)}{\Delta(\phi)},
\nonumber \eea
where
\bea \Delta(\phi) \equiv \det_{j,k}\left(e^{i(N-j)\phi_k}\right) =
\prod_{j<k}\left(e^{i\phi_j} - e^{i\phi_k}\right). \nn \eea
Also, the measure becomes
\bea [\dd U] = \frac{1}{N!}\,\left(\prod_{i=1}^N \frac{\dd
\phi_i}{2\pi}\right)\,
           \Delta(\phi)\,\Delta(\phi)^* \, \dd\Omega
\nn \eea
with the normalization for the angular integral as $\int \dd
\Omega = 1$. Substituting these into Eq.(\ref{zeta_inv}), after
some algebra, we arrive at
\bea \zeta_R(\gamma) = \prod_i \gamma^{m_i}\,
\det_{l,j}\frac{1}{(l-j+m_j)!} = \prod_i
\gamma^{m_i}\,\frac{1}{m_1!\cdots m_N!}\,
\prod_{j<l}\left(1-\frac{m_l}{m_j+l-j}\right), \nn\eea
and hence
\bea z(\gamma) = \prod_{m_1\geq  m_2 \geq \cdots \geq m_N \geq 0}
\,\gamma^{2m_i}\,\frac{1}{(m_1!)^2\cdots (m_N!)^2}\,
\prod_{j<l}\left(1-\frac{m_l}{m_j+l-j}\right)^2. \label{zgamma0}
\eea
\subsection{asymptotic behavior}
By computing the generating function $z(\gamma)$ explicitly up to
${\cal O} (\gamma^{10})$, we were convinced that it is expressible
as
\bea z(\gamma) = \sum_{n=0}^N\,\frac{\gamma^{2n}}{n!} +
\sum_{n=N+1}^{\infty}\, \frac{v_n}{n!}\, \gamma^{2n},
\label{zgamma} \eea
where $0 < v_n < 1$. In other words, the first $N$ terms coincide
with those of $e^{\gamma^2}$, but remaining higher terms are
suppressed compared to the corresponding terms of $e^{\gamma^2}$.
At ${\cal O}(\gamma^{2n})$, each term contributing to
Eq.(\ref{zgamma0}) corresponds to a way of partition of the number
$n$ into $N$. For $n\leq N$, the number of such terms is just the
number of partition of $n$, but as $n$ goes over $N$ it becomes
smaller than the partition number. This is a reason why the
suppression factor arises. In fact, the first few terms of $v_n$
takes the form:
\bea
v_{N+1} & = & 1 - \frac{1}{(N+1)!}, \nn \\
v_{N+2} & = & 1 - \frac{N+1}{(N+2)N!} - \frac{1}{(N+2)!}, \nn \\
v_{N+3} & = & 1 - \frac14\frac{(N+1)(N+2)}{(N+3)N!}
-\frac14\frac{N(N+3)}{(N+1)(N+2)(N-1)!} -\frac{N+2}{(N+3)(N+1)!} \nn \\
 & &  -\frac{1}{(N+3)!}. \nn
\eea
As increasing the order, $v_n$ gets smaller. In fact,
Eq.(\ref{zgamma}) is consistent with the large-$N$ result for the
strong coupling phase studied in \cite{GrossWitten}
\bea \lim_{N\limit\infty}\frac{1}{N^2}\ln z(N/\lambda)=
\frac{1}{\lambda^2} \qquad \mbox{for } \lambda > 2. \nn \eea

From the power-series Eq.(\ref{zgamma}), we find that
\bea \int [\dd U]\,\left(\tr\,U\tr\,U^\dagger\right)^n = \left\{
\begin{array}{cl} n!    & \mbox{for } n\leq N \\
                 v_n\, n! & \mbox{for } n \geq N+1,
\end{array} \right.
\nn\eea
and obtain the result
\bea C_{000}(\beta) = \sum_{n=0}^N \,n!\, (3\beta)^{2n} + \sum_{n
= N+1}^{\infty}\, (v_n)^3 \,n!\, (3\beta)^{2n}. \label{C000appC}
\eea
Due to the cubic power of the $U$-integral term in
Eq.(\ref{C000}), the expansion coefficient grows as $n!$ up to
${\cal O}(\beta^{2n})$. It is quite different from the ordinary
lattice gauge theory case, where the corresponding quantity in
$z(\gamma)$ has behavior of $1/n!$.

%%%%%%%%%%%%%%%%%%%% Table 1 %%%%%%%%%%%%%%%%%%%%%%%%%%%%%%%%%%%%
\begin{table}
\begin{center}
\begin{tabular}{|c|c||c|c|c|c|c|c|} \hline \hline
\multicolumn{2}{|c|}{$N=2$} &
\multicolumn{6}{c|}{$N=3$} \\
\cline{1-2} \cline{3-8}
$n$    & $(v_n)^3 n!$ & $n$ & $(v_n)^3 n!$ & $n$ & $(v_n)^3 n!$ &
$n$ & $(v_n)^3 n!$  \\ \cline{1-2} \cline{3-8}
$3$   & $3.47  $         & --    & ---       & $21$ &$ 698000 $ &
$39$ & $12.4$ \\
$4$  & $4.76   $         & $4$    &$ 21.1 $   & $22$ & $622000 $ &
$40$ & $4.20$ \\
$5$  & $5.15  $          & $5$     &$ 75.9 $   & $23$ & $518000 $&
$41$ & $1.37$ \\
$6$  & $4.44$            & $6$         &$ 224 $  & $24$ & $404000 $&
$42$ & $0.427$ \\
$7$  & $3.11$            & $7$         &$ 829 $   & $25$ & $296000 $&
$43$ & $0.128$ \\
$8$  & $1.80$            & $8$         &$ 2410 $   & $26$ & $205000$ &
$44$ & $0.0369$ \\
$9$  & $0.872$           & $9$         &$ 6380 $  & $27$ & $ 133000$ &
$45$ & $0.0102$ \\
$10$ & $0.360$           & $10$         &$ 15300 $ & $28$ & $81500$ &
$46$ & $0.00273$ \\
$11$ & $0.128$           & $11$         &$ 33400 $ & $29$ & $47300$ &
$47$ & $0.000700$ \\
$12$ & $0.0392$          & $12$         &$ 66400 $ & $30$ & $26000$ &
$48$ & $0.000173$ \\
$13$ & $0.0106$          & $13$         &$ 120000 $ & $31$ & $13500$ &
$49$ & $0.0000414$ \\
$14$ & $0.00252$         & $14$        &$ 199000 $  & $32$ & $6680$ &
$50$ & $9.53 \times 10^{-6}$\\
$15$ & $0.000533$        & $15$         &$ 302000 $  & $33$ & $3140$ &
  & \\
$16$ & $0.000101$        & $16$         &$ 422000 $ & $34$ & $1400$ &
  & \\
$17$ & $0.0000172$       & $17$         &$ 544000 $  & $35$ & $597$ &
  & \\
$18$ & $2.66\times 10^{-6}$   & $18$    &$ 647000 $ & $36$ & $242$ &
  & \\
$19$ & $3.73 \times 10^{-7}$  & $19$    &$ 713000 $ & $37$ & $94.0$ &
  & \\
$20$ & $4.78 \times 10^{-8}$  & $20$    &$ 731000 $ & $38$ & $34.9$ &
  & \\
\hline \hline
\end{tabular}
\end{center}
\caption{\sl Numerical values of the coefficients $(v_n)^3 n!$ in
$C_{000}(\beta)$ (\ref{C000appC}) up to the first 20 and 50 orders
for $N=2$ and 3, respectively. We took three significant figures
for each value. } \label{vncube}
\end{table}
%%%%%%%%%%%%%%%%%%%%%%%%%%%%%%%%%%%%%%%%%%%%%%%%%%%%%%%%%%%%%%%%%
\subsection{Convergence of $C_{000}(\beta)$}

To understand the large-order behavior of the second sum in
Eq.(\ref{C000appC}) for finite $N$, we examined the first 20 and
50 terms in the cases $N=2$ and 3, respectively. The result is
tabulated in Table \ref{vncube}.

For $N=2$, the value saturates at $n=5$ and then decreases.
Suppression rate becomes stronger with increasing order. For
$N=3$, the behavior is similar, and saturates at $n=20$. Passing
the saturation point, the value decrease rapidly. For both cases,
the asymptotic behavior shows clearly that that series converges.
However, the convergence should be considered typical only for
finite $N$. As $N$ gets larger, the saturation point shifts to a
larger $n$ quickly. For infinite $N$, the series is dominated by
the first sum, which is neither convergent nor Borel summable.

%%%%%%%%%%%%%%%%%%%%%%%%%%%%%%%%%%%%%%%%%%%%%%%%%%%%%%%%%%%%%%%%%%%%


\begin{thebibliography}{99}

\bibitem{kalbramond}
M.~Kalb and P.~Ramond,
{\em Classical Direct Interstring Action},
Phys.\ Rev.\ D {\bf 9} (1974) 2273.
%%CITATION = PHRVA,D9,2273;%%

\bibitem{polchinski}
J.~Polchinski, {\em Dirichlet-Branes and Ramond-Ramond Charges},
Phys.\ Rev.\ Lett.\  {\bf 75}, 4724 (1995) [arXiv:hep-th/9510017].
%%CITATION = HEP-TH 9510017;%%

\bibitem{alvarez} 
O. Alvarez, L.A. Ferreira and J.S\'anchez Guill\'en, 
{\em A new approach to integrable theories in any
dimension}, Nucl.\ Phys.\ B {\bf 529}, 689 (1998)
[arXiv:hep-th/9710147].
%%CITATION = HEP-TH 9710147;%%

\bibitem{stringnet}
S.~J.~Rey, {\em The Higgs Mechanism For Kalb-Ramond Gauge Field},
Phys.\ Rev.\ D {\bf 40}, 3396 (1989);\\
%%CITATION = PHRVA,D40,3396;%%
M.A. Levin and X.G. Wen, 
{\sl String net condensation: a physical mechanism for topological phases},
Phys. Rev. B {\bf 71}, 045110 (2005); ibid. {\sl Detecting topological order in a ground state wavefunction}, Phys. Rev. Lett. 100, 030502 (2008);\\
H. Bombin and M.A. Martin-Delgado,
{\sl Exact topological order in D=3 and beyond}, Phys. Rev. B {\bf 75}, 075103 (2007);\\
L. Fidkowski, M. Freedman, C. Nayak, K. Walker and Z. Wang, {\sl From string nets to nonabelions},
arXiv:cond-mat/0610583.

\bibitem{errorcorrection}
A. Kitaev, {\sl Fault-tolerant quantum computation by anyons}, Ann. Phys. {\bf 303}, 2 (2003);\\
M. Freedman, A. Kitaev, M. Larsen and Z. Wang, {\sl Topological quantum computation}, Bull. Amer. Math. Soc. (N.S.) {\bf 40}, 31 (2003);\\
H. Bombin and M.A. Martin-Delgado, {\sl Statistical mechanical models and topological color codes}, arXiv:0711.0468;\\
Y. Zhang, L.H. Kaufmann and M.L. Ge, {\sl Universal quantum gate, Yang-Baxterization and Hamiltonian}, 
Int. J. Quant. Inform. {\bf 3}, 669 (2005).
%Y. Zhang, {\sl Quantum error correction code in the Hamiltonian formulation}, arXiv:0801.2561.

\bibitem{quantumcomputation}
P.W. Shor, {\sl Schemes for reducing decoherence in quantum memory}, Phys. Rev. A {\bf 52}, 2493 (1995);\\
A. Calderbank, E. Rains, P.W. Shor and N. Sloane, {\sl Quantum error correction and orthogonal geometry}, Phys. Rev. Lett. {\bf 78}, 405 (1997);\\
J. Preskill, {\sl Reliable quantum computers}, Proc. R. Soc. London A {\bf 454}, 385 (1998);\\
E. Dennis, A. Kitaev, A. Landahl and J. Preskill, {\sl Topological quantum memory}, J. Math. Phys. {\bf 43}, 4452 (2002).

\bibitem{parisi}
H.~G.~Ballesteros, L.~A.~Fernandez, V.~Martin-Mayor, A.~Munoz
Sudupe, G.~Parisi and J.~J.~Ruiz-Lorenzo, {\em The four
dimensional site-diluted Ising model: A finite-size scaling
study}, Nucl.\ Phys.\ B {\bf 512}, 681 (1998)
[arXiv:hep-lat/9707017].
%%CITATION = HEP-LAT 9707017;%%

\bibitem{wittencp} E. Witten,
{\em Bound states of strings and p-branes}, Nucl.\ Phys.\ B {\bf
460}, 335 (1996) [arXiv:hep-th/9510135].
%%CITATION = HEP-TH 9510135;%%

\bibitem{20theory}
E.~Witten,
{\sl Some comments on string dynamics},
  arXiv:hep-th/9507121;
  %%CITATION = HEP-TH/9507121;%%
ibid.,
{\sl Five-brane effective action in M-theory},
  J.\ Geom.\ Phys.\  {\bf 22} (1997) 103
  [arXiv:hep-th/9610234].
  %%CITATION = JGPHE,22,103;%%

\bibitem{tensionless}
 A.~Strominger,
{\sl Open p-branes},
  Phys.\ Lett.\  B {\bf 383} (1996) 44
  [arXiv:hep-th/9512059];\\
A.~Losev, G.~W.~Moore and S.~L.~Shatashvili, {\em M \& m's},
Nucl.\ Phys.\ B {\bf 522} (1998) 105 [{\tt hep-th/9707250}];
\\
O.~Aharony,
{\em A brief review of 'little string theories'},
Class.\ Quant.\ Grav.\  {\bf 17} (2000) 929
[{\tt hep-th/9911147}];\\
%%CITATION = HEP-TH 9911147;%%
D.~Kutasov,
{\sl Introduction To Little String Theory},
Published in "Trieste 2001, Superstrings and related matters", pp. 165-209
(World Scientific, Singapore, 2001).

\bibitem{klebanovtseytlin}
I.~R.~Klebanov and A.~A.~Tseytlin,
{\sl Entropy of Near-Extremal Black p-branes},
  Nucl.\ Phys.\  B {\bf 475} (1996) 164
  [arXiv:hep-th/9604089].

\bibitem{mooreanomaly}
D.~Freed, J.~A.~Harvey, R.~Minasian and G.~W.~Moore,
{\sl Gravitational anomaly cancellation for M-theory fivebranes},
  Adv.\ Theor.\ Math.\ Phys.\  {\bf 2} (1998) 601
  [arXiv:hep-th/9803205];\\
  %%CITATION = 00203,2,601;%%
J.~A.~Harvey, R.~Minasian and G.~W.~Moore,
{\sl Non-abelian tensor-multiplet anomalies},
  JHEP {\bf 9809} (1998) 004
  [arXiv:hep-th/9808060].
  %%CITATION = JHEPA,9809,004;%%

\bibitem{n4thermo}
S.~S.~Gubser, I.~R.~Klebanov and A.~A.~Tseytlin,
{\sl Coupling constant dependence in the thermodynamics of N = 4  supersymmetric
Yang-Mills theory}, 
  Nucl.\ Phys.\  B {\bf 534} (1998) 202
  [arXiv:hep-th/9805156];\\
  %%CITATION = NUPHA,B534,202;%%
 A.~Fotopoulos and T.~R.~Taylor,
{\sl Comment on two-loop free energy in N = 4 supersymmetric Yang-Mills  theory
at finite temperature}, 
  Phys.\ Rev.\  D {\bf 59} (1999) 061701
  [arXiv:hep-th/9811224];\\
  %%CITATION = PHRVA,D59,061701;%%
C.~J.~Kim and S.~J.~Rey,
{\sl Thermodynamics of large-N super Yang-Mills theory and AdS/CFT
  correspondence},
  Nucl.\ Phys.\  B {\bf 564} (2000) 430
  [arXiv:hep-th/9905205].
  %%CITATION = NUPHA,B564,430;%%

\bibitem{Baez} R. Attal, {\em Combinatorics of nonabelian gerbes with
connection
and curvature}, arXiv:math-ph/0203056;\\
J.C. Baez, {\em Higher Yang-Mills theory}, arXiv:hep-th/0206130;\\
H. Pfeiffer, {\em Higher gauge theory and a nonabelian
genralization of 2-form electrodynamics}, arXiv:hep-th/0304074;\\
J.C. Baez and A.S. Crans, {\em Higher-Dimensional Algebra VI: Lie
2-Algebras}, arXiv:math.QA/0307263.;\\
R. Attal, {\em Homotopy and duality in non-abelian lattice gauge
theory}, arXiv:hep-th/0308100;\\
F. Girelli and H. Pfeiffer, {\em Higher gauge theory --
differential versus integral formulation}, arXiv:hep-th/0309173.

\bibitem{others}
E.~T.~Akhmedov,
{\sl Towards the theory of non-Abelian tensor fields. I},
  Theor.\ Math.\ Phys.\  {\bf 145} (2005) 1646
  [Teor.\ Mat.\ Fiz.\  {\bf 145} (2005) 321]
  [arXiv:hep-th/0503234];
  %%CITATION = TMFZA,145,321;%%
ibid.
{\sl Towards the theory of non-abelian tensor fields. II},
  Theor.\ Math.\ Phys.\  {\bf 147} (2006) 509
  [Teor.\ Mat.\ Fiz.\  {\bf 147} (2006) 73]
  [arXiv:hep-th/0506032];\\
  %%CITATION = TMFZA,147,73;%%
A.~Gustavsson,
{\sl The non-Abelian tensor multiplet in loop space},
  JHEP {\bf 0601} (2006) 165
  [arXiv:hep-th/0512341];
  %%CITATION = JHEPA,0601,165;%%
ibid.
{\sl Closed non-abelian strings},
  Nucl.\ Phys.\  B {\bf 814} (2009) 53
  [arXiv:0705.3369 [hep-th]].
  %%CITATION = NUPHA,B814,53;%%

\bibitem{wilson}
K.~G.~Wilson,
{\em Confinement Of Quarks},
Phys.\ Rev.\ D {\bf 10} (1974) 2445.
%%CITATION = PHRVA,D10,2445;%%

\bibitem{chanpaton}
J.~E.~Paton and H.~M.~Chan, {\em Generalized Veneziano Model With
Isospin}, Nucl.\ Phys.\ B {\bf 10} (1969) 516;
%%CITATION = NUPHA,B10,516;%%
\\
N. Marcus and A. Sagnotti, {\em Group theory from `quarks' at the
ends of strings}, Phys. Lett. {\bf B188} (1987) 58.


\bibitem{20theory}
E.~Witten, {\em Five-brane effective action in M-theory}, J.\
Geom.\ Phys.\  {\bf 22}, 103 (1997) [arXiv:hep-th/9610234]
%%CITATION = HEP-TH 9610234;%%
;\\
M. Henningson, Phys. Rev. Lett. {\bf 85} (2000) 5280; ibid. {\em
Commutation relations for surface operators in
six-dimensional (2,0) theory}, JHEP 0103 (2001) 011;\\
R. Dijkgraaf, E. Verlinde and M. Vonk, {\em On the partition sum
of the NS fivebrane}, [arXiv:hep-th/0205281].

\bibitem{rey1}
S.~J.~Rey, {\em The Higgs Mechanism For Kalb-Ramond Gauge Field},
Phys.\ Rev.\ D {\bf 40}, 3396 (1989);
%%CITATION = PHRVA,D40,3396;%%
ibid. {\em The Collective Dynamics And The
Correlations Of Wormholes In Quantum Gravity}, Nucl.\ Phys.\ B
{\bf 319}, 765 (1989);
%%CITATION = NUPHA,B319,765;%%
ibid. {\em The Axion Dynamics In Wormhole Background}, Phys.\
Rev.\ D {\bf 39}, 3185 (1989).
%%CITATION = PHRVA,D39,3185;%%

\bibitem{rey2}
S.~J.~Rey, {\em The Confining Phase Of Superstrings And Axionic
Strings}, Phys.\ Rev.\ D {\bf 43}, 526 (1991).
%%CITATION = PHRVA,D43,526;%%

\bibitem{Onogi}
P.~Orland, {\em Instantons And Disorder In Antisymmetric Tensor
Gauge Fields}, Nucl.\ Phys.\ B {\bf 205} (1982) 107;
%%CITATION = NUPHA,B205,107;%%
\\
R.~B.~Pearson, {\em Phase Structure Of Antisymmetric Tensor Gauge
Fields}, Phys.\ Rev.\ D {\bf 26} (1982) 2013;
%%CITATION = PHRVA,D26,2013;%%
\\
T.~Onogi and S.~Hashimoto,
{\em Gauge theory of antisymmetric tensor fields},
Phys.\ Lett.\ B {\bf 266} (1991) 107.
%%CITATION = PHLTA,B266,107;%%

\bibitem{Polyakov}
A.~M.~Polyakov, {\em Compact Gauge Fields And The Infrared
Catastrophe}, Phys.\ Lett.\ B {\bf 59} (1975) 82;
%%CITATION = PHLTA,B59,82;%%
ibid. {\em Quark Confinement And Topology Of Gauge Groups}, Nucl.\
Phys.\ B {\bf 120} (1977) 429.
%%CITATION = NUPHA,B120,429;%%


\bibitem{Drouffe} For example, see J.~M.~Drouffe and J.~B.~Zuber,
{\em Strong Coupling And Mean Field Methods In Lattice Gauge Theories},
Phys.\ Rept.\  {\bf 102} (1983) 1.
%%CITATION = PRPLC,102,1;%%

\bibitem{EguchiKawai}
T.~Eguchi and H.~Kawai,
{\em Reduction Of Dynamical Degrees Of Freedom In The Large N Gauge
Theory},
Phys.\ Rev.\ Lett.\  {\bf 48} (1982) 1063.
%%CITATION = PRLTA,48,1063;%%

%\bibitem{nishimurareysugino} J. Nishimura, S.-J. Rey, F. Sugino,
%{\em Supersymmetry on a Noncommutative Lattice}, JHEP

\bibitem{GrossWitten}
D.~J.~Gross and E.~Witten, {\em Possible Third Order Phase
Transition In The Large N Lattice Gauge Theory}, Phys.\ Rev.\ D
{\bf 21} (1980) 446;
%%CITATION = PHRVA,D21,446;%%
\\
S.~R.~Wadia, {\em N = Infinity Phase Transition In A Class Of
Exactly Soluble Model Lattice Gauge Theories}, Phys.\ Lett.\ B
{\bf 93}, 403 (1980).
%%CITATION = PHLTA,B93,403;%%

\bibitem{Bars}
Y.~Y.~Goldschmidt, {\em 1/N Expansion In Two-Dimensional Lattice
Gauge Theory}, J.\ Math.\ Phys.\  {\bf 21} (1980) 1842;
%%CITATION = JMAPA,21,1842;%%
\\
%I.~Bars,
%{\em U(N) Integral For Generating Functional In Lattice Gauge Theory},
%J.\ Math.\ Phys.\  {\bf 21} (1980) 2678.
S.~Samuel, {\em U(N) Integrals, 1/N and the Dewit-'t Hooft
Anomalies}, J.\ Math.\ Phys.\  {\bf 21} (1980) 2695.
%%CITATION = JMAPA,21,2695;%%

\bibitem{YB}
See, for example, M. Jimbo, {\em Introduction to the Yang-Baxter
Equation}, in `Braid Groups, Knot Theory and Statistical
Mechanics', Eds. C.N. Yang and M.L. Ge (1990, Singapore, World
Scientific Co.), and references therein.

\bibitem{QYB}
J.L. Gervais and A. Neveu, {\sl Novel triangle relation and
absence of tachyons in Liouville string field theory}, Nucl. Phys.
B {\bf 238} (1984) 125;\\
G. Felder, {\em Conformal field theory and integrable systems
associated with elliptic curves}, in proceedings of the ICM `94,
arXiv:hep-th/9407154;\\
P. Etingof and O. Schiffmann, {\sl Lectures on the dynamical
Yang-Baxter equations}, arXiv:math.QA/9908064;\\
P. Xu, {\em quantum dynamical Yang-Baxter equation over a
nonabelian base}, arXiv:math.QA/0104071;\\
P. Etingof, {\sl On the dynamical Yang-Baxter equation},
arXiv:math.QA/0207008;\\
L. Feher and I. Marshall, {\sl On a Poisson-Lie analogue of the
classical Yang-Baxter equation for self-dual Lie algebras},
arXiv:math.QA/0208159;\\
Y. Suris and A. Veselov, {\em Lax Matrices for Yang-Baxter Maps},
arXiv:math.QA/0304122;\\
J. Donin and A. Mudrov, {\em Dynamical Yang-Baxter equation and
quantum vector bundles}, arXiv:math.QA/0306028;\\
V.F.R. Jones, {\em In and around the origin of quantum groups},
arXiv:math.QA/0309199.


\bibitem{strongcoupling} K. Osterwalder and E. Seiler, {\em
Gauge field theories on the lattice}, Ann. Phys. (N.Y.) {\bf 110}
(1978) 440;\\
E. Seiler, "Gauge theories as a problem of constructive quantum
field theory and statistical mechanics", Lecture Notes in Physics
{\bf 159} (Springer, Berlin, 1982).

\bibitem{eguchikawai}
T.~Eguchi and H.~Kawai, {\em Reduction Of Dynamical Degrees Of
Freedom In The Large N Gauge Theory}, Phys.\ Rev.\ Lett.\  {\bf
48}, 1063 (1982).
%%CITATION = PRLTA,48,1063;%%

\bibitem{durhuus} B. Durhuus, {\em On the structure of gauge invariant
classical observables in lattice gauge theories}, Lett. Math.
Phys. {\bf 4} (1980) 515.

\bibitem{gaussian}
S.~Oda and F.~Sugino, {\em Gaussian and mean field approximations
for reduced Yang-Mills  integrals}, JHEP {\bf 0103}, 026 (2001) [arXiv:hep-th/0011175];\\
%%CITATION = HEP-TH 0011175;%%
F.~Sugino, {\em Gaussian and mean field approximations for reduced
4D supersymmetric  Yang-Mills integral},
JHEP {\bf 0107}, 014 (2001) [arXiv:hep-th/0105284];\\
%%CITATION = HEP-TH 0105284;%%
J.~Nishimura and F.~Sugino, {\em Dynamical generation of
four-dimensional space-time in the IIB matrix  model}, JHEP {\bf
0205}, 001 (2002) [arXiv:hep-th/0111102];\\
%%CITATION = HEP-TH 0111102;%%
J.~Nishimura, T.~Okubo and F.~Sugino, {\em Convergent Gaussian
expansion method: Demonstration in reduced  Yang-Mills integrals},
JHEP {\bf 0210}, 043 (2002) [arXiv:hep-th/0205253].
%%CITATION = HEP-TH 0205253;%%


\bibitem{deconstruction}
N.~Arkani-Hamed, A.~G.~Cohen, D.~B.~Kaplan, A.~Karch and L.~Motl,
{\sl Deconstructing (2,0) and little string theories},
  JHEP {\bf 0301} (2003) 083
  [arXiv:hep-th/0110146].
  %%CITATION = JHEPA,0301,083;%%



\bibitem{randomsurfaceentropy}
  A.~B.~Zamolodchikov,
{\sl On The Entropy Of Random Surfaces},
  Phys.\ Lett.\  B {\bf 117} (1982) 87.
  %%CITATION = PHLTA,B117,87;%%






\end{thebibliography}
\end{document}